\documentclass[
    reprint,
    amsmath,amssymb,
    aps,
    prb,
    superscriptaddress,
]{revtex4-2}

\usepackage{graphicx}
\graphicspath{{fig/}}

\usepackage[dvipsnames,table,xcdraw]{xcolor}
\definecolor{rmpblue}{HTML}{2e3092}
\usepackage[
    colorlinks=true,
    citecolor=rmpblue,
    linkcolor=rmpblue,
    filecolor=magenta,
    urlcolor=rmpblue,
]{hyperref}

\usepackage{booktabs}
\usepackage{tikz}
\usetikzlibrary{arrows.meta,calc,decorations.markings,decorations.pathreplacing}
\usepackage{tikz-3dplot}
\definecolor{capl}{HTML}{0072B2}    %
\definecolor{cthin}{HTML}{D55E00}   %
\definecolor{cspec}{HTML}{009E73}   %

\newcommand{\affilANU}{Nonlinear Physics Center, Research School of Physics, Australian National University, Canberra ACT 2601, Australia}

\newcommand{\affilDUT}{School of Optoelectronic Engineering and Instrumentation Science, Dalian University of Technology, Dalian 116024, P. R. China}

\newcommand{\affilUoM}{Department of Electrical and Electronic Engineering, The University of Melbourne, Victoria 3010, Australia}

\newcommand{\affilFin}{Photonics Laboratory, Physics Unit, Tampere University, Tampere, Finland}

\begin{document}

\title{Hopping of nanoparticles in optical tweezers governed by Mie resonances}

\author{Ivan Toftul}
\thanks{These authors contributed equally to this work.}
\email{toftul.ivan@gmail.com}
\affiliation{\affilANU}

\author{Libang Mao}
\thanks{These authors contributed equally to this work.}
\email{maolibang@dlut.edu.cn}
\affiliation{\affilDUT}
\affiliation{\affilANU}

\author{Sivacarendran Balendhran}
\affiliation{\affilUoM}

\author{Mohammad Taha}
\affiliation{\affilUoM}

\author{Yuri Kivshar}
\affiliation{\affilANU}

\author{Sergey Kruk}
\email{sergey.kruk@tuni.fi}
\affiliation{\affilANU}
\affiliation{\affilFin}

\date{\today}

\begin{abstract}
Optical tweezers have become a standard tool for manipulating microscale and nanoscale particles and probing their local environments. However, complex particle dynamics under optical forces typically require structured light fields, multi-beam traps, or engineered environments.
Here we achieve complex particle dynamics in a single Gaussian-beam optical tweezer. The effect originates from higher-order Mie resonances supported by wavelength-scale particles. In our optical tweezer, small particles in the regime of Rayleigh scattering or the lowest-order dipole-type Mie modes remain confined at the beam center.
By contrast, particles within the range of sizes corresponding to quadrupole-type Mie modes exhibit more complex behavior. In a linearly polarized Gaussian beam, these particles are trapped in a potential with two off-axis equilibria. We observe thermally driven hopping between these equilibria, with the hopping frequency controlled by the laser power. In a circularly polarized Gaussian beam, the particles are confined to a stable orbit and exhibit circular motion driven by the spin (circular-polarization) degree of freedom of the beam, with angular velocity dependent on the laser power.
These results reveal higher-order Mie resonances as an intrinsic mechanism behind complex optical forces. This establishes Mie-resonant nanophotonics as a flexible platform for inducing and controlling complex motion in optical tweezers for nanoparticle manipulation as well as sensing of local environments.
\end{abstract}

\maketitle

\section{Introduction}
\label{sec:intro}

Thermally driven particle hopping between two metastable states over a potential barrier is a fundamental process that underpins a wide range of phenomena such as chemical reactions~\cite{Garcia-Muller2008Oct,Pollak2023Aug}, protein folding~\cite{Chung2009PNAS,Pirchi2011NC,Chung2012S}, nucleation events~\cite{Zhu2023PRE} and mechanical system stability~\cite{Badzey2005N}.  Previous studies have utilized dual-beam optical traps~\cite{McCann1999N,Rondin2017NN,Wu2009PRL,Seol2009PRL,Babic2004EurLett,Simon1992PRL}, plasmonic nanoantennas~\cite{Yoon2018NC,Yoon2020NanoPhot}, and spatially modulated light fields~\cite{Chupeau2020PNAS} to investigate thermally activated particle hopping.
Bistable trapping configuration can also be achieved with the help of nonlinear effects such as optical Kerr nonlinearity~\cite{Jiang2010NatPhys,Zhang2018NL,Mirzaei-Ghormish2025PRA,Zhou2020OL}. Particle motion within optical traps established itself as a probe for local viscosity~\cite{Lee2012OE}, temperature~\cite{Romero-Gonzalez2023OptLaserTech}  as well as external fields~\cite{Wu2009PRL}.
However, these approaches often require complex setups and external modulation, limiting their simplicity and applicability.

\begin{figure}
\centering
  \includegraphics[width=\linewidth]{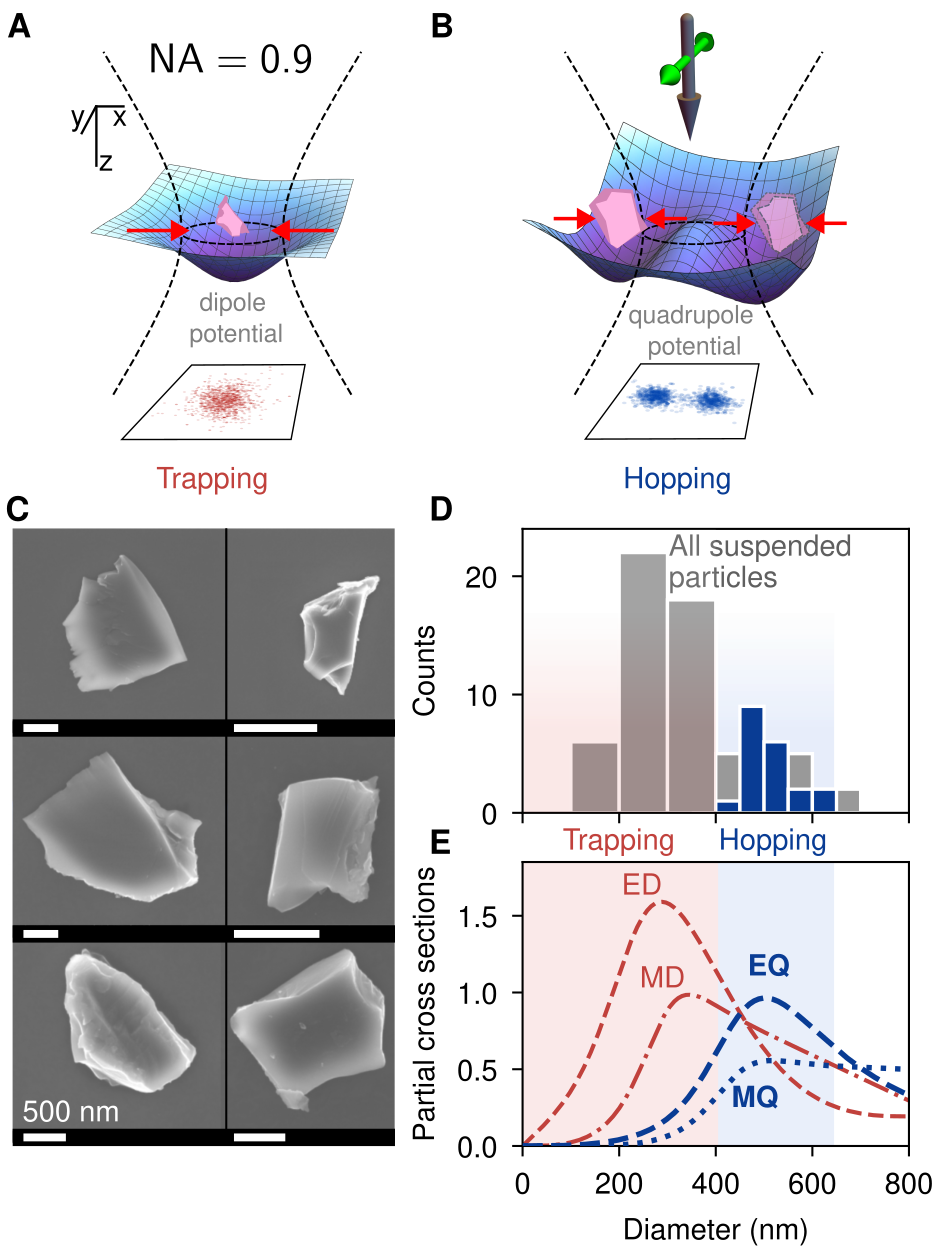}
  \caption{\textbf{Optical hopping of VO$_2$ nanoparticles.}
  (\textsf{\textbf{A}},\,\textsf{\textbf{B}}) Schematic of the two observed regimes. Some particles are confined in a single potential well at the intensity maximum (\emph{trapping}); others experience a quadrupolar potential and hop bistably between two off-axis minima under linear polarization. Point clouds show the recorded particle positions. (\textsf{\textbf{C}}) SEM images of the synthesized VO$_2$ nanoparticles.
  (\textsf{\textbf{D}}) Size distribution from  nanoparticle tracking analysis for all suspended particles and the hopping  subset. (\textsf{\textbf{E}}) Multipole decomposition of the  extinction cross section for an equivalent VO$_2$ sphere ($\varepsilon_{\text{VO}_2} = 6.55 + \mathrm{i}\,3.66$~\cite{Mao2025LRP}, $\lambda = 976$~nm) in ethanol ($\varepsilon_{\text{ethanol}} = 1.85$), normalized to the geometrical cross section. Here electric and magnetic dipole are labeled as ED and MD, electric and magnetic quadrupole are labeled as EQ and  MQ. The hopping-particle size range coincides  with the quadrupole resonances.}
    \label{fig:concept}
\end{figure}

A parallel area of interest is the optical transfer of angular momentum, where \textit{spin} angular momentum (SAM) of light is converted into \textit{orbital} angular momentum of particles, as it enables contactless, tunable nanoscale torques for micromechanical actuation and chiral sensing.
Traditionally, spin-to-orbital conversion has been demonstrated using structured light fields (e.g. vortex beams)~\cite{Zhang2018PRAppl,Gahagan1996OL, Gahagan1999JOSAB, ONeil2002PRL, Garces-Chavez2002PRA,Cao2016OE,Zhang2025NP,Gao2017LSA} or by coupling light to an engineered environment~\cite{Ni2022NL,Triolo2017ACSPhot,Tkachenko2020Optica,Valero2020AdvSci,Wang2011NC}. In these examples spin-to-orbital conversion arises either from the spin-orbit interaction inside the beam itself~\cite{Bliokh2015NP} or via scattering of the environment.
A related effect was reported by Svak \textit{et al.}~\cite{Svak2018NC}: in a weakly focused circularly polarized vacuum trap, a microsphere orbits the axis under a weak handedness-dependent azimuthal force, but with the radial confinement remaining a single on-axis well.
The dynamical stability and limit-cycle character of such orbiting around the on-axis equilibrium were further explored in vacuum traps~\cite{Brzobohaty2023NC,Arita2023CommunPhys}, and the spin-to-orbital momentum conversion was directly visualized through particle trajectories in structured beams~\cite{Arzola2019SciRep}. In all these cases the radial confinement remains a single well positioned on-axis with the orbit radius set by the laser power, whereas a static off-axis equilibrium has not been realized.

Mie resonance, a fundamental phenomenon in light-matter interactions, enables the excitation of high-order modes in high-index nanoparticles~\cite{Tzarouchis2018AS}, leading to complex optical force profiles. Resonances of such particles were observed early on in the radiation pressure on dielectric spheres~\cite{Ashkin1977PRL}. These high-order optical forces, which arise from the interference of multipole moments, can generate spatially varying potential landscapes~\cite{Cao2016Nanoscale,Kislov2021APR,Lepeshov2023PRL} and non-conservative force components~\cite{Xu2020LPR,Chen2015ACSnano,Zhou2023LPR,Nan2023NC,Kiselev2020OE}.
Previous studies have reported force sign changes~\cite{Chen2015ACSnano} and trapping--anti-trapping transitions~\cite{Rockstuhl2004OL,Kislov2021APR,Mao2025LRP} introduced via Mie resonances.
Past theoretical studies addressed the impact of Mie resonances on the trapping strength and stiffness~\cite{Stilgoe2008OE}, as well as the interplay of optical forces between a pair of particles supporting different Mie modes~\cite{Duan2022PRA}.
However, more complex motions such as stochastic hopping between two off-axis potential minima has not been revealed.

In this work, we introduce a novel approach to achieve particle hopping and spin--orbit coupling within an unmodulated single Gaussian beam optical trap by leveraging high-order optical forces arising from quadrupolar Mie resonances.
We identify a Mie-resonant regime of dielectric particles in which the same trap supports two polarization-selected dynamical modes: hopping under linear polarization and orbiting under circular polarization. We further link both behaviors to a unified multipolar-force framework. In the dipole regime the gradient force follows $\nabla|\mathbf{E}|^2$ and is polarization agnostic, giving a single symmetric well for any polarization. The polarization-selected hopping and orbiting we report are therefore a direct fingerprint of beyond-dipole, quadrupole Mie response.

To gain intuitive insight into how quadrupole multipoles in particle's scattering  reshape the trap,  we find that gradient optical trapping force, \( \mathbf{F}^{\text{grad}} = - \boldsymbol{\nabla} U \), in a tightly focused Gaussian beam can be approximated by the following potential (see derivation in Section~\ref{sec:theory} with further details in Appendix~\ref{app:all_terms_forces}):
\begin{equation}
U =  - \frac{U_0}{4} \left(A + B \frac{y^2}{w_0^2} + C \frac{x^2}{w_0^2} \right) e^{- \frac{2 \rho^2}{w_0^2}},
\label{eq:Utheory_intro}
\end{equation}
where $U_0$ depends linearly on the input power, $\rho = \sqrt{x^2 + y^2}$ is the distance to the beam center, $w_0$ is the beam waist, and dipole coefficient $A$ as well as two quadrupole coefficients $B$ and $C$ depend on the particle's shape and permittivity, as well as angle of linear polarization. Hopping direction is \textit{polarization dependent}, and can be either aligned with the beam polarization or transverse to it.
When switched to circular polarization, in addition to the conservative gradient potential, the azimuthal scattering force arises:
\begin{equation}
{F}^{\text{scat}}_{\varphi} \propto   \sigma  \rho \, e^{ - 2 \rho^2/w_0^2},
\label{eq:azimuthalF}
\end{equation}
where $\sigma$ denotes the degree of circular polarization of the input beam. This term shows that for elliptical polarization ($\sigma\neq0$), a nonzero \textit{orbiting torque} drives the particle into orbiting motion around the beam axis. Together, Eqs.~\eqref{eq:Utheory_intro} and \eqref{eq:azimuthalF} capture the essential physics behind polarization dependent hopping and orbiting behavior in a single-beam trap.

In Section~\ref{sec:exp}, we report experimental observation of optical hopping and orbiting of VO$_2$ nanoparticles in a single-beam optical trap. We demonstrate that the type of trapping: either confinement at the beam center or a more complex hopping and orbiting is linked to the particles' sizes, which we further link to their multipolar composition of Mie modes showing single-well trapping for particles supporting dipole resonances and double-well hopping for the particles supporting quadrupole resonances. Through quantitative analysis of particle trajectories, we characterize the dependence of trapping stiffness, transition rate, and barrier height on laser power. We investigate experimentally the polarization-dependent nature of the hopping and orbiting.
In Section~\ref{sec:theory}, we proceed with analytical description of the observed phenomena via high-order optical forces arising from quadrupole Mie resonances supported by the particles. The novelty of this work is the identification and quantitative analysis of a polarization-switchable dynamical regime in a single Gaussian trap, enabled by Mie-resonant multipolar optical forces: linear polarization yields stochastic hopping in an off-axis bistable landscape, whereas circular polarization drives orbiting through a handedness-dependent azimuthal force.

\section{Observation of Mie-resonant hopping and orbiting}
\label{sec:exp}

We optically trap an individual VO$_2$ nanoparticle in ethanol solutions using a tightly focused 976-nm laser. The experimental setup (shown in Fig.~\ref{fig:setup}) and particle synthesis methodology are detailed in Appendices~\ref{app:setup} and~\ref{app:synthesis}. Fig.~\ref{fig:concept}\textsf{\textbf{A}} and Fig.~\ref{fig:concept}\textsf{\textbf{B}} illustrate our key observations. Smaller particles were confined to a single trapping spot at the center of the laser focus (Fig.~\ref{fig:concept}\textsf{\textbf{A}}). However, larger particles exhibited bistable hopping behavior under linear polarization (Fig.~\ref{fig:concept}\textsf{\textbf{B}}).
To study the sizes of particles we first spin-coated a solution with nanoparticles onto a substrate and allowed it to dry out for imaging under a scanning electron microscope (SEM).
Figure~\ref{fig:concept}\textsf{\textbf{C}} shows examples of SEM images of typical VO$_2$ particles.
We proceed with gathering statistical data on particle sizes in a solution using a commercial nanoparticle tracking analysis instrument, ZetaView. The obtained size distribution is shown in Fig.~\ref{fig:concept}\textsf{\textbf{D}} in grey color histogram.
In our optical trapping experiment, we additionally measure the size of each individual particle in the tweezers. For this, we temporarily turn the trapping laser off and record the particle's Brownian motion. The trajectory is analyzed with NanoTrackJ, an ImageJ plugin that extracts the diffusion coefficient and converts it to a hydrodynamic diameter via the Stokes--Einstein relation~\cite{Wagner2014JNanopartRes}.
We show size estimates of the hopping particles as blue histogram in Fig.~\ref{fig:concept}\textsf{\textbf{D}}. We observe the hopping behavior only in a certain size range (approx. $400$~nm to $625$~nm). Smaller particles with sizes less than $400$~nm show consistently stable trapping at the beam center. Particles larger than about 625 nm are consistently repelled off the laser beam.

In Fig.~\ref{fig:concept}\textsf{\textbf{E}} we provide theoretical multipolar decomposition of isotropic VO$_2$ nanoparticles of different sizes using Mie theory~\cite{Bohren1984}. In our multipolar decomposition, we approximate particles as spheres here for simplicity while noting that the underlying Mie resonances are volumetric modes and similar mode compositions are expected for non-spherical geometries (see e.g. Ref.~\cite{Mao2025LRP} for comparison of multipolar composition and optical forces for spherical vs cubic particle and Appendix~\ref{app:shape}). The multipolar analysis reveals that the single-well trapping behavior corresponds to particle sizes supporting dipole-type Mie modes (below $400$~nm size), and the double-well hopping behavior aligns with the size range where quadrupole modes peak ($400$~nm to $625$~nm). The absence of stable trapping for particles larger than 625 nm coincides with the size region beyond the peaks of both the dipolar and quadrupolar Mie modes. We note that similar repelling behavior for VO$_2$ particles larger than 625 nm also agrees with previously reported observations~\cite{Mao2025LRP}.

\begin{figure*}
\centering
  \includegraphics[width=\linewidth]{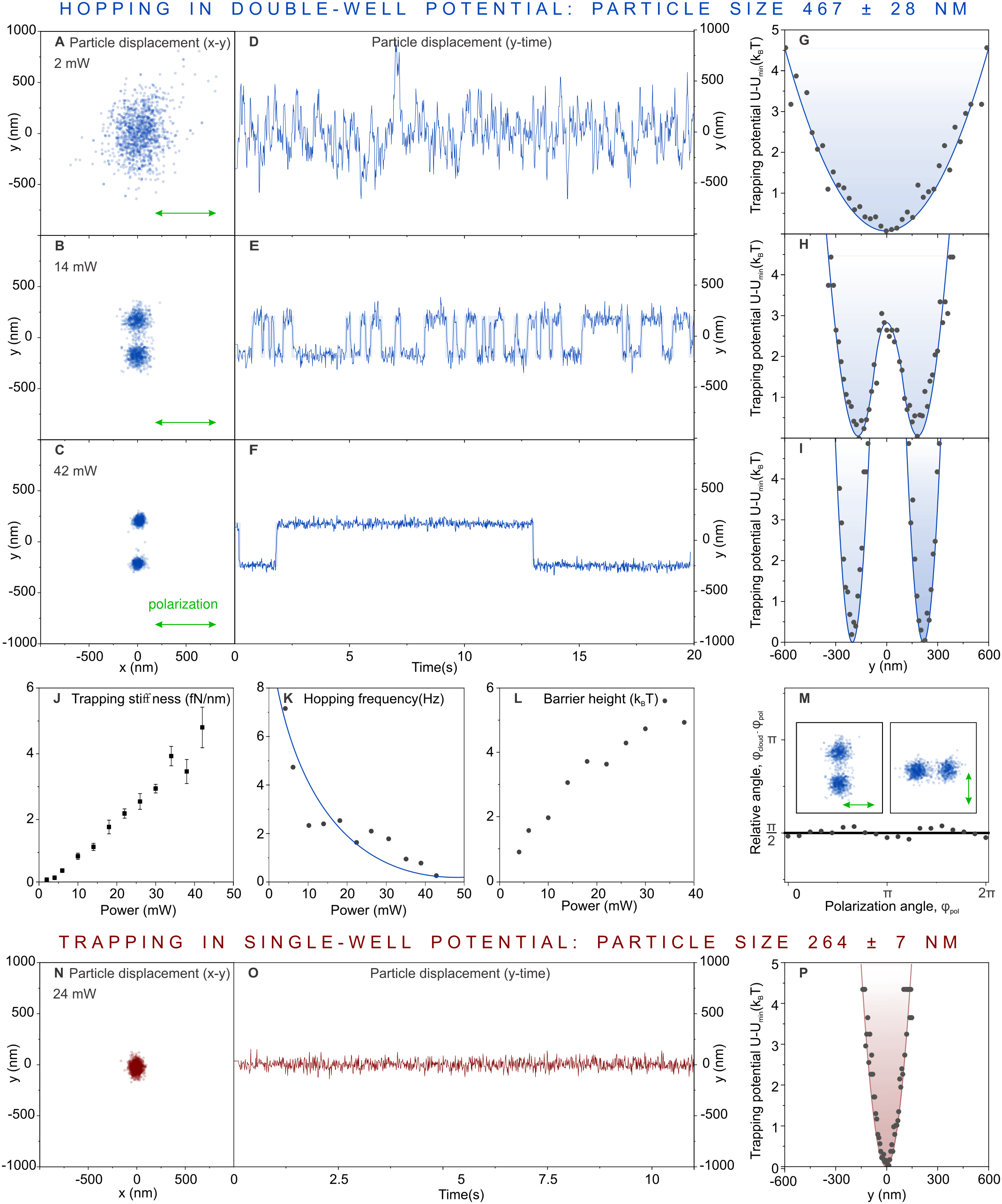}
  \caption{\textbf{Hopping of a larger particle versus stable trapping of a smaller particle}. (\textsf{\textbf{A}}-\textsf{\textbf{C}}) Scatter plots of trapped VO$_2$ particle positions under laser powers of $2$~mW, $14$~mW, and $42$~mW, respectively. The particle size is estimated to be $467 \pm 28$~nm by the nanoparticle tracking analysis. (\textsf{\textbf{D}}-\textsf{\textbf{F}}) Time traces of particle displacement along the y-axis under the corresponding laser powers. (\textsf{\textbf{G}}-\textsf{\textbf{I}}) Trapping potentials along the y-axis for the corresponding laser powers (lines are guides for an eye).
  (\textsf{\textbf{J}}-\textsf{\textbf{L}}) Laser power dependence of trapping stiffness (\textsf{\textbf{J}}), transition rate (\textsf{\textbf{K}}), and barrier height (\textsf{\textbf{L}}). (\textsf{\textbf{M}}) Angle difference between hopping orientation and polarization as a function of polarization angle, showing a consistent difference close to $\pi/2$. The insets are scatter plots when the polarization angle is $0$ and $\pi/2$. (\textsf{\textbf{N}}-\textsf{\textbf{P}}) Control experiment with a smaller particle ($264 \pm 7$~nm) showing no hopping.}
    \label{fig:experiment_transition_to_hopping}
\end{figure*}

To quantitatively analyze the particle hopping phenomenon, we recorded videos (Appendix~\ref{app:movies}) of movements of particles under varying laser powers and tracked their positions frame-by-frame.

We focus on the detailed study of two particles with different sizes and contrasting behavior: a particle with estimated size of $467 \pm 28$~nm which is in the range of quadrupole Mie modes (Fig.~\ref{fig:experiment_transition_to_hopping}\textsf{\textbf{A}}-\textsf{\textbf{M}}) versus a particle with an estimated size of $264 \pm 7$~nm which can support only the lowest-order dipole Mie modes (Fig.~\ref{fig:experiment_transition_to_hopping}\textsf{\textbf{N}}-\textsf{\textbf{P}}).
Fig.~\ref{fig:experiment_transition_to_hopping}\textsf{\textbf{A}}-\textsf{\textbf{C}} depict the bistable trapping via particle position distributions at different laser powers.
At $2$~mW (Fig.~\ref{fig:experiment_transition_to_hopping}\textsf{\textbf{A}}), the particle exhibited Brownian motion around large area forming a single cloud. Under these low-power conditions, the bi-stable trapping is not yet observed. Increasing the laser power to $14$~mW (Fig.~\ref{fig:experiment_transition_to_hopping}\textsf{\textbf{B}}) resulted in the formation of two distinct clouds, corresponding to two trapping points. Further increasing the power to $42$~mW (Fig.~\ref{fig:experiment_transition_to_hopping}\textsf{\textbf{C}}) led to more condensed clouds, suggesting enhanced trapping stiffness. As shown in Fig.~\ref{fig:experiment_transition_to_hopping}\textsf{\textbf{J}}, trapping stiffness increased monotonously with laser power, correlating with particle position variance (see Appendix~\ref{app:stiffness}).

\begin{figure*}
\centering
  \includegraphics[width=0.7\linewidth]{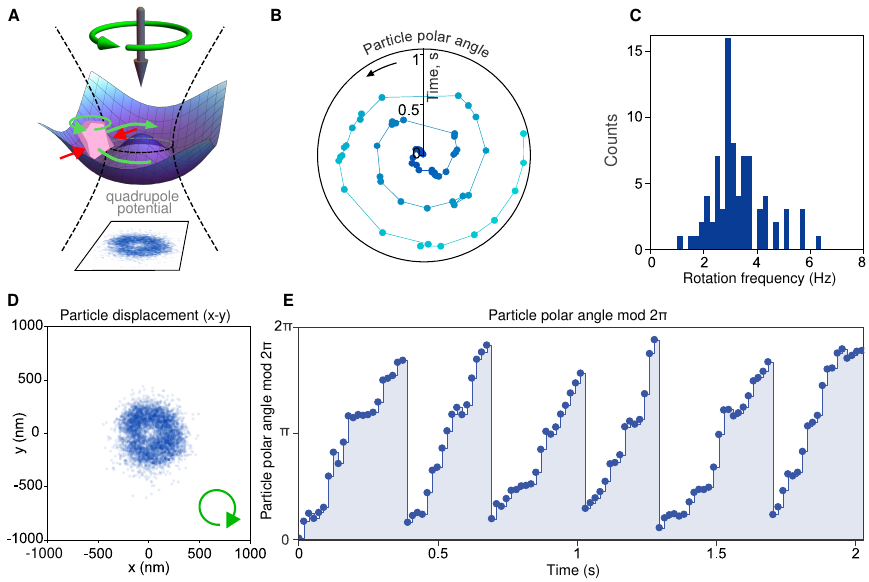}
  \caption{\textbf{Spin--orbit coupling}.
  \textsf{\textbf{A}} Schematics of the orbiting scenario. Particle is trapped in an azimuthally symmetric quadrupole potential under a circularly polarized beam.
  \textsf{\textbf{D}} Scatter plot of the particle under circular polarization, forming a donut-shaped distribution. Particle size is $452\pm33$~nm, laser power is $14$~mW.
  \textsf{\textbf{B}},\textsf{\textbf{E}} Polar angle of the particle as a function of time.
  \textsf{\textbf{C}} Histogram of rotation frequency with an average value of $3.1$~Hz.}
    \label{fig:lin_circ_pol_v1}
\end{figure*}

Fig.~\ref{fig:experiment_transition_to_hopping}\textsf{\textbf{D}}-\textsf{\textbf{F}} present time traces of particle displacement along the $y$-axis. At $2$~mW, the trace varies randomly. At $14$~mW, the trace oscillated between two positions ($y=\pm250$~nm) with frequent transitions, indicating particle hopping. At $42$~mW, while the oscillation remained between the same positions, the average hopping frequency decreased. Fig.~\ref{fig:experiment_transition_to_hopping}\textsf{\textbf{G}}-\textsf{\textbf{I}} illustrate the optical potentials along the y-axis under different laser powers, showing the emergence of the double potential wells with increasing power. The narrowing potential width and increasing barrier height between the potential wells correspond to increased trapping stiffness and decreased transition rate, respectively, which are shown in Fig.~\ref{fig:experiment_transition_to_hopping}\textsf{\textbf{K}}-\textsf{\textbf{L}}.

These changes in hopping behavior with power are captured by the Kramers theory of thermally driven transitions, which predicts that the transition rate \(W^{\text{K}}\) decays exponentially with increasing energy barrier~\cite{Kramers1940, Landauer1961PhysRev, Melnikov1991PhysRep, Rondin2017NN, McCann1999N}:
\begin{equation}
    W^{\text{K}} = W^{\text{K}}_0 \exp\left(- \frac{U_{\text{barr}} - U_{\text{min}}}{k_{\text{B}} T}\right),
    \label{eq:WK}
\end{equation}
where \(U_{\text{barr}} - U_{\text{min}}\) is the potential barrier height, and \(W^{\text{K}}_0\) is a prefactor that depends on the trap curvature at equilibrium points are of a saddle point in the middle~\cite{Landauer1961PhysRev}. In our experiment, the barrier height increases with laser power (Fig.~\ref{fig:experiment_transition_to_hopping}\textsf{\textbf{L}}), resulting in an exponential suppression of the hopping rate, as shown in Fig.~\ref{fig:experiment_transition_to_hopping}\textsf{\textbf{K}}, in agreement with the Kramers model. Kramers transition rates were obtained from the mean dwell time~\cite{McCann1999N} when the particle is trapped in $y<0$ region. The barrier heights were determined through curve fitting using quadrupole potential given by Eq.~\eqref{eq:Utheory} derived in the theory section.

Furthermore, we observed that the hopping direction was dependent on laser polarization. As demonstrated in Fig.~\ref{fig:experiment_transition_to_hopping}\textsf{\textbf{M}}, the angular difference between hopping orientation and polarization remained close to $\pi/2$ through a full rotation of the polarization angle ($0$ to $2\pi$). The insets show that the hopping direction consistently remained perpendicular to the linear polarization orientation. The polarization-induced asymmetry of a tightly focused spot cannot account for this: at high numerical aperture a dipolar particle still sees a single well, slightly elongated \textit{along} the polarization, whereas the observed minima appear \textit{across} it. Both the bistability and its orientation require beyond-dipole multipoles.

Fig.~\ref{fig:experiment_transition_to_hopping}\textsf{\textbf{N}}-\textsf{\textbf{P}} is a control experiment that contrasts the hopping behavior of a larger, quadrupolar particle with the behavior of its smaller counterpart supporting only dipole modes. For the $264 \pm 7$~nm size particle we observe no hopping behavior under $24$~mW laser power. We further varied the laser power from 10 mW to 30 mW confirming the single-well trapping and the absence of hopping behavior. These types of observations were consistent through our experiment where the hopping behavior only appears when the particle size lies in $400$~nm to $625$~nm range (also highlighted in Fig.~\ref{fig:concept}\textsf{\textbf{D}}), which is an evident characteristic of Mie resonant optical forces.

We next study the trapping of quadrupolar particles under circular polarization. In Figure~\ref{fig:lin_circ_pol_v1} we present observations of a particle with size $452 \pm 33$~nm which belongs to the range of sizes with quadrupole Mie resonances. At $14$~mW laser power, the particle position cloud formed a donut shape with a sparse center (see Fig.~\ref{fig:lin_circ_pol_v1}\textsf{\textbf{A}} for schematics and Fig.~\ref{fig:lin_circ_pol_v1}\textsf{\textbf{D}} for the experimental position cloud). Figures~\ref{fig:lin_circ_pol_v1}\textsf{\textbf{B}},\textsf{\textbf{E}} show that the particle polar angle changes consistently, indicating continuous orbital rotation  (Fig.~\ref{fig:lin_circ_pol_v1}\textsf{\textbf{B}} -- polar coordinate representation and Fig.~\ref{fig:lin_circ_pol_v1}\textsf{\textbf{E}} -- Cartesian representation). The rotational direction is directly linked to the handedness of the electric field and is inverted for the circular polarization of the opposite handedness. 
The average rotation frequency at 14 mW power is calculated to be 3.1~Hz shown in the histogram in Fig.~\ref{fig:lin_circ_pol_v1}\textsf{\textbf{C}}.
In the next section and in Appendix~\ref{app:materials} we show that it is possible to have both positive and negative orbiting with respect to the electric field rotation in time, as the only requirement is the conservation of the total angular momentum: negative orbiting results in the positive spinning of the particle around its center of mass.

\section{Multipolar optical forces and spin-orbit coupling mechanism}
\label{sec:theory}

We begin our theoretical analysis with a model based on the paraxial approximation for a Gaussian beam. Although this model is a simplification of our experimental conditions, where we use a high numerical aperture lens, we find it to be in a sufficiently good agreement with the experimental data.

We assume that particle trapping occurs approximately at the focal $xy$-plane located at $z = 0$. Paraxial approximation of the Gaussian monochromatic beam which propagates in $+z$ direction at that plane reads as~\cite{Novotny2012}
\begin{equation}
    \mathbf{E} \propto  \mathbf{u} \, e^{-\frac{\rho^2}{w_0^2}}, \qquad
    \mathbf{H} \propto \mathbf{\hat{z}} \times \mathbf{u} \, e^{-\frac{\rho^2}{w_0^2}},
    \label{eq:gauss}
\end{equation}
where $w_0$ is the beam waist, $\rho^2 = x^2 + y^2$ is the radial distance from the beam center, $\mathbf{\hat{z}}$ is the Cartesian unit vector, and $\mathbf{u}$ is the complex unit polarization vector ($|\mathbf{u}|^2 = 1$).
Polarization state of the beam can be fully characterized by three normalized Stokes parameters ($\tau^2 + \chi^2 + \sigma^2 = 1$)~\cite{Collett2005}
\begin{align}
\begin{split}
    \tau &= |u_x|^2 - |u_y|^2, \\
    \chi &= 2\operatorname{Re}(u_x^* u_y), \\
    \sigma &= 2 \operatorname{Im}(u_x^* u_y).
\end{split}
\label{eq:stokes}
\end{align}
Here and below the time dependence is $e^{-i \omega t}$, so that $\mathbf{u} = (\hat{\mathbf{x}} + i \hat{\mathbf{y}})/\sqrt{2}$ gives $\sigma = +1$, e.g. the electric field spins from $x$ to $y$, which is left-circular polarization and total angular momentum along $+z$.
Parameters~\eqref{eq:stokes} describe the
degrees of the vertical/horizontal, $\pm45^{\circ}$ diagonal, and
right-hand/left-hand circular polarizations, respectively.
Two important cases considered in this work are: (i) linearly polarized beam with angle $\phi$ with respect to the $x$-axis ($\tau = \cos (2\phi)$, $\chi = \sin (2 \phi)$, $\sigma = 0$) and (ii) right/left circularly polarized beam ($\tau = \chi = 0$, $\sigma = \mp 1$).

Mie particles --- particles beyond Rayleigh approximation but not large enough such that ray approximation is not applicable --- can be efficiently described using \textit{multipolar decomposition}~\cite{Bohren1984}.
Consequently, optical force can be also represented in terms of different multipolar contributions~\cite{Yu2019PRA, Zhou2023LPR, Jiang2015arXiv, Chen2011NP,Toftul2023PRL,Toftul2025ACSPhot}.
In general, the behavior of the trapped Mie particles is rather complex as it involves a number of different terms to consider: apart from the multipole-field interaction terms, there are also \textit{recoil} terms which describe  the interference between the neighboring $\ell$-th and $(\ell \pm 1)$-th multipoles~\cite{Toftul2026RMP}.
However, on resonance the corresponding terms can dominate.

Lowest multipole contribution, i.e. electric and magnetic dipoles, have simple radial dependence in the focal plane $z=0$. For instance for the dipole moment $\mathbf{e}$ we have
\begin{align}
    \mathbf{F}_{e} &= \frac{1}{2}\operatorname{Re} \left[ \mathbf{e}^{*} \cdot (\boldsymbol{\nabla}) \mathbf{E}\right]  \propto -  \mathbf{\hat{\boldsymbol{\rho}}} \operatorname{Re}(\alpha_e) \rho \exp\left(-2 \rho^2 /w_0^2\right),
    \label{eq:Fe}
\end{align}
where $\mathbf{e} = \varepsilon \alpha_e  \mathbf{E}$ is the induced dipole moment of the particle with dipole polarizability $\alpha_e$, and $\mathbf{e}^* \cdot (\boldsymbol{\nabla}) \mathbf{E} \equiv \sum_{i=x,y,z}e_i^* \boldsymbol{\nabla} E_i$. In Eq.~\eqref{eq:Fe} we have assumed that gradient trapping forces in $z$ direction are greater than scattering pressure force, thus the latter one is negligible.
This manifests in trapping or anti-trapping behavior depending on the corresponding sign of the real part of dipole polarizability~\cite{Toftul2026RMP,Mao2025LRP}. Magnetic counterpart is going to have similar functional dependence. We stress that dipole contributions are \textit{polarization agnostic}, i.e. there is no dependence on  $\tau$, $\chi$, and $\sigma$ defined in Eq.~\eqref{eq:stokes}.

The situation becomes drastically different for the higher order multipoles.

\subsection{Magnetic quadrupole contribution analyses}

Here we consider only magnetic quadrupole force and show that it has several important features: (i) it is polarization dependent, (ii) it can produce equilibrium points \textit{not} in the beam center, and (iii) it can couple spin degree of freedom of Gaussian beam to the orbital motion of the trapped particle.

Magnetic quadrupole term reads as~\cite{Kislov2021APR,Mao2025LRP}
\begin{align}
    {F}_{Q_{\text{m}}, i} &= \frac{1}{4} \operatorname{Re} \left[ \hat{Q}^{*}_{\text{m},jk} \nabla_{i} \nabla_{k} H_{j} \right], \label{eq:force_quad} \\
    Q_{\text{m},jk} &= \frac{\mu}{2} \alpha_{Q_m} \left( \nabla_j H_k + \nabla_k H_j\right), \label{eq:quad}
\end{align}
where \(\hat{Q}_{\text{m},jk}\) is the magnetic quadrupole moment, \(\mu\) is the magnetic permeability, \(\alpha_{Q_m}\) is the complex magnetic quadrupole polarizability, and \(\mathbf{H}\) is the local magnetic field.
In Eq.~\eqref{eq:force_quad} Einstein summation convention is used.
The force \(\mathbf{F}_{Q_{\text{m}}}\) arises from the change of the momentum flux calculated using interference terms between incident and scattered by the quadrupole fields~\cite{Toftul2026RMP}.
Its decomposition into gradient and scattering components is governed by the real and imaginary parts of \(\alpha_{Q_m}\), respectively.
Specifically, \(F^{\text{grad}}_{Q_{\text{m}}} \propto \operatorname{Re}(\alpha_{Q_m})\) is the conservative part of the force, while the scattering force \(F^{\text{scat}}_{Q_{\text{m}}} \propto \operatorname{Im}(\alpha_{Q_m})\) emerges from momentum transfer due to `radiation friction'.

For the further analysis, we assume a weak \(z\)-dependence for a tightly focused beam with \(k w_0 \approx 2\), i.e., all \(|\mathrm{d} A / \mathrm{d} z| \ll |\mathrm{d} A / \mathrm{d} x|, \: |\mathrm{d} A / \mathrm{d} y |\), where \(A\) is any component of the electric or magnetic fields.
In this rather simple approximation for the incident field given by Eq.~\eqref{eq:gauss}, we find a compact expression for the gradient and scattering parts of the force:
\begin{align}
    \begin{split}
    \mathbf{F}^{\text{grad}}_{Q_{\text{m}}} \propto& \: \: \: \    \mathbf{\hat{\boldsymbol{\rho}}} \, 3\rho (w_0^2- 2 \rho^2)e^{ - \frac{2\rho^2}{w_0^2}}  \\
    & +  \mathbf{\hat{x}}\left[  x (-w_0^2 + 2x^2 - 2y^2) \tau + y(4x^2 - w_0^2) \chi \right] e^{ - \frac{2\rho^2}{w_0^2}} \\
    & + \mathbf{\hat{y}} \left[ y (w_0^2 + 2x^2 - 2y^2) \tau  + x(4y^2  - w_0^2)\chi \right] e^{ - \frac{2\rho^2}{w_0^2}},
    \end{split} \label{eq:FQgrad} \\
    \mathbf{F}^{\text{scat}}_{Q_{\text{m}}} \propto&  \: \: \   -\mathbf{\hat{\boldsymbol{\varphi}}}  \sigma  \rho \, e^{ - 2 \rho^2/w_0^2}, \label{eq:FQscat}
\end{align}
where we have neglected a scattering pressure contribution along $z$-axis.
In Appendix~\ref{app:approx} we compare this result with the true paraxial approximation  of Appendix~\ref{app:gaussian}, and find small discrepancies.
The gradient force \(\mathbf{F}^{\text{grad}}_{Q_{\text{m}}}\) exhibits a complex dependence on the Stokes parameters \(\tau\) and \(\chi\), which characterize the \textit{linear} polarization state of the beam, reflecting its spin-independent nature.
In contrast, the scattering force \(\mathbf{F}^{\text{scat}}_{Q_{\text{m}}}\) is proportional and opposite to \(\sigma\), the degree of circular polarization, which drives the orbital motion of the particle. 
We use the term spin--orbit coupling here in an effective sense: spinning of the electric field vector produces orbital motion of the particle, i.e. it is coupled to the the circular-polarization degree of freedom

\begin{figure}
    \centering
    \includegraphics[width=\linewidth]{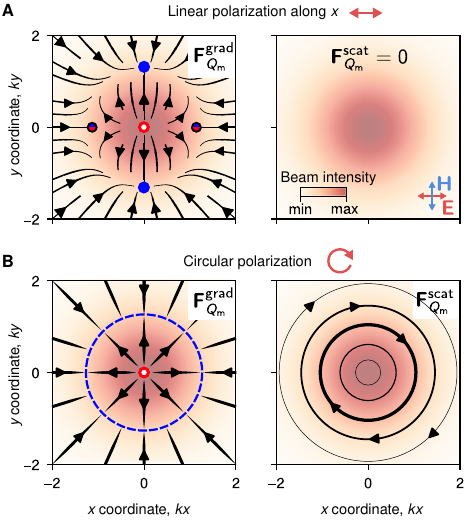}
    \caption{\textbf{Magnetic quadrupole force}. Black vector field shows gradient and scattering contributions of the magnetic quadrupole force in a linearly polarized Gaussian beam along the \(x\)-axis (\textsf{\textbf{A}}) and in a circularly polarized beam (\textsf{\textbf{B}}). Parameters were used \(k w_0 = 2\) and $\operatorname{Re}(\alpha_{Q_m}) > 0$. Background gradient red color shows the beam intensity. }
    \label{fig:quad_force_theory}
\end{figure}

In Fig.~\ref{fig:quad_force_theory} we plot magnetic quadrupole force~\eqref{eq:FQgrad}--\eqref{eq:FQscat} for the two distinct cases assuming $\operatorname{Re} \alpha_{Q_m} > 0$. Fig.~\ref{fig:quad_force_theory}\textsf{\textbf{A}} shows gradient and scattering contributions for the linearly polarized Gaussian beam along $x$-axis, i.e. $\tau = 1$, $\chi=\sigma=0$. Gradient force exhibits two stable trapping positions along the $y$-axis (marked by blue dots), one unstable position in the beam center (marked by a red circle), and two saddle points along the $x$-axis (marked by red-blue points). The scattering part is zero.  Fig.~\ref{fig:quad_force_theory}\textsf{\textbf{B}} shows force distribution for the circularly polarized Gaussian beam with $\tau = \chi = 0$ and $\sigma = 1$. We see that in this case force is azimuthally symmetric for both gradient and scattering parts. Gradient force exhibits an unstable point in the beam center and an infinite number of zeros at $\rho_{\text{eq}} = w_0 / \sqrt{2}$. The scattering force has only azimuthal component and is responsible for the orbital motion. Importantly, in Appendix~\ref{app:aperture} we also show that in a weakly focused beam two distinct trapping positions merge into one, thus a \textit{high-NA lens is a requirement} for this behavior.

The equilibrium radial distance \(\rho_{\text{eq}}\) derived from the balance of these forces yields the characteristic hopping distance based on Eq.~\eqref{eq:FQgrad}:
\begin{equation}
    D_{\text{hopping}} = 2 \rho_{\text{eq}} \approx \frac{2 w_0}{\sqrt{2}}.
    \label{eq:Dhopping}
\end{equation}
For the experimental conditions we find that $w_0 \approx \lambda / (\pi \cdot \mathrm{NA}) \approx 350$~nm, thus  $D_{\text{hopping}} \approx 490$~nm, which is in agreement with experimental observations (Fig.~\ref{fig:experiment_transition_to_hopping}\textsf{\textbf{B}},\textsf{\textbf{C}} and Fig.~\ref{fig:lin_circ_pol_v1}\textsf{\textbf{A}}).

\subsection{Interplay between multipoles in the quadrupole approximation}

Dominance of a single multipole is rather a rare occasion for the Mie particles. The multipolar series of the optical force up to a quadrupole reads as~\cite{Kislov2021APR,Chen2011NP,Jiang2015arXiv,Yu2019PRA,Jiang2016arXiv}:
\begin{equation}
    \label{eq:force_all}
    \mathbf{F} = \mathbf{F}^{\text{ext}} + \mathbf{F}^{\text{rec}}
    \simeq \mathbf{F}_{e} + \mathbf{F}_{m} + \mathbf{F}_{Q_e} + \mathbf{F}_{Q_m} + \mathbf{F}^{\text{rec}}.
\end{equation}
Here recoil term \( \mathbf{F}^{\text{rec}}  \) is associated with the momentum carried away by the scattered field alone~\cite{Toftul2026RMP}. Other terms, \(\mathbf{F}^{\text{ext}}\),  represent the result of multipole interaction with the incident field.
The explicit form of the Eq.~\eqref{eq:force_all} can be found in Appendix~\ref{app:all_terms_forces}. We neglect recoil terms $\mathbf{F}^{\text{rec}}$ as their contributions are generally weaker once there are strong field gradients.

For the linearly polarized Gaussian beam given by Eq.~\eqref{eq:gauss} with $\tau = 1$ and $\chi = \sigma = 0$ (linearly polarized along $x$-axis) we find a compact form of the Eq.~\eqref{eq:force_all}:
\begin{align}
\begin{split}
F^{\text{ext}}_x \simeq &   - \frac{U_0}{w_0}  \frac{x}{w_0}\left[A + B \frac{y^2}{w_0^2}  +  C \left(\frac{x^2}{w_0^2} - \frac{1}{2}\right)   \right] e^{- \frac{2 \rho^2}{w_0^2}}, \\
F^{\text{ext}}_y \simeq &   -   \frac{U_0}{w_0}  \frac{y}{w_0} \left[A + C \frac{x^2}{w_0^2}  + B \left( \frac{y^2}{w_0^2} - \frac{1}{2}\right)    \right] e^{- \frac{2 \rho^2}{w_0^2}},
\end{split}
\label{eq:Fext_lin}
\end{align}
where $U_0 = w_0^3 \varepsilon |E_0|^2$ is a constant with a dimension of energy, $E_0$ is the amplitude of the incident electric field~\eqref{eq:gauss} at the beam center, real dimensionless coefficients $A$, $B$, and $C$ depend on the polarizabilities as
\begin{align}
\begin{split}
    A & = w_0^{-3}  \left(\operatorname{Re} \alpha_{e}  + \operatorname{Re} \alpha_{m} \right) \simeq - \frac{6 \pi}{k^3 w_0^3} \operatorname{Im}(a_1 + b_1), \\
    B & = w_0^{-5} \left(\operatorname{Re} \alpha_{Q_e} + 2 \operatorname{Re} \alpha_{Q_m} \right) \simeq - \frac{40 \pi}{k^5 w_0^5} \operatorname{Im}(a_2 + 2 b_2), \\
    C & = w_0^{-5}\left(2 \operatorname{Re} \alpha_{Q_e} + \operatorname{Re} \alpha_{Q_m} \right) \simeq - \frac{40 \pi}{k^5 w_0^5} \operatorname{Im}(2 a_2 + b_2),
\end{split}
\label{eq:ABC}
\end{align}
where the right hand side of~\eqref{eq:ABC} are written for the isotropic sphere using Mie coefficients $a_\ell$ and $b_\ell$ (see Eq.~\eqref{eq:Mie_coef}).

We can assign a trapping potential \(U(\mathbf{r}) \) to the force~\eqref{eq:Fext_lin} such that \( \mathbf{F}^{\text{ext}} = - \boldsymbol{\nabla} U \). Integration of~\eqref{eq:Fext_lin} gives
\begin{equation}
    U =  - \frac{U_0}{4} \left(A + B \frac{y^2}{w_0^2} + C \frac{x^2}{w_0^2} \right) e^{- \frac{2 \rho^2}{w_0^2}},
    \label{eq:Utheory}
\end{equation}
where $U_0$ is the dimensional coefficient which scales linearly with the intensity.
Analyses of Eq.~\eqref{eq:Utheory} gives the following conditions for the force distribution with two equilibrium points. First of all, the dipole contribution should be low such that \( |A| \ll |B|,\: |C| \). Next, two distinct cases are possible: (i) \( B > 0 \) and \(B > C\) would give two stable positions along the $y$-axis and (ii) \( C > 0 \) and \(C > B\) would give two stable positions along the $x$-axis.
For the dipole dominant behavior, $|A| \gg |B|, |C|$, there is only one stable position located at the origin.

In the absence of thermal fluctuations, for instance in vacuum, a particle would remain indefinitely in one of the local minima of the potential~\eqref{eq:Utheory}. However, in a fluid environment at finite temperature, thermal energy allows the particle to overcome the barrier between the two wells, leading to stochastic hopping between them (see Appendix~\ref{app:stochastic}). This process is well described by Kramers' theory of thermally activated escape~\cite{Kramers1940, McCann1999N, Rondin2017NN}, which predicts the hopping rate to scale as dictated by Eq.~\eqref{eq:WK}.
Theoretical expression in Eq.~\eqref{eq:Utheory} predicts the emergence of bistability under suitable conditions and enables the estimation of \(\Delta U\) as a function of laser power and polarization.

For the circularly polarized beam (\(\tau = \chi = 0\), \(\sigma = \pm 1\)) the same procedure gives, keeping now the recoil terms as well, since for this beam they are as compact as the rest:
\begin{align}
\begin{split}
    F_\rho \simeq & - \frac{U_0}{w_0} \frac{\rho}{w_0} \left[ A - D + \frac{B + C}{2} \left( \frac{\rho^2}{w_0^2} - \frac{1}{2} \right) \right] e^{-\frac{2 \rho^2}{w_0^2}}, \\
    F_\varphi \simeq & - \frac{U_0}{w_0} \frac{\rho}{w_0} \, \sigma S \, e^{-\frac{2 \rho^2}{w_0^2}},
\end{split}
\label{eq:F_circ}
\end{align}
with $A$, $B$, $C$ as in Eq.~\eqref{eq:ABC} and two new real coefficients
\begin{align}
\begin{split}
    D & = \frac{3 k^5}{80 \pi w_0^3} \operatorname{Im} \left( \alpha_{Q_e} \alpha_e^* + \alpha_{Q_m} \alpha_m^* \right) \\
      & \simeq \frac{9 \pi}{k^3 w_0^3} \operatorname{Im} \left( a_2 a_1^* + b_2 b_1^* \right), \\
    S & = \frac{1}{4 w_0^5} \operatorname{Im} \left( \alpha_{Q_e} + \alpha_{Q_m} \right) \\
      & \quad - \frac{k^5}{80 \pi w_0^3} \operatorname{Re} \left( \alpha_{Q_e} \alpha_e^* + \alpha_{Q_m} \alpha_m^* \right) \\
      & \simeq \frac{10 \pi}{k^5 w_0^5} \operatorname{Re} \left( a_2 + b_2 \right) - \frac{3 \pi}{k^3 w_0^3} \operatorname{Re} \left( a_2 a_1^* + b_2 b_1^* \right).
\end{split}
\label{eq:DS}
\end{align}
The force~\eqref{eq:F_circ} depends only on $\rho$ and is isotropic in $\varphi$. Its radial part carries the quadrupoles through the single combination $B + C = 3 w_0^{-5} \operatorname{Re}(\alpha_{Q_e} + \alpha_{Q_m})$, and the recoil enters it only through $D$, which shifts the balance between dipole and quadrupole rather than adding a feature of its own.
Radial force changes sign at most once (see Appendix~\ref{app:regimes}), and the stable equilibria form a ring at
\begin{equation}
    \frac{\rho_{\rm eq}^2}{w_0^2} = \frac{1}{2} + \frac{2 (D - A)}{B + C},
    \label{eq:rho_eq}
\end{equation}
which exists when $B + C > 0$ and $A - D < (B + C)/4$. In the dipole-free limit $|A|, |D| \ll B + C$ it returns $\rho_{\rm eq} = w_0/\sqrt{2}$, see also Eq.~\eqref{eq:Dhopping}.
Only the azimuthal force depends on the handedness $\sigma$. 
Orbiting direction depends on the sign of $\sigma S$.
For $S > 0$ the particle circulates against the direction in which $\mathbf{E}$ spins in time, while for $S < 0$ it circulates with it. The first term of $S$ is the quadrupole extinction, $\operatorname{Im}(\alpha_{Q_e} + \alpha_{Q_m})$. It is positive for any passive sphere, so it always drives counter-rotation. The second term is the dipole--quadrupole recoil, which can reverse the direction. 
Both terms are already present in the purely transverse field~\eqref{eq:gauss}. Negative orbiting rotation suggests strong positive spinning of the trapped particle according to the angular momentum conservation law (see Appendix~\ref{app:materials} for the analyses of different materials). This however was not possible to directly observe in our experiments due to the experimental setup limitations. 
It therefore does not rely on the longitudinal field component of a focused beam, in contrast to dipole-level descriptions that require the weak transverse momentum carried by such components~\cite{Svak2018NC}. For wavelength-scale particles these higher-multipole contributions dominate the conversion. See also Appendix~\ref{app:all_terms_forces} for all the terms up to the octupole.

From this analyses we see that the presence of the dominant quadrupole response gives rise to several important consequences for the trapping in Gaussian beam:
\begin{enumerate}
    \item[(i)] trapping becomes polarization dependent;
    \item[(ii)]  equilibrium points may occur \textit{off}-center from the beam axis;
    \item[(iii)] spin degree of freedom of Gaussian beam couples to the orbital motion of the trapped particle.
\end{enumerate}
These features correlate with the experimental observations for the VO$_2$ particles with diameters in the range of $400$~nm to $625$~nm.

Finally, we note that the angular anisotropy of the force response from higher-order multipoles arises from the lack of dual symmetry in the particle's electromagnetic properties, i.e., \(\varepsilon_{\text{par}} \neq \mu_{\text{par}}\). In contrast, a dual-symmetric isotropic particle has identical electric and magnetic polarizabilities at each multipole order, resulting in an isotropic optical force response with respect to the azimuthal angle even for a linearly polarized Gaussian beam.

\section{Conclusion}
\label{sec:conc}

We have demonstrated polarization-dependent multipolar trapping regime in which off-axis bistability (hopping) and handedness-controlled azimuthal transport (orbiting) coexist and can be achieved within the same Gaussian trap.  The observed dynamics depends on the size of the particles and has been consistently observed for particles in the 400-625 nm range.
Specifically, in our experiments and calculations we observed for the same Gaussian beam optical trap (i) particles with sizes less than 400 nm that support dipole Mie resonances being trapped in a single-well potential at the beam axis, (ii) in the 400-625 nm size range that support quadrupolar Mie modes hopping in a double-well potential (linearly polarized beam) or orbiting around beam axis (circular polarization), and (iii) particles larger than 625 nm sizes beyond the peaks of both the dipolar and quadrupolar Mie modes being consistently repelled off the beam centre.
We describe analytically the effect of hopping and orbiting using quadrupole optical forces.
This highlights the importance of detailed structural and size-dependent studies, and motivates further experimental and theoretical exploration of effects of Mie resonances in diverse material systems as well as for different multipoles and regimes of interference between several modes. The complex motions of Mie-resonant particles suggest potential opportunities for probing local environmental parameters via measurements of hopping frequencies or angular velocities of such particles.

\begin{acknowledgments}
Authors thank Dmitry Pidgayko,  Mikael K\"all, and Mihail Petrov for fruitful discussions and suggestions.

\textbf{Funding.} The work was partially supported by the Australian Research Council (Grant No. DP210101292).
Vanadium oxide nanoparticle research was supported by the Defence Science Institute, an initiative of the State Government of Victoria.
Dr. Mohammad Taha is the recipient of an Office of National Intelligence Postdoctoral Grant (project number NIPG202502) funded by the Australian Government.

\textbf{Author contributions.} I.T. and L.M. contributed equally to this work.
I.T. developed the theoretical model, carried out numerical simulations, and interpreted experimental results.
L.M. designed and conducted optical trapping experiments, analyzed particle trajectories, and extracted quantitative parameters.
S.B. and M.T. synthesized VO$_2$ nanoparticles.
Y.K. and S.K. supervised the project.
I.T. and L.M. prepared the first draft of the manuscript. All authors discussed the results and contributed to writing the manuscript.

\textbf{Competing interests.} There are no competing interests to declare.

\textbf{Data availability.} Data related to this project can be found in the public repository~\cite{zenodo_archive}.
\end{acknowledgments}

\appendix
\onecolumngrid
\newpage
\section{Optical tweezer setup}
\label{app:setup}

\begin{figure}[h]
    \centering
    \includegraphics[width=0.5\linewidth]{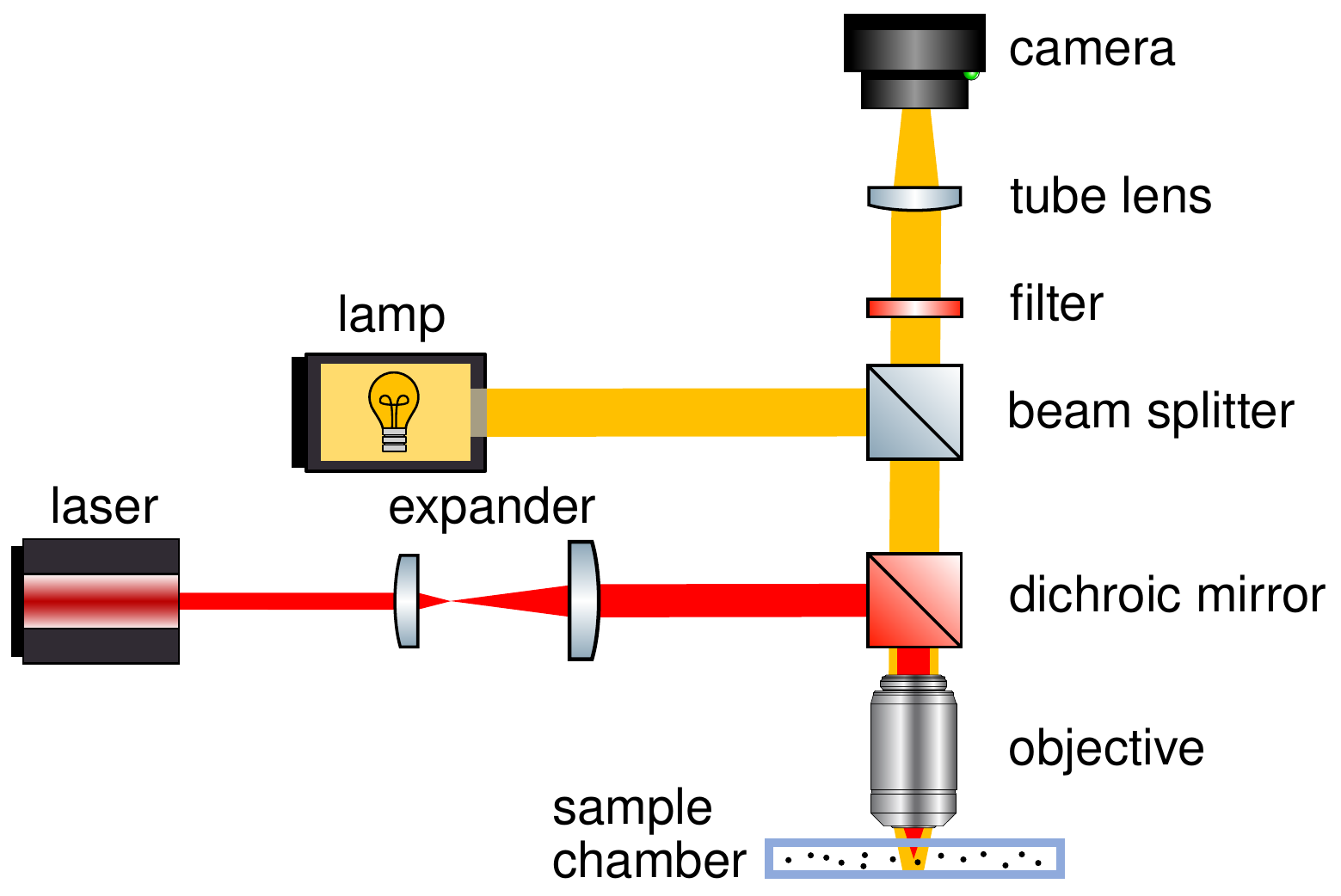}
    \caption{\textbf{Experimental setup.} Schematic of the optical trapping setup.}
    \label{fig:setup}
\end{figure}

The optical trapping system (Fig.~\ref{fig:setup}) utilized a near-infrared (NIR) fiber-coupled laser (Thorlabs BL976-PAG500, $976$~nm) as the trapping beam, which was focused through a high numerical aperture (NA) objective lens (Olympus MPlanFLN 100X, $\mathrm{NA} = 0.9$). To facilitate stable trapping along the optical axis, glass cover slips were used for creating a counter-propagating reflected beam. Particle dynamics were captured using a high-resolution CMOS camera (Thorlabs DCC3260M). All experiments were conducted at room temperature ($23 ^{\circ}$C), and the laser power was measured after transmission through the objective lens.

\section{Synthesis of VO$_2$ Nanoparticles}
\label{app:synthesis}

VO$_2$ nanoparticles were synthesized via a reflux reaction in a hyperhydrated environment using vanadium pentoxide (V$_2$O$_5$) and oxalic acid dihydrate as precursors. Briefly, the reaction mixture was heated to  $200 ^{\circ}$ under vigorous stirring (450 rpm) to ensure uniform heat distribution and reactant mixing. The reaction was maintained at a reflux temperature of $\sim 100^{\circ}$C for 48 hours, enabling controlled kinetics and consistent product quality without the need for conventional hydrothermal autoclave processes. Post-reaction, the products were washed, centrifuged, filtered, and dried under ambient conditions to remove impurities and excess water. The final product consisted of sub-stoichiometric, hydrated VO$_2$ nanoparticles. Detailed material characterization and compositional analyses are described in previous work~\cite{Taha2023Apr}. This benchtop synthesis method offers a simpler, less energy-intensive alternative to traditional approaches while minimizing the risk of over-hydration of metal oxides.

\section{Trap Stiffness Measurement}
\label{app:stiffness}

The two-dimensional (2D) trap stiffness of optically trapped particles was determined using video tracking analysis. Assuming a linear optical force near the trapping center $\mathbf{r}_0$, the force can be expressed as $\mathbf{F}(\mathbf{r}) \approx - \kappa (\mathbf{r} - \mathbf{r}_0)$, where $\kappa$ represents the trap stiffness. According to the Equipartition Theorem, each degree of freedom in the system possesses an average thermal energy of $k_{\text{B}} T /2$, where $T$ is the absolute temperature and $k_{\text{B}}$ is the Boltzmann constant. For the x-direction, the average potential energy of the trapped particle is given by $\kappa_x \langle\left( x - x_0 \right)^2 \rangle /2 = k_{\text{B}}T / 2$, where $\langle\left( x - x_0 \right)^2 \rangle$ is the position variance. Thus, the trap stiffness in the x- and y-directions can be calculated as $\kappa_{x,y} =  k_{\text{B}} T/\langle\sigma_{x,y}^2\rangle$, where $\langle\sigma_{x}^2\rangle$ is the variance of the particle's position. The 2D position of the trapped particle was determined by localizing its center in each frame. The recorded video was divided into 200-frame segments, and the trap stiffness was calculated for each segment. The standard deviation of the trap stiffness values, derived from these segments, is presented as error bars in the results.

\section{Gaussian beams}
\label{app:gaussian}

\subsection{Paraxial Gaussian beam}

The paraxial approximation of a monochromatic Gaussian beam reads~\cite{Novotny2012}
\begin{align}
\begin{split}
    \sqrt{\varepsilon}\mathbf{E} &= \mathcal{A}_0 \: \mathbf{u} \frac{w_0}{w(z)} e^{- \frac{\rho^2}{w^2(z)}} e^{i \left[ k z - \eta(z) + R^{-1}(z) k \rho^2 / 2 \right]}, \\
    \sqrt{\mu}\mathbf{H} &= \mathbf{\hat{z}} \times \sqrt{\varepsilon}\mathbf{E},
\end{split}
\label{eq:parax_true}
\end{align}
where $\mathcal{A}_0 / \sqrt{\varepsilon}$ is the electric field amplitude, $\mathbf{u}$ is the unit complex polarization vector, $w(z) = w_0 \sqrt{1 + z^2/z_0^2}$ is the beam radius, $R^{-1}(z) = z/(z^2 + z_0^2)$ is the inverse wavefront radius, $\eta(z) = \arctan(z/z_0)$ is the phase correction, and $z_0 = k w_0^2 / 2$ is the Rayleigh range.
For polarization parametrization it is convenient to use the circular polarization unit vectors 
\begin{equation}
    \mathbf{\hat{e}}_\sigma=(\mathbf{\hat{x}}+i \sigma\mathbf{\hat{y}})/\sqrt{2},
    \label{eq:esigma}
\end{equation}
with the handedness $\sigma=\pm1$. 

\subsection{Non-paraxial Gaussian beams}

A high-NA lens creates a strongly focused beam, and the paraxial approximation no longer holds.
The distinctive feature of a tightly focused beam is the longitudinal field component.
Equation~\eqref{eq:parax_true} solves the paraxial wave equation, not Maxwell's equations, and it needs $kw_0\gg2$. Beyond this limit no closed-form Gaussian beam exists, and the name ``Gaussian beam'' no longer defines a unique field. One must say where the Gaussian is. It can be in the angular spectrum, in the pupil of an aplanatic lens, or in the pupil of a thin lens.
This section is based on Refs.~\cite{Wolf1959,RichardsWolf1959,Novotny2012,Lenton2020OTT,Mizuyama2018COMSOLBlog,COMSOL63WaveOptics}. The code that calculates the fields is in Ref.~\cite{toftul2026github_nonparax}.

\begin{figure}[t]
\centering
\tdplotsetmaincoords{65}{115}
\begin{tikzpicture}[tdplot_main_coords, >=Latex, font=\footnotesize, scale=1.8,
                    baseline=(current bounding box.north)]
  \def\th{40}\def\ph{55}
  \pgfmathsetmacro{\ct}{cos(\th)}\pgfmathsetmacro{\st}{sin(\th)}
  \pgfmathsetmacro{\cp}{cos(\ph)}\pgfmathsetmacro{\sp}{sin(\ph)}
  \coordinate (O) at (0,0,0);
  \coordinate (K) at ({\st*\cp},{\st*\sp},{\ct});
  \coordinate (Kxy) at ({\st*\cp},{\st*\sp},0);
  \draw[->] (O) -- (1.25,0,0) node[anchor=north east] {$x$};
  \draw[->] (O) -- (0,1.25,0) node[anchor=north west] {$y$};
  \draw[->] (O) -- (0,0,1.3) node[anchor=south] {$z$};
  \draw[dotted] plot[domain=-10:100,samples=40] ({cos(\x)},{sin(\x)},0);
  \draw[dotted] plot[domain=0:95,samples=40] ({sin(\x)*\cp},{sin(\x)*\sp},{cos(\x)});
  \fill[cspec,opacity=0.12] (O) -- (0,0,1) -- plot[domain=0:\th,samples=20] ({sin(\x)*\cp},{sin(\x)*\sp},{cos(\x)}) -- cycle;
  \draw[->,very thick] (O) -- (K) node[anchor=south east] {$\mathbf{\hat{k}}$};
  \draw[dashed] (K) -- (Kxy) (O) -- (Kxy);
  \draw[->,capl] plot[domain=0:\ph,samples=20] ({0.35*cos(\x)},{0.35*sin(\x)},0);
  \node[capl] at ({0.48*cos(\ph/2)},{0.48*sin(\ph/2)},0) {$\phi$};
  \draw[->,capl] plot[domain=0:\th,samples=20] ({0.45*sin(\x)*\cp},{0.45*sin(\x)*\sp},{0.45*cos(\x)});
  \node[capl] at ({0.58*sin(\th/2)*\cp},{0.58*sin(\th/2)*\sp},{0.58*cos(\th/2)}) {$\theta$};
  \draw[->,thick,cthin] (K) -- ++({0.4*\ct*\cp},{0.4*\ct*\sp},{-0.4*\st}) node[anchor=west] {$\mathbf{\hat{\boldsymbol{\theta}}}$};
  \draw[->,thick,cthin] (K) -- ++({-0.4*\sp},{0.4*\cp},0) node[anchor=west] {$\mathbf{\hat{\boldsymbol{\phi}}}$};
  \fill (O) circle (0.6pt);
  \node[anchor=north west, inner sep=0pt, font=\sffamily\bfseries\small]
    at (current bounding box.north west) {A};
\end{tikzpicture}%
\hspace{1.5em}%
\begin{tikzpicture}[>=Latex, font=\footnotesize, scale=1.3,
                    baseline=(current bounding box.north)]
  \def\c{1.5}\def\wt{40}
  \pgfmathsetmacro{\u}{\c/sqrt(1+\c*\c)}\pgfmathsetmacro{\v}{1/sqrt(1+\c*\c)}
  \draw[->] (-1.2,0) -- (1.3,0) node[below] {$\mathbf{\hat{\boldsymbol{\theta}}}$};
  \draw[->] (0,-0.62) -- (0,1.0) node[left] {$\mathbf{\hat{\boldsymbol{\phi}}}$};
  \draw[dotted] (0,0) circle ({1/sqrt(2)});
  \draw[very thick,capl] (0,0) ellipse ({\u} and {\v});
  \draw[->,very thick,capl] (100:{\u} and {\v}) -- (101:{\u} and {\v});
  \draw[->,thick] (0,0) -- ({\u*cos(\wt)},{\v*sin(\wt)}) node[above right=-2pt] {$\operatorname{Re}\mathbf p\,e^{-i\omega t}$};
  \draw[dotted] (\u,0) -- (\u,{-\v-0.25}) (0,\v) -- ({-\u-0.25},\v);
  \draw[<->] (0,{-\v-0.18}) -- (\u,{-\v-0.18}) node[below,pos=0.5] {$c/\sqrt{1+c^2}$};
  \draw[<->] ({-\u-0.18},0) -- ({-\u-0.18},\v) node[left,pos=0.5] {$1/\sqrt{1+c^2}$};
  \draw (0,0) circle (2.2pt); \fill (0,0) circle (0.7pt);
  \node[above left] at (0,0) {$\mathbf{\hat{{k}}}$};
  \node[anchor=north west, inner sep=0pt, font=\sffamily\bfseries\small]
    at (current bounding box.north west) {B};
\end{tikzpicture}
\caption{\textsf{\textbf{A}}~Angles and unit vectors of Eqs.~\eqref{eq:debye}--\eqref{eq:rule}. A plane wave travels along $\mathbf{\hat{k}}$.
\textsf{\textbf{B}}~The unit polarization vector $\mathbf p$ of one plane wave, Eq.~\eqref{eq:rule}, for $\sigma=+1$. 
In time, the real field $\operatorname{Re}\mathbf p\,e^{-i\omega t}$ traces an ellipse (blue) with the semi-axes $c/\sqrt{1+c^2}$ along $\mathbf{\hat{\boldsymbol{\theta}}}$ and $1/\sqrt{1+c^2}$ along $\mathbf{\hat{\boldsymbol{\phi}}}$, here $c=1.5$. }
\label{fig:angles_and_polarization}
\end{figure}

A monochromatic beam radiated by sources at $z\to-\infty$ is a sum of propagating plane waves,
\begin{equation}
    \mathbf E(\mathbf r)=\int \limits_{\theta<\theta_{\max}}\mathrm{d}\Omega\;
    \mathbf{E}_\infty(\theta,\phi)\,e^{i\mathbf k\cdot\mathbf r},
    \label{eq:debye}
\end{equation}
with $\mathbf k=k\mathbf{\hat{k}}$, $\mathbf{\hat{k}} = \cos\phi \sin\theta\, \mathbf{\hat{x}} +\sin\phi \sin\theta\, \mathbf{\hat{y}} + \cos\theta\, \mathbf{\hat{z}}$ (Fig.~\ref{fig:angles_and_polarization}\textsf{\textbf{A}}), $\mathrm{d}\Omega=\sin\theta\mathrm{d}\theta\mathrm{d}\phi$, $0\le\phi<2\pi$ and $\mathbf{\hat{k}}\cdot\mathbf{E}_\infty=0$. The limit is $\theta_{\max}=\pi/2$, unless a stop sets $\theta_{\max}=\arcsin(\mathrm{NA}/n)$. The beam contains no evanescent waves, because they decay away from the sources. Each plane wave solves Maxwell's equations, so Eq.~\eqref{eq:debye} is exact for every transverse $\mathbf{E}_\infty$. The amplitude $\mathbf{E}_\infty$ is the plane-wave amplitude per solid angle. The magnetic amplitude is $\sqrt{\mu}\,\mathbf H_\infty=\mathbf{\hat{k}}\times\sqrt{\varepsilon}\,\mathbf E_\infty$. The magnetic field $\mathbf H(\mathbf r)$ is Eq.~\eqref{eq:debye} with $\mathbf E_\infty$ replaced by $\mathbf H_\infty$. A real $\mathbf E_\infty$ puts the focus at $\mathbf r=0$.

A non-paraxial Gaussian beam is a choice of $\mathbf{E}_\infty$. The choice has two parts, an envelope $a(\theta)$ and a unit polarization vector $\mathbf p$:
\begin{equation}
\sqrt{\varepsilon}\,\mathbf{E}_\infty(\theta,\phi)=\mathcal A\,a(\theta)\,\mathbf{p}(\theta,\phi),
\qquad
\sqrt{\mu}\,\mathbf{H}_\infty(\theta,\phi) =\mathcal A\,a(\theta)\,\mathbf{\hat{{k}}}\times\mathbf{p}(\theta,\phi),
\label{eq:ap}
\end{equation}
with $a(0)=1$, $|\mathbf p|=1$ and $\mathbf p(0,\phi)=\mathbf{\hat{e}}_\sigma$, so that the plane wave along $z$ has the polarization of the paraxial beam~\eqref{eq:parax_true}. 
Every rule in this Appendix has the form
\begin{equation}
\mathbf p=e^{i\sigma\phi}\,\frac{c(\theta)\,\mathbf{\hat{\boldsymbol{\theta}}}+ i \sigma\,\mathbf{\hat{\boldsymbol{\phi}}}}{\sqrt{1+c^2(\theta)}},
\label{eq:rule}
\end{equation}
where $\mathbf{\hat{\boldsymbol{\theta}}}= \cos\phi \cos\theta \mathbf{\hat{x}} + \sin\phi \cos\theta \mathbf{\hat{y}} -\sin\theta \mathbf{\hat{z}}$ and  $\mathbf{\hat{\boldsymbol{\phi}}}= -\sin\phi \mathbf{\hat{x}} + \cos\phi \mathbf{\hat{y}}$
are the meridional and azimuthal unit vectors. With $\mathbf{\hat{{k}}}$ they form the right-handed triad $\mathbf{\hat{\boldsymbol{\theta}}}\times\mathbf{\hat{\boldsymbol{\phi}}}=\mathbf{\hat{{k}}}$. At $\theta=0$ the pair $(\mathbf{\hat{\boldsymbol{\theta}}},\mathbf{\hat{\boldsymbol{\phi}}})$ is $(\mathbf{\hat{x}},\mathbf{\hat{y}})$ turned by $\phi$ about $z$. The phase $e^{i\sigma\phi}$ cancels this turn, so that $\mathbf p(0,\phi)=\mathbf{\hat{e}}_\sigma$ for every $\phi$. The real weight $c(\theta)$ is the ratio of the meridional and the azimuthal components of $\mathbf p$, with $c(0)=1$. A real $c$ means that the system adds no phase between these two components. 
Figure~\ref{fig:angles_and_polarization}\textsf{\textbf{B}} shows the direction of $\mathbf p$ and the role of $c$.
The integral over $\phi$ in Eq.~\eqref{eq:debye} can be calculated analytically. It gives Bessel functions, which can simplify the numerical evaluation~\cite{toftul2026github_nonparax}.

The amplitude $\mathcal{A}$ in Eq.~\eqref{eq:ap} can be uniquely connected to the power of the beam
\begin{equation}
    P =  \int \limits_{x,y} \frac{1}{2} \operatorname{Re} \left( \mathbf{E}^{*} \times \mathbf{H} \right) \cdot \mathrm{d}  \mathbf{S}=\frac{4\pi^3|\mathcal A|^2}{k^2\sqrt{\varepsilon\mu}}\int \limits_0^{\theta_{\max}}a^2(\theta)\sin\theta\mathrm{d}\theta.
    \label{eq:power}
\end{equation}
It is connected to the paraxial limit amplitude in Eq.~\eqref{eq:parax_true} as $\mathcal A_0=4\pi\mathcal A/(kw_0)^2$.

The two circular polarizations $\mathbf{\hat{e}}_\pm$ form a basis, so every input polarization is $\mathbf{\hat{e}}=\alpha\,\mathbf{\hat{e}}_++\beta\,\mathbf{\hat{e}}_-$ with $|\alpha|^2+|\beta|^2=1$. Maxwell's equations are linear, so this input gives the beam
\begin{equation}
	\begin{split}
		\mathbf E&=\alpha\,\mathbf E^{(+)}+\beta\,\mathbf E^{(-)}, \\
		\mathbf H&=\alpha\,\mathbf H^{(+)}+\beta\,\mathbf H^{(-)},
	\end{split}
\label{eq:arbitrary}
\end{equation}
where $\mathbf E^{(\sigma)}$ and $\mathbf H^{(\sigma)}$ are the fields of the beam with the handedness $\sigma$ and the same $a$ and $c$. In practice, all possible polarizations can be parametrized by one complex number $m$ as
\begin{equation}
    \alpha=\frac{1-i m}{\sqrt{2(1+|m|^2)}},\qquad
    \beta=\frac{1+i m}{\sqrt{2(1+|m|^2)}}.
    \label{eq:m}
\end{equation}

In principle, there are many ways to parametrize a Gaussian beam by the envelope $a(\theta)$ and the weight $c(\theta)$. Each choice corresponds to a specific physical construction of the beam. The main choices are listed in Table~\ref{tab:env}.

Figure~\ref{fig:constructions} shows where the Gaussian lies. For rules A and D it lies in the angular spectrum: the plane-wave amplitudes are $e^{-(kw_0\sin\theta)^2/4}$, the Fourier transform of the focal-plane field $e^{-\rho^2/w_0^2}$ (Fig.~\ref{fig:constructions}\textsf{\textbf{B}}). For the lens rules it lies in the pupil, where the collimated input beam is~\cite[Eq.~(3.52)]{Novotny2012}
\begin{equation}
    \mathbf E_\mathrm{inc}(h,\phi)=E_0\,\mathbf{\hat e}_\sigma\,e^{-h^2/w_\mathrm{ap}^2},
    \qquad w_\mathrm{ap}=\frac{2f}{kw_0}.
    \label{eq:Einc}
\end{equation}
Here $h$ is the distance from the axis, $f$ is the focal length, and $w_\mathrm{ap}$ is the input radius (called $w_0$ in Ref.~\cite{Novotny2012}), which the lens focuses to the waist $w_0$. The lens sends the ray at the height $h$ into the direction $\theta$, with $h=f\sin\theta$ for an aplanatic lens and $h=f\tan\theta$ for a thin lens (Fig.~\ref{fig:constructions}\textsf{\textbf{A}}). It turns the radial component of $\mathbf E_\mathrm{inc}$ into the meridional one and keeps the azimuthal one, so $c=1$~\cite[Eq.~(3.39)]{Novotny2012}. A lossless lens keeps the power of each ray: the input power $\tfrac12\sqrt{\varepsilon_1/\mu_1}\,|\mathbf E_\mathrm{inc}|^2\,h\,\mathrm{d}h\,\mathrm{d}\phi$, with $\varepsilon_1,\mu_1$ of the input medium, equals the power in $\mathrm{d}\Omega$ in Eq.~\eqref{eq:power}. This gives
\begin{equation}
    a(\theta)=\sqrt{\frac{h\,h'}{f^2\sin\theta}}\;e^{-h^2(\theta)/w_\mathrm{ap}^2},
    \qquad
    \mathcal A=\frac{kf}{2\pi}\Big(\frac{\varepsilon\varepsilon_1\mu}{\mu_1}\Big)^{1/4}E_0,
    \label{eq:lens}
\end{equation}
with $h'=\mathrm{d}h/\mathrm{d}\theta$. The aplanatic lens has $hh'=f^2\sin\theta\cos\theta$ and the thin lens has $hh'=f^2\sin\theta/\cos^3\theta$, which gives rows B and C of Table~\ref{tab:env}. Rules E and F keep only the exponential.

\begin{figure}[t]
\centering
\begin{tikzpicture}[>=Latex, font=\footnotesize, scale=0.78, baseline=(current bounding box.north),
    ray/.style={thick, postaction={decorate},
      decoration={markings, mark=at position 0.3 with {\arrow{>}}}},
    dim/.style={<->, thin}]
  \def\f{3.0}      %
  \def\xp{-4.9}    %
  \def\h{2.4}      %
  \def\dh{0.3}     %
  \def\R{2.85}     %
  \def\Lk{2.0}     %
  \pgfmathsetmacro{\thB}{asin(\h/\f)}
  \pgfmathsetmacro{\thBb}{asin((\h+\dh)/\f)}
  \pgfmathsetmacro{\thC}{atan(\h/\f)}
  \pgfmathsetmacro{\zB}{-\f*cos(\thB)}
  \pgfmathsetmacro{\zBb}{-\f*cos(\thBb)}
  \pgfmathsetmacro{\thmB}{asin(\R/\f)}
  \pgfmathsetmacro{\thmC}{atan(\R/\f)}
  \coordinate (F) at (0,0);
  \coordinate (P) at (\zB,\h);
  \coordinate (L) at (-\f,-\h);
  \fill[capl,opacity=0.15] (\xp,\h) -- (P) -- (F) -- (\zBb,\h+\dh) -- (\xp,\h+\dh) -- cycle;
  \fill[cthin,opacity=0.15] (\xp,-\h) -- (L) -- (F) -- (-\f,-\h-\dh) -- (\xp,-\h-\dh) -- cycle;
  \draw[->] (\xp-2.4,0) -- (0.9,0) node[below] {$z$};
  \fill[gray!10] plot[domain=-3.6:3.6,samples=100] ({\xp-0.8*exp(-(\x/\h)^2)},\x) -- (\xp,3.6) -- (\xp,-3.6) -- cycle;
  \fill[gray!40] plot[domain=-\R:\R,samples=100] ({\xp-0.8*exp(-(\x/\h)^2)},\x) -- (\xp,\R) -- (\xp,-\R) -- cycle;
  \draw[thick] plot[domain=-\R:\R,samples=100] ({\xp-0.8*exp(-(\x/\h)^2)},\x);
  \draw[thin,densely dashed] plot[domain=\R:3.6,samples=20] ({\xp-0.8*exp(-(\x/\h)^2)},\x)
                             plot[domain=-3.6:-\R,samples=20] ({\xp-0.8*exp(-(\x/\h)^2)},\x);
  \draw[dashed] (\xp,-\R) -- (\xp,\R);
  \draw[line width=2.2pt] (\xp,\R) -- (\xp,3.75) (\xp,-\R) -- (\xp,-3.75);
  \node[below] at (\xp,-3.75) {stop};
  \node[anchor=south east] at (\xp-0.1,3.6) {$|\mathbf E_\mathrm{inc}(h)|\propto e^{-h^2/w_\mathrm{ap}^2}$};
  \draw[dim] (\xp-2.05,0) -- (\xp-2.05,\h) node[pos=0.5,left] {$w_\mathrm{ap}$};
  \draw[dim] (\xp-2.05,0) -- (\xp-2.05,-\R) node[pos=0.5,left] {$R$};
  \draw[densely dotted] (\xp-2.15,-\R) -- (\xp,-\R);
  \foreach \y in {-1.8,-0.9,0.9,1.8}
    \draw[->,thick] (\xp-1.75,\y) -- (\xp-1.0,\y);
  \draw[decorate,decoration={brace,amplitude=2pt}] (\xp+0.1,\h+\dh) -- (\xp+0.1,\h)
    node[pos=0.5,right=1pt] {$\mathrm{d}h$};
  \draw[capl,very thick] (90:\f) arc (90:180:\f);
  \node[capl,anchor=south east] at (100:\f) {aplanatic lens};
  \draw (\xp,\R) -- ({180-\thmB}:\f) -- (F);                 %
  \draw[ray,capl] (\xp,\h) -- (P) -- (F);
  \fill[capl] (P) circle (1.4pt);
  \path (P) -- (F) node[capl,pos=0.5,below left=-2pt] {$f$};
  \draw[densely dotted] (P) -- (\zB,0);
  \draw[dim] ({\zB-0.2},0) -- ({\zB-0.2},\h) node[pos=0.5,left] {$h$};
  \draw[capl] (180:0.5) arc (180:{180-\thB}:0.5);
  \node[capl] at ({180-\thB/2}:0.7) {$\theta$};
  \draw (180:0.8) arc (180:{180-\thmB}:0.8);
  \node[anchor=south west,inner sep=1pt] at ({180-\thmB}:0.8) {$\theta_{\max}$};
  \coordinate (S) at (0.5,1.9);
  \coordinate (Sh) at ($(S)+({\Lk*cos(\thB)},0)$);
  \coordinate (St) at ($(Sh)+(0,{-\Lk*sin(\thB)})$);
  \draw[dashed,thick] (S) -- (Sh) node[pos=0.5,above] {$k\cos\theta$};
  \draw[dashed,thick] (Sh) -- (St) node[pos=0.5,right] {$k\sin\theta$};
  \draw[->,very thick,capl] (S) -- (St) node[pos=0.5,below left=-2pt] {$\mathbf k$};
  \draw[gray,dotted] (180:\f) arc (180:270:\f);
  \draw[cthin,line width=1.2pt,<->] (-\f,-0.2) -- (-\f,-3.45) node[below] {thin lens};
  \draw (\xp,-\R) -- (-\f,-\R) -- (F);                       %
  \draw[ray,cthin] (\xp,-\h) -- (L) -- (F);
  \fill[cthin] (L) circle (1.4pt);
  \fill[cthin] ({180+\thC}:\f) circle (1.4pt);
  \draw[dim] ({-\f-0.2},0) -- ({-\f-0.2},-\h) node[pos=0.5,left] {$h$};
  \draw[cthin] (180:0.9) arc (180:{180+\thC}:0.9);
  \node[cthin] at ({180+\thC/2}:1.1) {$\theta$};
  \draw (180:0.55) arc (180:{180+\thmC}:0.55);
  \node[anchor=north west,inner sep=1pt] at ({180+\thmC}:0.55) {$\theta_{\max}$};
  \draw[dim] (-\f,-4.1) -- (F|-0,-4.1) node[pos=0.5,below] {$f$};
  \draw[densely dotted] (F) -- (0,-4.1);
  \fill (F) circle (1.3pt) node[above right=-1pt] {F};
  \node[capl,anchor=west] at (0.3,2.95) {B, F: $h=f\sin\theta$};
  \node[cthin,anchor=west] at (0.3,-2.6) {C, E: $h=f\tan\theta$};
  \node[anchor=north west,inner sep=0pt,font=\sffamily\bfseries\small]
    at ($(current bounding box.north west)+(-0.6,0)$) {A};
\end{tikzpicture}%
\hfill
\begin{tikzpicture}[>=Latex, font=\footnotesize, scale=0.8, baseline=(current bounding box.north),
                    dim/.style={<->, thin}]
  \def\k{2.3}      %
  \def\kb{1.84}    %
  \def\wz{0.8}     %
  \def\yr{-4.6}    %
  \pgfmathsetmacro{\th}{asin(\kb/\k)}
  \pgfmathsetmacro{\ct}{cos(\th)}
  \coordinate (O) at (0,0);
  \coordinate (P) at ({\k*\ct},\kb);
  \fill[cspec!8] plot[domain=0:2.9,samples=80] ({-0.7*exp(-(\x/\kb)^2)},\x) -- (0,2.9) -- (0,0) -- cycle;
  \fill[cspec!35] plot[domain=0:\k,samples=80] ({-0.7*exp(-(\x/\kb)^2)},\x) -- (0,\k) -- (0,0) -- cycle;
  \draw[thick,cspec!70!black] plot[domain=0:\k,samples=80] ({-0.7*exp(-(\x/\kb)^2)},\x);
  \draw[thin,densely dashed,cspec!70!black] plot[domain=\k:2.9,samples=20] ({-0.7*exp(-(\x/\kb)^2)},\x);
  \node[cspec!70!black,anchor=south east] at (-0.1,2.85) {$e^{-(w_0\,k\sin\theta/2)^2}$};
  \draw[->] (-1.1,0) -- (2.9,0) node[below] {$k\cos\theta$};
  \draw[->] (0,0) -- (0,3.1) node[right] {$k\sin\theta$};
  \draw[cspec,very thick] (0:\k) arc (0:90:\k);
  \node[cspec,anchor=south west] at (28:\k) {$|\mathbf k|=k$};
  \draw[densely dotted] (-0.6,\k) -- (0,\k);
  \node[gray,anchor=east] at (-0.45,2.5) {evanescent};
  \draw[dim] (-1.0,0) -- (-1.0,\kb) node[pos=0.5,left] {$2/w_0$};
  \draw[densely dotted] (0,\kb) -- (P) -- ({\k*\ct},0);
  \draw[->,very thick] (O) -- (P) node[pos=0.55,below right=-2pt] {$\mathbf k$};
  \draw (0.55,0) arc (0:\th:0.55); \node at ({\th/2}:0.75) {$\theta$};
  \fill (O) circle (1.2pt);
  \node[cspec,anchor=south west] at (1.2,2.6) {A, D: no lens};
  \draw[<->,thick] (0,-0.75) -- (0,-2.05) node[pos=0.5,right,align=left] {2D Fourier\\transform};
  \draw[->] (-0.3,\yr) -- (1.4,\yr);
  \draw[->] (0,\yr) -- (0,{\yr+2.2}) node[left] {$\rho$};
  \fill[cspec!35] plot[domain=0:2.0,samples=60] ({0.8*exp(-(\x/\wz)^2)},{\yr+\x}) -- (0,{\yr+2.0}) -- (0,\yr) -- cycle;
  \draw[thick,cspec!70!black] plot[domain=0:2.0,samples=60] ({0.8*exp(-(\x/\wz)^2)},{\yr+\x});
  \draw[dim] (1.05,\yr) -- (1.05,{\yr+\wz}) node[pos=0.5,right] {$w_0$};
  \node[below] at (0.5,\yr) {focal plane $z=0$};
  \node[cspec!70!black,anchor=west] at (0.3,{\yr+1.9}) {$|\mathbf E_\perp(\rho,0)|\propto e^{-\rho^2/w_0^2}$};
  \node[anchor=north west,inner sep=0pt,font=\sffamily\bfseries\small]
    at ($(current bounding box.north west)+(-0.3,0)$) {B};
\end{tikzpicture}
\caption{\textbf{Construction of the non-paraxial Gaussian beams.} Where the Gaussian of each rule in Table~\ref{tab:env} lies, for $kw_0=2.5$.
\textsf{\textbf{A}}~Rules B, C, E, F. The input field~\eqref{eq:Einc} is bent towards the focus F by an aplanatic lens, $h=f\sin\theta$ (upper half), or by a thin lens, $h=f\tan\theta$ (lower half). The drawn rays are at $h=w_\mathrm{ap}$, which is the $1/e$ angle $\theta_b$ of rules E, F. The shaded tubes carry the power of the same strip $\mathrm{d}h$, which gives the prefactor in Eq.~\eqref{eq:lens}; E and F omit it. A stop of radius $R$ blocks $|h|>R$ and sets $\theta_{\max}$, with $\mathrm{NA}=n\sin\theta_{\max}$.
\textsf{\textbf{B}}~Rules A, D. The focal-plane field $e^{-\rho^2/w_0^2}$ (bottom) has the spectrum $e^{-(w_0k\sin\theta/2)^2}$ over the transverse wavevector $k\sin\theta$ (top), the same function of $\theta$ as for the aplanatic lens. The part $k\sin\theta>k$ is evanescent and not in Eq.~\eqref{eq:debye}.}
\label{fig:constructions}
\end{figure}

\begin{table}[t]
	\caption{\textbf{Choices of the non-paraxial Gaussian beam parametrization.} Each beam follows Eq.~\eqref{eq:debye}. 
    For E and F, $\theta_b=\arcsin(\mathrm{NA}/n)$ is the angle at which the amplitude falls to $1/e$ (it is not a stop).
    }
	\label{tab:env}
	\begin{tabular*}{\textwidth}{@{\extracolsep{\fill}}lllll@{}}
		\toprule
		label & beam & parameters & envelope $a(\theta)$ & $c(\theta)$\\
		\midrule
		A & Gaussian spectrum & $w_0$ & $\sqrt{(1+\cos^2\theta)/2}\;e^{-(kw_0\sin\theta)^2 / 4}$ & $1/\cos\theta$\\
		B & aplanatic lens & $w_0$ ($+$ NA of a stop) & $\cos^{1/2}\theta\;e^{-(kw_0\sin\theta)^2/4}$ & $1$\\
		C & thin lens & $w_0$ ($+$ NA of a stop) & $\cos^{-3/2}\theta\;e^{-(kw_0\tan\theta)^2/4}$ & $1$\\
		D& COMSOL plane wave expansion & $w_0$ & $\cos\theta\sqrt{(1+\cos^2\theta)/2}\;e^{-(kw_0\sin\theta)^2/4}$ & $\cos\theta$\\
		E & OTT \texttt{tantheta} (default) & NA & $e^{-(kw_0\tan\theta)^2/4}$, $kw_0=2/\tan\theta_b$ & $1$\\
		F&  OTT \texttt{sintheta} & NA & $e^{-(kw_0\sin\theta)^2/4}$, $kw_0=2n/\mathrm{NA}$ & $1$\\
		\bottomrule
	\end{tabular*}
\end{table}

\section{All optical force terms up to an octupole order}
\label{app:all_terms_forces}

The multipolar series of the optical force~\cite{Kislov2021APR,Chen2011NP,Jiang2015arXiv,Yu2019PRA,Jiang2016arXiv}, continued here to the octupole order, reads as:
\begin{align}
    \label{eq:force_all_app}
    \begin{split}
        &\mathbf{F} = \mathbf{F}^{\text{ext}} + \mathbf{F}^{\text{rec}}, \\
        &\mathbf{F}^{\text{ext}} \simeq \mathbf{F}_{e} + \mathbf{F}_{m} + \mathbf{F}_{Q_e} + \mathbf{F}_{Q_m} + \mathbf{F}_{O_e} + \mathbf{F}_{O_m}, \\
        &\mathbf{F}^{\text{rec}} \simeq \mathbf{F}^{\text{rec}}_{em} +  \mathbf{F}^{\text{rec}}_{e Q_e} + \mathbf{F}^{\text{rec}}_{m Q_m} + \mathbf{F}^{\text{rec}}_{Q_e Q_m} + \mathbf{F}^{\text{rec}}_{Q_e O_e} + \mathbf{F}^{\text{rec}}_{Q_m O_m} + \mathbf{F}^{\text{rec}}_{O_e O_m},
    \end{split}
\end{align}
where term with superscript ``rec'' (recoil term) arise from the asymmetry of the
scattered momentum flux by interfering multipoles indicated by the corresponding superscript~\cite{Toftul2026RMP}.
Other terms, \(\mathbf{F}^{\text{ext}}\),  represent the result of multipole interaction with the incident field. Explicitly these terms are written as such~\footnote{octupole terms were derived with the help of Claude Code Opus 5 model developed by Anthropic}:
\begin{align}
    \mathbf{F}_{e} &= \frac{1}{2} \operatorname{Re} \left[ {e}_i^* \cdot (\boldsymbol{\nabla}) {E}_i \right], \\
    \mathbf{F}_{m} &= \frac{1}{2} \operatorname{Re} \left[ {m}_i^* \cdot (\boldsymbol{\nabla}) {H}_i \right], \\
    \mathbf{F}^{\text{rec}}_{em} &= - \frac{k^4}{12 \pi \sqrt{\varepsilon \mu}} \operatorname{Re} \left( \mathbf{e}^* \times \mathbf{m}\right), \\
    {F}_{Q_e, i} &= \frac{1}{4} \operatorname{Re} \left[ (Q_e^*)_{jk} \nabla_i \nabla_k E_j \right], \\
    {F}_{Q_m, i} &= \frac{1}{4} \operatorname{Re} \left[ (Q_m^*)_{jk} \nabla_i \nabla_k H_j \right] , \\
    \mathbf{F}^{\text{rec}}_{e Q_e} &= - \frac{k^5}{ 40 \pi \varepsilon} \operatorname{Im} \left( \hat{\mathbf{Q}}_{e} \mathbf{e}^* \right), \\
    \mathbf{F}^{\text{rec}}_{m Q_m} &= - \frac{k^5}{ 40 \pi \mu} \operatorname{Im} \left( \hat{\mathbf{Q}}_{m} \mathbf{m}^* \right), \\
    {F}^{\text{rec}}_{Q_e Q_m, i} &= - \frac{k^6}{240 \pi \sqrt{\varepsilon \mu}} \operatorname{Re} \left( \epsilon_{ijk} Q_{e,lj}^* Q_{m, lk} \right), \\
    {F}_{O_e, i} &= \frac{1}{12} \operatorname{Re} \left[ (O_e^*)_{jkl} \nabla_i \nabla_k \nabla_l E_j \right], \\
    {F}_{O_m, i} &= \frac{1}{12} \operatorname{Re} \left[ (O_m^*)_{jkl} \nabla_i \nabla_k \nabla_l H_j \right], \\
    {F}^{\text{rec}}_{Q_e O_e, i} &= - \frac{k^7}{630 \pi \varepsilon} \operatorname{Im} \left( O_{e,ijk} Q^*_{e,jk} \right), \\
    {F}^{\text{rec}}_{Q_m O_m, i} &= - \frac{k^7}{630 \pi \mu} \operatorname{Im} \left( O_{m,ijk} Q^*_{m,jk} \right), \\
    {F}^{\text{rec}}_{O_e O_m, i} &= - \frac{k^8}{7560 \pi \sqrt{\varepsilon \mu}} \operatorname{Re} \left( \epsilon_{ijk} O_{e,lmj}^* O_{m, lmk} \right),
\end{align}
where \( \epsilon_{ijk}\) is Levi-Civita symbol, indices $i,j,k,l,m$ can be $x,y,z$. The extinction prefactors $1/2$, $1/4$ and $1/12$ are the Taylor coefficients $1/(2 \, \ell!)$ of the incident field expanded about the particle center. The induced multipoles for the isotropic particles are
\begin{equation}
     \mathbf{e} = \varepsilon \alpha_e \mathbf{E}, \qquad
    \mathbf{m} = \mu \alpha_m \mathbf{H},
\end{equation}
\begin{align}
    Q_{e, ij} &= \frac{1}{2}\varepsilon \alpha_{Q_e} \left( \nabla_i E_j + \nabla_j E_i\right) , \\
    Q_{m, ij} &= \frac{1}{2}\mu \alpha_{Q_m} \left( \nabla_i H_j + \nabla_j H_i\right), \\
    O_{e, ijk} &= \frac{1}{3}\varepsilon \alpha_{O_e} \left( \nabla_i \nabla_j E_k + \nabla_j \nabla_k E_i + \nabla_k \nabla_i E_j\right)^{(0)}, \\
    O_{m, ijk} &= \frac{1}{3}\mu \alpha_{O_m} \left( \nabla_i \nabla_j H_k + \nabla_j \nabla_k H_i + \nabla_k \nabla_i H_j\right)^{(0)}.
\end{align}
Here \( (\cdot)^{(0)} \) is the traceless part of a symmetric tensor, \( T^{(0)}_{ijk} = T_{ijk} - (\delta_{ij} t_k + \delta_{jk} t_i + \delta_{ki} t_j)/5 \) with \( t_i = T_{ill} \). Unlike the quadrupole, the octupole bracket is not traceless on its own: \( \boldsymbol{\nabla} \cdot \mathbf{E} = 0 \) and the Helmholtz equation give \( t_i = -k^2 E_i \). Keeping that trace would add a term with the shape of \( \mathbf{F}_{e} \).
We write these expressions in a slightly different way that can be found in literature in favor of more symmetric form, in particular magnetic dipole and magnetic quadrupole moments are defined to have dimensions  \( \left[ \mathbf{m} \right] = \left[ \text{kg}\:\text{m}^{3}\:\text{A}^{-1}\:\text{s}^{-2} \right] \) and $\left[ \hat{\mathbf{Q}}_{m} \right] = \left[\text{kg}\:\text{m}^{4}\:\text{A}^{-1}\:\text{s}^{-2} \right]$.
Polarizabilities for spherical particles are given by the Mie theory
\begin{align}
\begin{split}
    &\alpha_e = i \frac{6 \pi}{k^3} a_1, \quad
    \alpha_m = i \frac{6 \pi}{k^3} b_1, \\
    &\alpha_{Q_e} = i \frac{40 \pi}{k^5} a_2, \quad
    \alpha_{Q_m} = i \frac{40 \pi}{k^5} b_2, \\
    &\alpha_{O_e} = i \frac{315 \pi}{k^7} a_3, \quad
    \alpha_{O_m} = i \frac{315 \pi}{k^7} b_3,
\end{split}
\end{align}
where Mie coefficients are~\cite{Bohren1984}
\begin{align}
\begin{split}
    a_{\ell} &= \frac{\sqrt{\bar{\varepsilon}}\, \psi_\ell(k_{\text{par}} R) \psi^{\prime}_\ell(k R) - \sqrt{\bar{\mu}}\, \psi_\ell(k R) \psi^{\prime}_\ell(k_{\text{par}} R)}{\sqrt{\bar{\varepsilon}}\, \psi_\ell(k_{\text{par}} R) \xi^{\prime}_\ell(k R) - \sqrt{\bar{\mu}}\, \xi_\ell(k R) \psi^{\prime}_\ell(k_{\text{par}} R)}, \\
    b_{\ell} &= \frac{\sqrt{\bar{\mu}}\, \psi_\ell(k_{\text{par}} R) \psi^{\prime}_\ell(k R) - \sqrt{\bar{\varepsilon}}\, \psi_\ell(k R) \psi^{\prime}_\ell(k_{\text{par}} R)}{\sqrt{\bar{\mu}}\, \psi_\ell(k_{\text{par}} R) \xi^{\prime}_\ell(k R) - \sqrt{\bar{\varepsilon}}\, \xi_\ell(k R) \psi^{\prime}_\ell(k_{\text{par}} R)},
\end{split}
\label{eq:Mie_coef}
\end{align}
where $\psi_\ell(x) = x j_\ell(x)$ and $\xi_\ell(x) = x h_\ell^{(1)}(x)$ are the Riccati-Bessel functions,
$j_\ell$ is the spherical Bessel function, $h_\ell^{(1)}$ is the spherical Hankel function of the first kind,
the prime denotes derivative with respect to the argument, $\bar{\varepsilon} = \varepsilon_{\text{par}} /\varepsilon$ and $\bar{\mu} = \mu_{\text{par}} /\mu$ are the relative permittivity and permeability, and  $k_{\text{par}} = \omega \sqrt{\varepsilon_{\text{par}} \mu_{\text{par}}}$ is the wavenumber inside the sphere  of radius $R$.

\subsection{Force benchmarks and verification}

\begin{figure}
    \centering
    \includegraphics[width=0.55\linewidth]{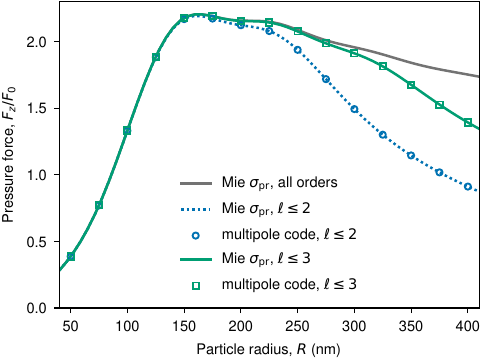}
    \caption{\textbf{Plane wave pressure.} Force on a VO$_2$ sphere ($\varepsilon = 6.55 + 3.66 i$) in ethanol at $\lambda = 976$~nm. Lines are the Mie series for $\sigma_{\rm pr}$ with all orders and truncated at $\ell_{\max} = 2$ and $3$; symbols are the force terms~\eqref{eq:force_all_app} evaluated on the same plane wave up to the quadrupole ($\ell \leq 2$) and the octupole ($\ell \leq 3$).}
    \label{fig:plane_wave}
\end{figure}

\begin{figure}
    \centering
    \includegraphics[width=\linewidth]{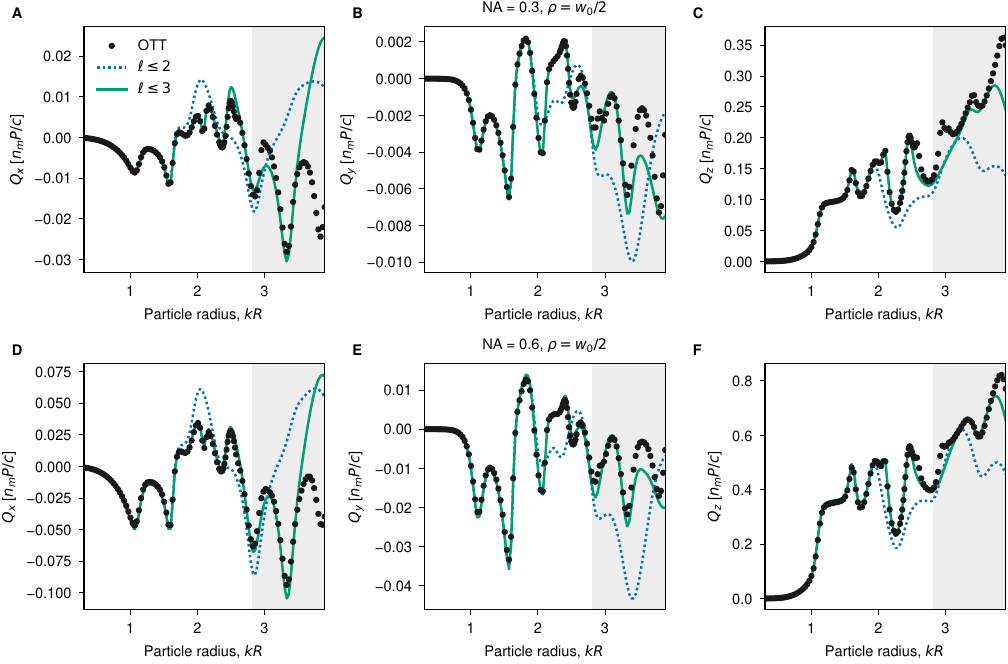}
    \caption{\textbf{Benchmark against the T-matrix method.} Force components $Q_x$ (\textsf{\textbf{A}}, \textsf{\textbf{D}}), $Q_y$ (\textsf{\textbf{B}}, \textsf{\textbf{E}}) and $Q_z$ (\textsf{\textbf{C}}, \textsf{\textbf{F}}) on a sphere with $n_p = 3.5 + 0.05 i$ in water ($n_m = 1.33$) at $\rho = w_0/2$ in the focal plane of a circularly polarized beam of $\mathrm{NA} = 0.3$ (\textsf{\textbf{A}}--\textsf{\textbf{C}}) and $\mathrm{NA} = 0.6$ (\textsf{\textbf{D}}--\textsf{\textbf{F}}), against the size parameter $kR$, $k = n_m \omega / c$. %
    The beam has the envelope of OTT, row E of Table~\ref{tab:env}, with $w_0 = 2/(k \tan\theta_b)$, but each plane wave is polarized with $c = 1/\cos\theta$ instead of $c = 1$. Forces are normalized as $\mathbf{Q} = \mathbf{F} c / (n_m P)$, where the $P$ is beam power \eqref{eq:power}. Grey marks $kR > 2.8$, where $|a_4|$ and/or $|b_4|$ are no longer small.}
    \label{fig:octupole_benchmark}
\end{figure}

We verify multipole decomposition of the force by testing it against analytical solution for the plane wave pressure~\cite{Bohren1984} (Fig.~\ref{fig:plane_wave}) as well as against the existing implementation using T-matrix approach, namely Optical Tweezers Toolbox~\cite{Lenton2020OTT} (Fig.~\ref{fig:octupole_benchmark}).

Plane wave pressure on an isotropic particle is~\cite{Bohren1984}
\begin{equation}
    F_z =  F_0 \: \frac{\sigma_{\rm pr}}{  \sigma_{\text{geom}}},
\qquad
\sigma_{\rm pr} = \sigma_{\rm ext} - \langle \cos\theta \rangle \sigma_{\rm sca},
\end{equation}
where $\sigma_{\text{geom}} = \pi R^2$ is the geometrical cross section, pressure cross section $\sigma_{\rm pr}$ expressed via
\begin{align}
    \sigma_{\rm ext} &= \frac{2\pi}{k^2} \sum_{\ell=1}^{\ell_{\max}} (2\ell+1)\, \operatorname{Re}(a_\ell + b_\ell), \\
    \langle \cos\theta \rangle \sigma_{\rm sca}
&= \frac{4\pi}{k^2} \left[
    \sum_{\ell=1}^{\ell_{\max}-1} \frac{\ell(\ell+2)}{\ell+1}\, \operatorname{Re}\!\left(a_\ell a_{\ell+1}^* + b_\ell b_{\ell+1}^*\right)
  + \sum_{\ell=1}^{\ell_{\max}} \frac{2\ell+1}{\ell(\ell+1)}\, \operatorname{Re}\!\left(a_\ell b_\ell^*\right)
  \right],
\end{align}
with $a_\ell$, $b_\ell$ the Mie coefficients \eqref{eq:Mie_coef}. The normalization force $F_0$ is the force on  a perfectly absorbing disk with area $\sigma_{\text{geom}}$:
\begin{equation}
    F_0 =\sigma_{\text{geom}} \varepsilon_0 \varepsilon_{\rm h} |E_0|^2 /2.
\end{equation}
Truncating the series at $\ell_{\max} = 2$ or $3$ gives the pressure that the force terms~\eqref{eq:force_all_app} up to the quadrupole or the octupole must reproduce, and they do to $10^{-4}$ (Fig.~\ref{fig:plane_wave}). The gap to the full series is the truncation error of the multipole expansion itself.

For a focused beam there is no closed form, so we compare with the T-matrix calculation of the Optical Tweezers Toolbox~\cite{Lenton2020OTT} for a sphere with $n_p = 3.5 + 0.05 i$ in water swept through its Mie resonances (Fig.~\ref{fig:octupole_benchmark}). The particle sits at $\rho = w_0/2$ in the focal plane of a circularly polarized beam, which makes all three force components non-zero. The octupole terms are what keeps the series on the T-matrix result above $R \approx 0.3 \lambda_m$, up to where $a_4$ and $b_4$ become noticeable.

\subsection{Comparison of different approximations}
\label{app:approx}

\begin{figure}
    \centering
    \includegraphics[width=0.9\linewidth]{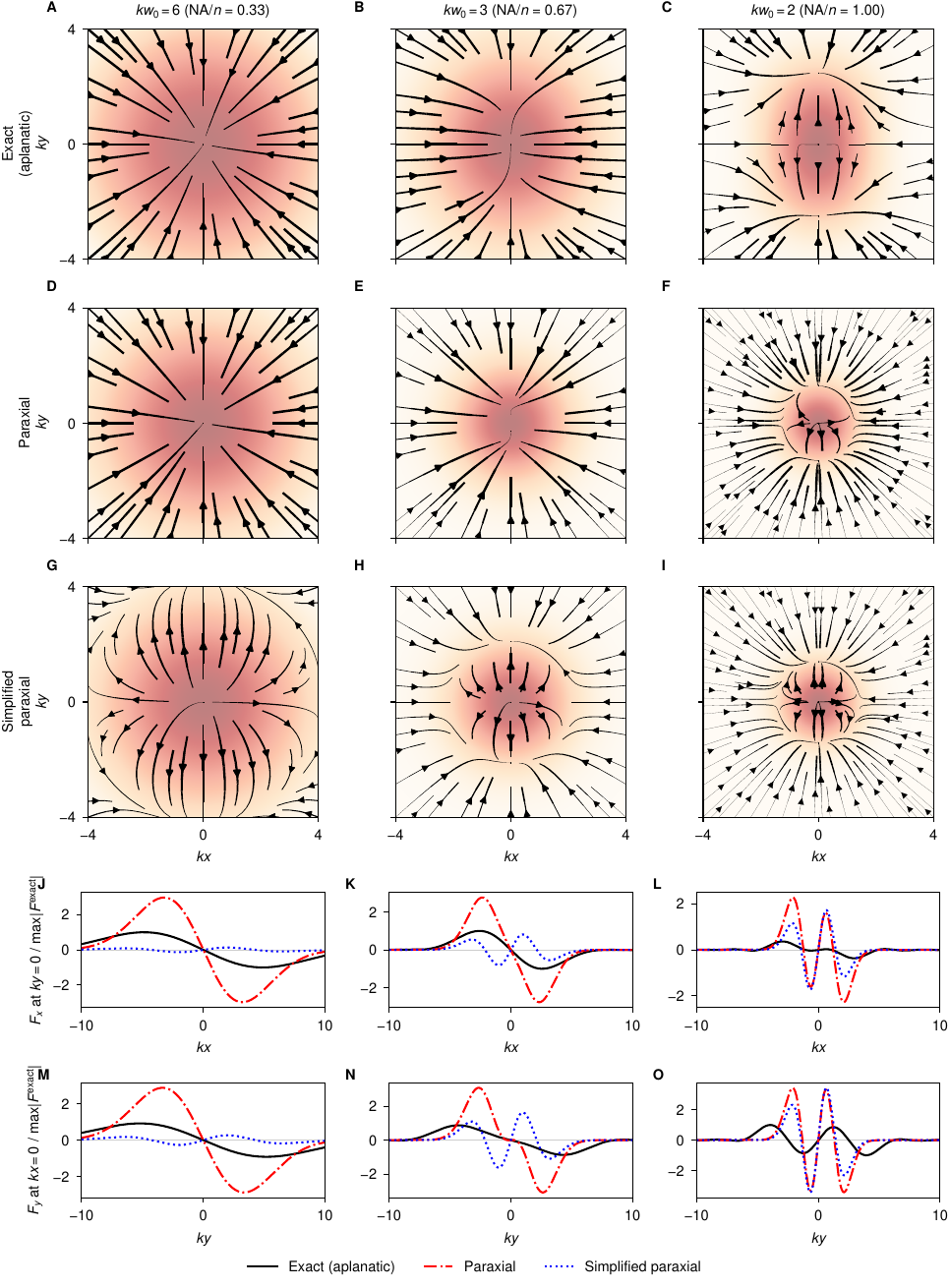}
    \caption{\textbf{Comparison of the beam approximations.} Streamlines of the magnetic quadrupole force $\mathbf{F}_{Q_{m}}$, Eq.~\eqref{eq:force_quad}, in the focal plane over the beam intensity, for the aplanatic beam B of Table~\ref{tab:env} with a stop at $\mathrm{NA}/n = 2/(kw_0)$ (\textsf{\textbf{A}}--\textsf{\textbf{C}}), the paraxial beam~\eqref{eq:parax_true} (\textsf{\textbf{D}}--\textsf{\textbf{F}}) and the simplified paraxial beam~\eqref{eq:gauss} (\textsf{\textbf{G}}--\textsf{\textbf{I}}), for $k w_0 = 6, 3, 2$; $\mathrm{NA}/n = 2/(k w_0)$. Line width follows $|\mathbf{F}_{Q_m}|$ within each panel, $\alpha_{Q_m}$ is real. (\textsf{\textbf{J}}--\textsf{\textbf{L}}) $F_x$ along the $x$ axis and (\textsf{\textbf{M}}--\textsf{\textbf{O}}) $F_y$ along the $y$ axis for the same three beams, each column normalized to the peak of the non-paraxial beam over both cuts. Beam is linearly polarized along the $x$-axis. 
    All beams carry the same power, that of the paraxial beam~\eqref{eq:parax_true} with the focal amplitude $E_0$, as in Fig.~\ref{fig:model_comparison}.}
    \label{fig:parax_vs_simple_parax}
\end{figure}

In Fig.~\ref{fig:parax_vs_simple_parax} we compare $\mathbf{F}_{Q_{m}}$ of Eq.~\eqref{eq:force_quad} for the aplanatic beam B of Table~\ref{tab:env} with a stop at $\mathrm{NA}/n = 2/(kw_0)$ (Fig.~\ref{fig:parax_vs_simple_parax}\textsf{\textbf{A}}--\textsf{\textbf{C}}), the paraxial beam~\eqref{eq:parax_true} (Fig.~\ref{fig:parax_vs_simple_parax}\textsf{\textbf{D}}--\textsf{\textbf{F}}) and the simplified paraxial beam~\eqref{eq:gauss} used in the main text (Fig.~\ref{fig:parax_vs_simple_parax}\textsf{\textbf{G}}--\textsf{\textbf{I}}), with cuts along the two axes on one scale (Fig.~\ref{fig:parax_vs_simple_parax}\textsf{\textbf{J}}--\textsf{\textbf{O}}).

\begin{figure*}
    \centering
    \includegraphics[width=\linewidth]{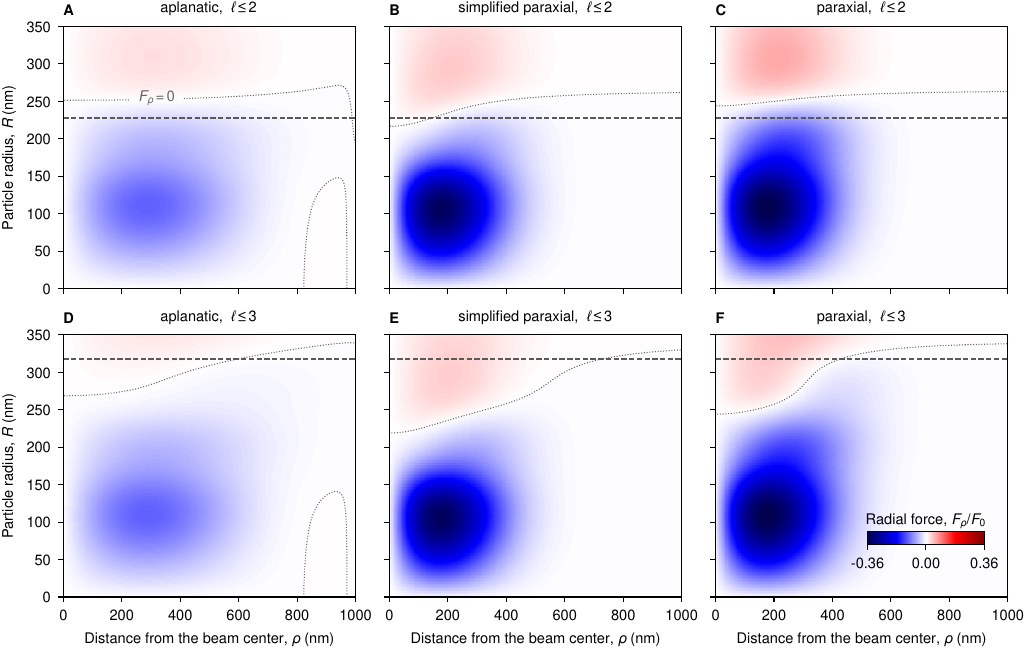}
    \caption{\textbf{Beam model against multipole order.} Radial force $F_\rho / F_0$ on a VO$_2$ sphere ($\varepsilon = 6.55 + 3.66 i$) in ethanol ($\varepsilon_{\rm h} = 1.85$) at $\lambda = 976$~nm in a circularly polarized beam of $\mathrm{NA} = 0.9$.  
    Columns are the aplanatic beam B of Table~\ref{tab:env} with a stop at $\mathrm{NA}$, the simplified paraxial beam~\eqref{eq:gauss} and the paraxial beam~\eqref{eq:parax_true}, all with the waist $w_0 = \lambda/(\pi\,\mathrm{NA}) = 346$~nm.
    Rows are the multipole series~\eqref{eq:force_all_app} up to the quadrupole ($\ell \leq 2$) and the octupole ($\ell \leq 3$). 
    $F_0 = \sigma_{\text{geom}} \varepsilon_0 \varepsilon_{\rm h} |E_0|^2 / 2$, where $E_0$ is fixed by the beam power, $P = \pi w_0^2 \sqrt{\varepsilon/\mu}\,|E_0|^2/4$, the power of the paraxial beam~\eqref{eq:parax_true} with the focal amplitude $E_0$; the stop leaves the focal field of beam B at $0.68\,E_0$. Dotted line is $F_\rho = 0$. 
    Dashed line is the radius above which the first neglected order carries more than $0.05$: $|a_3|$ or $|b_3|$ in the top row, $|a_4|$ or $|b_4|$ in the bottom.}
    \label{fig:model_comparison}
\end{figure*}

Figure~\ref{fig:model_comparison} separates the two approximations behind Eq.~\eqref{eq:Fext_lin}, the beam and the truncation of the multipole series, for the VO$_2$ particle of the experiment. The columns change only the beam and the rows only the series, so the zero contour of the radial force, which is where the particle sits off the axis, can be read across both. %
Off-center trapping sets in at $R = 252$~nm ($\ell \leq 2$) and $270$~nm ($\ell \leq 3$) in the aplanatic beam, at $244$ and $246$~nm in the paraxial beam and at $218$ and $220$~nm in the simplified beam. Above $R \approx 265$~nm the series truncated at the quadrupole predicts expulsion from the beam for every beam model, whereas the octupole terms keep the stable ring up to $R \approx 330$--$340$~nm. 

\subsection{Role of numerical aperture}
\label{app:aperture}

\begin{figure*}
    \centering
    \includegraphics[width=\linewidth]{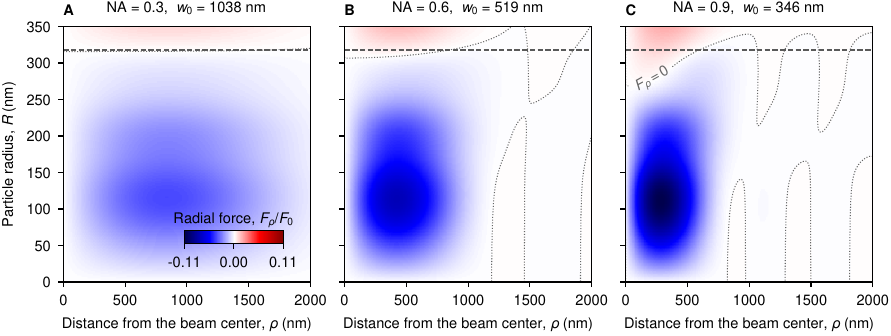}
    \caption{\textbf{Role of the numerical aperture.} Radial force $F_\rho / F_0$ on the VO$_2$ sphere of Fig.~\ref{fig:model_comparison} 
    in the aplanatic beam B of Table~\ref{tab:env} with a stop at the aperture and all terms to the octupole, for $\mathrm{NA} = 0.3$ (\textsf{\textbf{A}}), $0.6$ (\textsf{\textbf{B}}) and $0.9$ (\textsf{\textbf{C}}), with the waist $w_0 = \lambda/(\pi\,\mathrm{NA})$. 
    Same normalization and lines as in Fig.~\ref{fig:model_comparison}, one colour scale for all three panels.}
    \label{fig:aperture_maps}
\end{figure*}

Figure~\ref{fig:aperture_maps} shows the same particle in the same beam at three apertures. %
The zero contour leaves the axis at $R = 316$, $308$ and $270$~nm for $\mathrm{NA} = 0.3$, $0.6$ and $0.9$, and that radius is the smallest particle that sits off the beam center. Opening the aperture brings it down into the size range of the experiment: at $\mathrm{NA} = 0.3$ every particle we observed hopping would sit on the axis. 

\subsection{Trapping regimes of the quadrupole model}
\label{app:regimes}

\begin{figure}
    \centering
    \includegraphics[width=\linewidth]{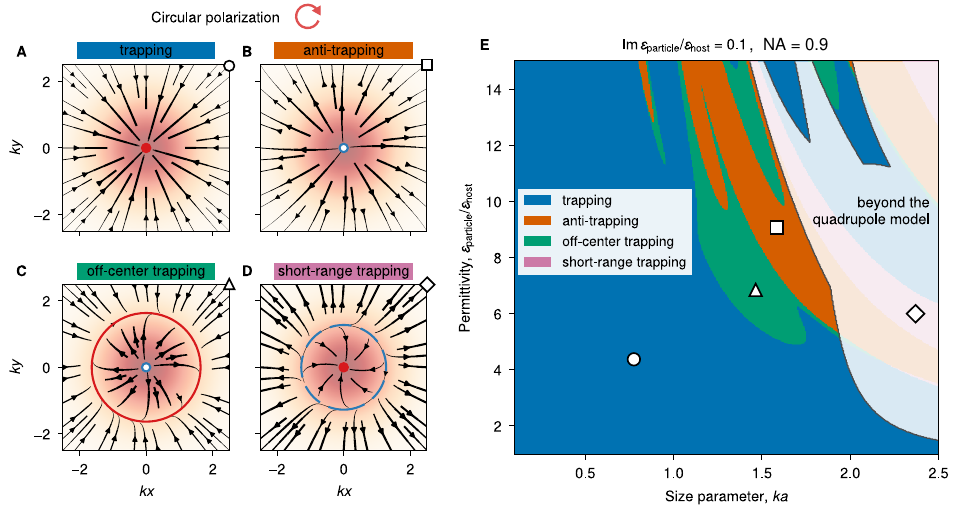}
    \caption{\textbf{Trapping regimes of the quadrupole model.} (\textsf{\textbf{A}}--\textsf{\textbf{D}}) Streamlines of the transverse force on one sphere of each regime in the focal plane of the circularly polarized beam~\eqref{eq:gauss}, over the beam intensity, with the series~\eqref{eq:force_all_app} stopped at the quadrupole; the line width follows the force. Red marks a stable equilibrium, a point or a ring, blue dashed an unstable one. (\textsf{\textbf{E}}) The regime against size parameter $ka$ and permittivity for $\mathrm{NA} = 0.9$ in ethanol, $k w_0 = 2.27$, $\operatorname{Im} \varepsilon_{\rm particle} / \varepsilon_{\rm host} = 0.1$, with the four examples marked. The whited-out region is where $|a_3|$ or $|b_3|$ exceeds $0.05$.}
    \label{fig:trapping_phases_quad}
\end{figure}

In the beam~\eqref{eq:gauss} with the series stopped at the quadrupole, the radial force under circular polarization is $F_{\rho} \propto \rho \, (a + b \rho^2) \, e^{-2\rho^2/w_0^2}$ ($a$ and $b$ are some constants), Eq.~\eqref{eq:F_circ}, and the recoil terms keep that form. It changes sign at most once, so only four regimes are possible:
\begin{enumerate}
    \item the particle is pulled to the axis at every distance (trapping),
    \item pushed away at every distance (anti-trapping),
    \item pushed away near the axis and pulled in further out, which is a stable ring (off-center trapping),
    \item pulled in near the axis and pushed away beyond a finite radius (short-range trapping).
\end{enumerate}
Figure~\ref{fig:trapping_phases_quad} shows one example of each and the regime of every sphere with $ka \leq 2.5$ and $\varepsilon_{\rm particle} / \varepsilon_{\rm host} \leq 15$ in the beam~\eqref{eq:gauss} of $\mathrm{NA} = 0.9$ with the waist $w_0 = 2/(k\tan\theta_{\max}) = 259$~nm. The sign is read only where the force exceeds $10\%$ of its peak, which stands for the thermal floor. Where $|a_3|$ or $|b_3|$ exceeds $0.05$ the quadrupole model has nothing to say and the map is whited out; short-range trapping lives almost entirely there. 

\subsection{Other trapping materials}
\label{app:materials}

\begin{figure}
    \centering
    \includegraphics[width=\linewidth]{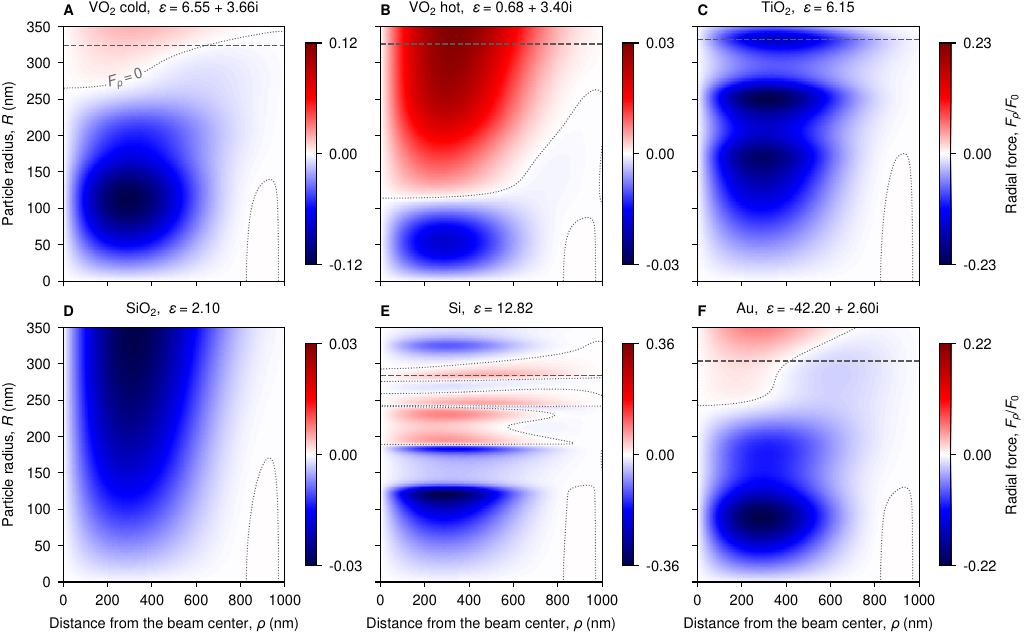}
    \caption{\textbf{Six materials in the trap.} Radial force $F_\rho / F_0$ on spheres made of different materials submerged in water ($n = 1.33$) at $\lambda = 976$~nm 
    in the circularly polarized aplanatic beam B of Table~\ref{tab:env} with a stop at $\mathrm{NA} = 0.9$, $w_0 = \lambda/(\pi\,\mathrm{NA}) = 346$~nm, with all terms to the octupole. Normalization and lines as in Fig.~\ref{fig:model_comparison}. Dashed line is where $|a_4|$ or $|b_4|$ exceeds $0.05$.}
    \label{fig:material_maps}
\end{figure}

\begin{figure}
    \centering
    \includegraphics[width=\linewidth]{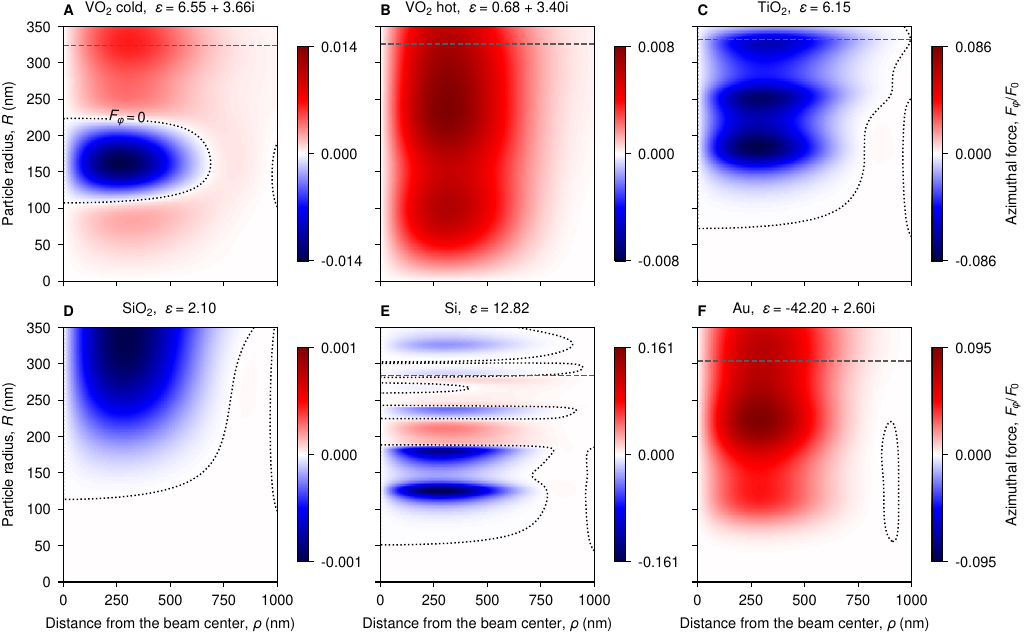}
    \caption{\textbf{Which way six materials orbit.} Azimuthal force $F_\varphi / F_0$ on the spheres of Fig.~\ref{fig:material_maps} in the same beam, $\sigma = +1$. Red ($F_\varphi > 0$) is orbiting in the sense in which the field spins in time, blue against it. Dotted is $F_\varphi = 0$, dashed is where $|a_4|$ or $|b_4|$ exceeds $0.05$. Normalization as in Fig.~\ref{fig:model_comparison}.}
    \label{fig:material_maps_orbiting}
\end{figure}

Figure~\ref{fig:material_maps} repeats the $\mathrm{NA} = 0.9$ map for six materials common in optical trapping, in water at $\lambda = 976$~nm, with the relative permittivity of each in the panel title. %
Off-centre trapping is not particular to VO$_2$ and is possible with other materials. Our numerical observations show that the soft requirements are
\begin{enumerate}
    \item high numerical aperture, as lower in-plane field gradients flatten the $F_{\rho}$ lines in Fig.~\ref{fig:aperture_maps};
    \item lossy and/or high index materials are more prominent to the appearance of the stable rings (Fig.~\ref{fig:material_maps}).
\end{enumerate}
We note that the orbiting direction in principle can arbitrary. Figure~\ref{fig:material_maps_orbiting} shows the azimuthal force on the same spheres, which sets the sense of orbiting on the ring.
Negative orbiting rotation suggests strong positive spinning of the trapped particle according to the angular momentum conservation law. To describe these effectes the gradeint of the fields has to be taken into account in the torque calculation routines in the multipolar approach~\cite{Xu2024NC,Wei2022PRB}.

\subsection{Particle shape}
\label{app:shape}

\begin{figure}
    \centering
    \includegraphics[width=0.7\linewidth]{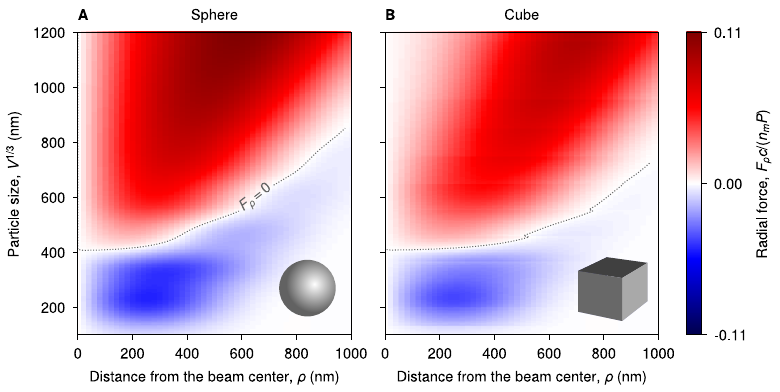}
    \caption{\textbf{Sphere and cube of equal volume.} Radial force $Q_\rho = F_\rho c/(n_m P)$ on a VO$_2$ sphere (\textsf{\textbf{A}}) and a VO$_2$ cube (\textsf{\textbf{B}}), $\varepsilon_{\rm VO_2} = 6.55 + 3.66i$, in ethanol ($\varepsilon = 1.85$) at $\lambda = 976$~nm, against the distance $\rho$ from the beam axis in the focal plane and the cube root of the particle volume $V^{1/3}$. The beam is the circularly polarized beam E of Table~\ref{tab:env} with $\mathrm{NA} = 0.9$, $w_0 = 2/(k \tan\theta_{\max}) = 259$~nm. The cube faces are normal to $x$, $y$ and $z$, and the cube moves along a face normal. Forces are computed with the Optical Tweezers Toolbox~\cite{Lenton2020OTT} (the cube T-matrix by point matching) and normalized to the beam power $P$. Red is outward, blue inward, one colour scale for both panels. Dotted line shows $F_\rho = 0$.}
    \label{fig:shape_maps}
\end{figure}

The synthesized particles are not spheres (Fig.~\ref{fig:concept}). Figure~\ref{fig:shape_maps} tests whether the off-centre ring depends on the spherical shape. We compare a sphere with a cube of the same volume, both computed by the T-matrix method of the Optical Tweezers Toolbox~\cite{Lenton2020OTT}, the cube by point matching. The vertical axis is the cube root of the volume, $V^{1/3}$, which is $(4\pi/3)^{1/3} R \approx 1.61 R$ for a sphere and the edge length for a cube. For both shapes the beam centre becomes unstable at $V^{1/3} = 410$~nm, which is a sphere of diameter $510$~nm, inside the experimental window of $400$--$625$~nm. Above this size the zero line of $F_\rho$, where the particle sits on a ring, runs at nearly the same distance from the axis for the two shapes, and the peak force on the cube is $20\%$ smaller. 
The size window of off-centre trapping is therefore set rather by the particle volume, not by its shape.

\section{Stochastic simulations}
\label{app:stochastic}

\begin{figure}
    \centering
    \includegraphics[width=0.8\linewidth]{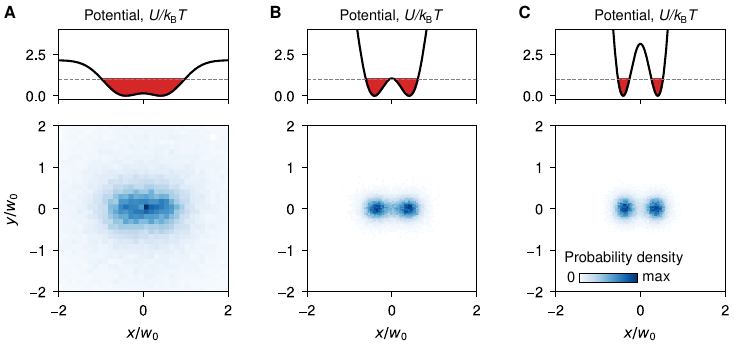}
    \caption{\textbf{Stochastic simulation results.} Parameters were used: $A = 0.5$, $B = 0$, $C = 1.5$, and $U_0 = 16k_{\text{B}} T$ (\textsf{\textbf{A}}), $U_0 = 112 k_{\text{B}} T$ (\textsf{\textbf{B}}), and $U_0 = 336 k_{\text{B}} T$ (\textsf{\textbf{C}}).}
    \label{fig:stochastic_sim}
\end{figure}

We simulate the Brownian motion of a particle of mass $m$ in the optical force vector field $\mathbf{F}$ using Newton's equations of motion~\cite{Pesce2020EPJP}:
\begin{equation}
    m \ddot{\mathbf{r}} = \mathbf{F} - \gamma \dot{\mathbf{r}}  + \mathbf{F}_{\text{st}}.
    \label{eq:langevine}
\end{equation}
Here $\gamma$ is the friction coefficient, $\mathbf{F}_{\text{st}}$ is the stochastic force,
Stochastic force is such that $\langle\mathbf{F}_{\text{st}} (t)\rangle_t = 0$ and $\langle{F}_{\text{st},i} (t) {F}_{\text{st},j} (t+T) \rangle_t = 2 S_F \delta_{ij} \delta(T)$, i.e. it is a white noise with the force-noise strength $S_F$. Both friction force and stochastic force originate from numerous collisions with smaller particles in the host fluid.
It is possible to derive that $S_F = k_{\text{B}} T \gamma$ (e.g. see Chapter 5 in Ref.~\cite{Jacobs2013}); the diffusion constant of the particle is then $k_{\text{B}} T / \gamma$.
So once there is a viscosity force which leads to dissipation, there is a
stochastic force always present.
These two terms are fundamentally interrelated by the fluctuation-dissipation theorem~\cite{Pesce2020EPJP,Jacobs2013}.

In the low-Reynolds-number regime, it is possible to drop the inertial term in Eq.~\eqref{eq:langevine}, obtaining the \textit{overdamped Langevin equation}:
\begin{equation}
    \dot{\mathbf{r}} = \gamma^{-1}\mathbf{F} + \gamma^{-1} \mathbf{F}_{\text{st}}.
\end{equation}
After a non-dimensionalization procedure and using finite difference iteration method, we arrive at the following iterative rule:
\begin{equation}
    \bar{\mathbf{r}}_{i+1} = \bar{\mathbf{r}}_{i} +  \bar{\mathbf{F}}(\bar{\mathbf{r}}_{i})  \Delta \bar{t} + \sqrt{2 \Delta \bar{t}} \: \mathbf{w}_{i}
\end{equation}
where $\bar{\mathbf{r}} = \mathbf{r} / w_0$, $\bar{t} = t / \tau$, $\Delta \bar{t}$ is the simulation time step, $ \bar{\mathbf{F}} =- \partial \bar{U} / \partial \bar{\mathbf{r}}$, and $\bar{U} = U / (k_{\text{B}} T)$. Here $\mathbf{w}_i = (w_{i,x}, w_{i,y})$, where $w_{i,x}$ and $w_{i,y}$ are independent Gaussian random numbers with zero mean and unitary variance.
We note that the dispersion of the white noise depends on square root of the time step~\cite{Pesce2020EPJP}.
In the overdamped regime there is only one critical parameter, which is a relative potential depth in the units of $k_{\text{B}}T$.
The characteristic time is $\tau = w_0^2 \gamma / (k_{\text{B}} T)$, where for spherical particle  of radius $a$ in a liquid with dynamic viscosity $\eta$ we have $\gamma = 6\pi a \eta$.

Fig.~\ref{fig:stochastic_sim} shows the steady-state probability densities obtained from stochastic simulations in three representative potentials of increasing depth, i.e. varying trapping laser intensity. The top panels display the dimensionless potential energy landscapes $U/k_{\text{B}}T$, while the bottom panels show the corresponding two-dimensional probability density maps. The latter is computed by collecting the positions of the particle over a long trajectory and binning them into a two-dimensional histogram. This reflects the likelihood of finding the particle at a given position under thermal fluctuations in the specified potential landscape.

In Fig.~\ref{fig:stochastic_sim}\textsf{\textbf{A}}, the potential well is shallow ($U_0 = 16 k_{\text{B}} T$), and the particle distribution remains broad and centered. 
In Fig.~\ref{fig:stochastic_sim}\textsf{\textbf{B}}, for a moderate potential depth ($U_0 = 112 k_{\text{B}} T$), the landscape develops a double-well structure, leading to a bimodal distribution. 
In Fig.~\ref{fig:stochastic_sim}\textsf{\textbf{C}}, for a deep potential ($U_0 = 336 k_{\text{B}} T$), the probability density is strongly localized near the two minima, indicating reduced thermal hopping between wells. The parameters used for Eq.~\eqref{eq:Utheory} from the main text in all cases are $A = 0.5$, $B = 0$, and $C = 1.5$.

\section{Quantitative comparison of the Kramers prefactor}
\label{app:prefactors}

To further quantify the thermally activated hopping dynamics, we evaluate the Kramers prefactor from the experimentally reconstructed trapping potential. For an overdamped particle escaping from a one-dimensional potential well, the Kramers transition rate is
\[
W_{\mathrm K}
=
W_0
\exp\left(-\frac{\Delta U}{k_{\mathrm B}T}\right),
\tag{G1}
\]
where \(\Delta U=U(y_b)-U(y_{\min})\) is the barrier height and the prefactor is given by
\[
W_0^{\mathrm{th}}
=
\frac{
\sqrt{
U''(y_{\min})
\left|U''(y_b)\right|
}
}
{2\pi\gamma}.
\tag{G2}
\]
Here, \(y_{\min}\) denotes the position of the stable potential minimum, \(y_b\) is the position of the central barrier, and \(\gamma\) is the viscous damping coefficient. In the present analysis, we use
\[
\gamma=4.41\times10^{-9}\ {\rm kg\,s^{-1}}.
\]
The one-dimensional trapping potential along the hopping direction was reconstructed from the experimentally measured particle-position probability density \(P(y)\) according to
\[
U_{\mathrm{exp}}(y)
=
-k_{\mathrm B}T\ln P(y)+C,
\tag{G3}
\]
where \(C\) is an arbitrary energy offset. The reconstructed potential was subsequently fitted using the one-dimensional form of the quadrupolar potential in Eq. (15). An additive energy offset and a small displacement of the potential center were allowed in the fitting to account for the arbitrary energy reference and experimental coordinate origin.
Importantly, both the barrier height and the local curvatures entering Eq. (G2) were obtained from the same fitted potential. This procedure avoids direct numerical evaluation of the second derivative of the experimentally reconstructed potential, which is particularly sensitive to statistical fluctuations in the position histogram. The curvature \(U''(y_{\min})\) was evaluated at the potential minimum corresponding to the well used for the experimental dwell-time analysis, while \(U''(y_b)\) was evaluated at the central barrier.
The experimentally inferred prefactor was independently obtained from the measured hopping rate \(W_{\mathrm{exp}}\) as
\[
W_0^{\mathrm{exp}}
=
W_{\mathrm{exp}}
\exp\left(\frac{\Delta U}{k_{\mathrm B}T}\right).
\tag{G4}
\]
Fig. 18 compares the experimentally inferred prefactors with the theoretical values calculated from Eq. (G2). The two quantities are found to lie on the same frequency scale and show the same overall increasing tendency. The agreement is semi-quantitative rather than exact. This is expected because the Kramers prefactor depends explicitly on the local curvatures of the potential and is therefore more sensitive than the barrier height to experimental uncertainty in the reconstructed potential landscape.
Additional deviations may arise from the irregular morphology of the VO\(_2\) particles, uncertainty in their effective hydrodynamic drag, particle–surface interactions, and the use of the simplified analytical potential. Nevertheless, the experimentally inferred and theoretically calculated prefactors remain of the same order of magnitude over the investigated range, providing an additional quantitative validation of the Kramers description of the observed hopping dynamics.

\begin{figure}
    \centering
    \includegraphics[width=0.8\linewidth]{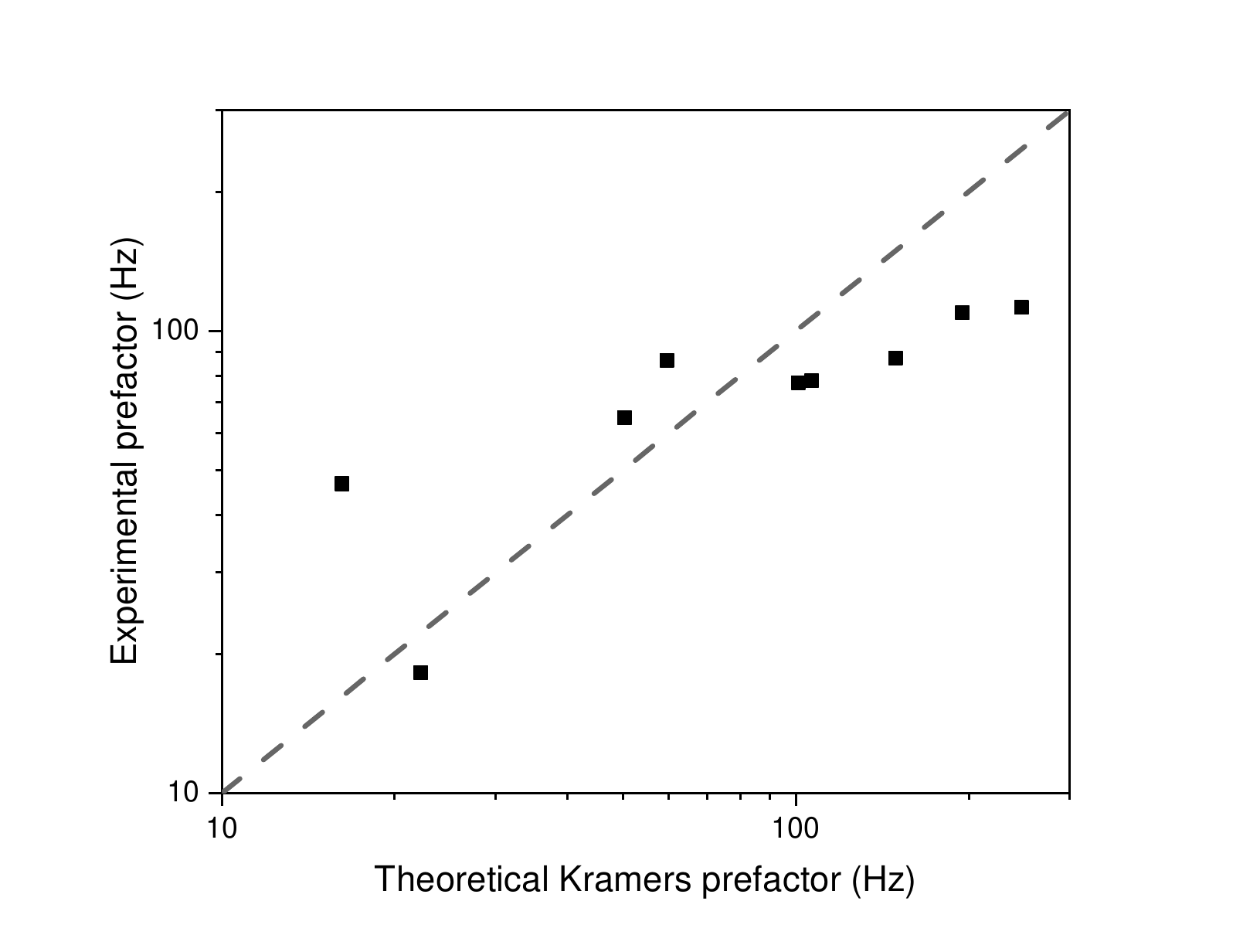}
    \caption{\textbf{Comparison of the experimentally inferred and theoretically calculated Kramers prefactors.} The experimental prefactor \(W_0^{\mathrm{exp}}\) is obtained from the measured hopping rate and barrier height using Eq. (G4), whereas the theoretical prefactor \(W_0^{\mathrm{th}}\) is calculated from the local curvatures of the fitted double-well potential and the viscous damping coefficient using Eq. (G2). Both axes are shown on logarithmic scales. The dashed line indicates \(W_0^{\mathrm{exp}}=W_0^{\mathrm{th}}\).}
    \label{fig:prefactors}
\end{figure}

\section{Supplementary movies}
\label{app:movies}

\textbf{Movie S1.} Original recorded video of a trapped particle, which is analyzed in Fig.~\ref{fig:experiment_transition_to_hopping}\textsf{\textbf{A}}.

\textbf{Movie S2.} Original recorded video of a trapped particle, which is analyzed in Fig.~\ref{fig:experiment_transition_to_hopping}\textsf{\textbf{B}}.

\textbf{Movie S3.} Original recorded video of a trapped particle, which is analyzed in Fig.~\ref{fig:experiment_transition_to_hopping}\textsf{\textbf{C}}.

\textbf{Movie S4.} Original recorded video of a trapped particle for vertical polarization, which is analyzed in Fig.~\ref{fig:experiment_transition_to_hopping}\textsf{\textbf{M}}.

\textbf{Movie S5.} Original recorded video of a trapped particle, which is analyzed in Fig.~\ref{fig:experiment_transition_to_hopping}\textsf{\textbf{N}}.

\textbf{Movie S6.} Original recorded video of a trapped particle, which is analyzed in Fig.~\ref{fig:lin_circ_pol_v1}\textsf{\textbf{A}}.

\bibliography{VO2_hopping_refs}

\begin{thebibliography}{81}%
\makeatletter
\providecommand \@ifxundefined [1]{%
 \@ifx{#1\undefined}
}%
\providecommand \@ifnum [1]{%
 \ifnum #1\expandafter \@firstoftwo
 \else \expandafter \@secondoftwo
 \fi
}%
\providecommand \@ifx [1]{%
 \ifx #1\expandafter \@firstoftwo
 \else \expandafter \@secondoftwo
 \fi
}%
\providecommand \natexlab [1]{#1}%
\providecommand \enquote  [1]{``#1''}%
\providecommand \bibnamefont  [1]{#1}%
\providecommand \bibfnamefont [1]{#1}%
\providecommand \citenamefont [1]{#1}%
\providecommand \href@noop [0]{\@secondoftwo}%
\providecommand \href [0]{\begingroup \@sanitize@url \@href}%
\providecommand \@href[1]{\@@startlink{#1}\@@href}%
\providecommand \@@href[1]{\endgroup#1\@@endlink}%
\providecommand \@sanitize@url [0]{\catcode `\\12\catcode `\$12\catcode
  `\&12\catcode `\#12\catcode `\^12\catcode `\_12\catcode `\%12\relax}%
\providecommand \@@startlink[1]{}%
\providecommand \@@endlink[0]{}%
\providecommand \url  [0]{\begingroup\@sanitize@url \@url }%
\providecommand \@url [1]{\endgroup\@href {#1}{\urlprefix }}%
\providecommand \urlprefix  [0]{URL }%
\providecommand \Eprint [0]{\href }%
\providecommand \doibase [0]{https://doi.org/}%
\providecommand \selectlanguage [0]{\@gobble}%
\providecommand \bibinfo  [0]{\@secondoftwo}%
\providecommand \bibfield  [0]{\@secondoftwo}%
\providecommand \translation [1]{[#1]}%
\providecommand \BibitemOpen [0]{}%
\providecommand \bibitemStop [0]{}%
\providecommand \bibitemNoStop [0]{.\EOS\space}%
\providecommand \EOS [0]{\spacefactor3000\relax}%
\providecommand \BibitemShut  [1]{\csname bibitem#1\endcsname}%
\let\auto@bib@innerbib\@empty
\bibitem [{\citenamefont
  {Garc{\ifmmode\acute{\imath}\else\'{\i}\fi}a-M{\ifmmode\ddot{u}\else\"{u}\fi}ller}\
  \emph {et~al.}(2008)\citenamefont
  {Garc{\ifmmode\acute{\imath}\else\'{\i}\fi}a-M{\ifmmode\ddot{u}\else\"{u}\fi}ller},
  \citenamefont {Borondo}, \citenamefont {Hernandez},\ and\ \citenamefont
  {Benito}}]{Garcia-Muller2008Oct}%
  \BibitemOpen
  \bibfield  {author} {\bibinfo {author} {\bibfnamefont {P.~L.}\ \bibnamefont
  {Garc{\ifmmode\acute{\imath}\else\'{\i}\fi}a-M{\ifmmode\ddot{u}\else\"{u}\fi}ller}},
  \bibinfo {author} {\bibfnamefont {F.}~\bibnamefont {Borondo}}, \bibinfo
  {author} {\bibfnamefont {R.}~\bibnamefont {Hernandez}},\ and\ \bibinfo
  {author} {\bibfnamefont {R.~M.}\ \bibnamefont {Benito}},\ }\bibfield  {title}
  {\bibinfo {title} {{Solvent-Induced Acceleration of the Rate of Activation of
  a Molecular Reaction}},\ }\href
  {https://doi.org/10.1103/PhysRevLett.101.178302} {\bibfield  {journal}
  {\bibinfo  {journal} {Phys. Rev. Lett.}\ }\textbf {\bibinfo {volume} {101}},\
  \bibinfo {pages} {178302} (\bibinfo {year} {2008})}\BibitemShut {NoStop}%
\bibitem [{\citenamefont {Pollak}\ and\ \citenamefont
  {Miret-Art{\ifmmode\acute{e}\else\'{e}\fi}s}(2023)}]{Pollak2023Aug}%
  \BibitemOpen
  \bibfield  {author} {\bibinfo {author} {\bibfnamefont {E.}~\bibnamefont
  {Pollak}}\ and\ \bibinfo {author} {\bibfnamefont {S.}~\bibnamefont
  {Miret-Art{\ifmmode\acute{e}\else\'{e}\fi}s}},\ }\bibfield  {title} {\bibinfo
  {title} {{Recent Developments in Kramers{'} Theory of Reaction Rates}},\
  }\href {https://doi.org/10.1002/cphc.202300272} {\bibfield  {journal}
  {\bibinfo  {journal} {ChemPhysChem}\ }\textbf {\bibinfo {volume} {24}},\
  \bibinfo {pages} {e202300272} (\bibinfo {year} {2023})}\BibitemShut {NoStop}%
\bibitem [{\citenamefont {Chung}\ \emph {et~al.}(2009)\citenamefont {Chung},
  \citenamefont {Louis},\ and\ \citenamefont {Eaton}}]{Chung2009PNAS}%
  \BibitemOpen
  \bibfield  {author} {\bibinfo {author} {\bibfnamefont {H.~S.}\ \bibnamefont
  {Chung}}, \bibinfo {author} {\bibfnamefont {J.~M.}\ \bibnamefont {Louis}},\
  and\ \bibinfo {author} {\bibfnamefont {W.~A.}\ \bibnamefont {Eaton}},\
  }\bibfield  {title} {\bibinfo {title} {{Experimental determination of upper
  bound for transition path times in protein folding from single-molecule
  photon-by-photon trajectories}},\ }\href
  {https://doi.org/10.1073/pnas.0901178106} {\bibfield  {journal} {\bibinfo
  {journal} {Proc. Natl. Acad. Sci. U.S.A.}\ }\textbf {\bibinfo {volume}
  {106}},\ \bibinfo {pages} {11837} (\bibinfo {year} {2009})}\BibitemShut
  {NoStop}%
\bibitem [{\citenamefont {Pirchi}\ \emph {et~al.}(2011)\citenamefont {Pirchi},
  \citenamefont {Ziv}, \citenamefont {Riven}, \citenamefont {Cohen},
  \citenamefont {Zohar}, \citenamefont {Barak},\ and\ \citenamefont
  {Haran}}]{Pirchi2011NC}%
  \BibitemOpen
  \bibfield  {author} {\bibinfo {author} {\bibfnamefont {M.}~\bibnamefont
  {Pirchi}}, \bibinfo {author} {\bibfnamefont {G.}~\bibnamefont {Ziv}},
  \bibinfo {author} {\bibfnamefont {I.}~\bibnamefont {Riven}}, \bibinfo
  {author} {\bibfnamefont {S.~S.}\ \bibnamefont {Cohen}}, \bibinfo {author}
  {\bibfnamefont {N.}~\bibnamefont {Zohar}}, \bibinfo {author} {\bibfnamefont
  {Y.}~\bibnamefont {Barak}},\ and\ \bibinfo {author} {\bibfnamefont
  {G.}~\bibnamefont {Haran}},\ }\bibfield  {title} {\bibinfo {title}
  {{Single-molecule fluorescence spectroscopy maps the folding landscape of a
  large protein}},\ }\href {https://doi.org/10.1038/ncomms1504} {\bibfield
  {journal} {\bibinfo  {journal} {Nat. Commun.}\ }\textbf {\bibinfo {volume}
  {2}},\ \bibinfo {pages} {1} (\bibinfo {year} {2011})}\BibitemShut {NoStop}%
\bibitem [{\citenamefont {Chung}\ \emph {et~al.}(2012)\citenamefont {Chung},
  \citenamefont {McHale}, \citenamefont {Louis},\ and\ \citenamefont
  {Eaton}}]{Chung2012S}%
  \BibitemOpen
  \bibfield  {author} {\bibinfo {author} {\bibfnamefont {H.~S.}\ \bibnamefont
  {Chung}}, \bibinfo {author} {\bibfnamefont {K.}~\bibnamefont {McHale}},
  \bibinfo {author} {\bibfnamefont {J.~M.}\ \bibnamefont {Louis}},\ and\
  \bibinfo {author} {\bibfnamefont {W.~A.}\ \bibnamefont {Eaton}},\ }\bibfield
  {title} {\bibinfo {title} {{Single-Molecule Fluorescence Experiments
  Determine Protein Folding Transition Path Times}},\ }\href
  {https://doi.org/10.1126/science.1215768} {\bibfield  {journal} {\bibinfo
  {journal} {Science}\ }\textbf {\bibinfo {volume} {335}},\ \bibinfo {pages}
  {981} (\bibinfo {year} {2012})}\BibitemShut {NoStop}%
\bibitem [{\citenamefont {Zhu}\ and\ \citenamefont {Su}(2023)}]{Zhu2023PRE}%
  \BibitemOpen
  \bibfield  {author} {\bibinfo {author} {\bibfnamefont {K.}~\bibnamefont
  {Zhu}}\ and\ \bibinfo {author} {\bibfnamefont {H.}~\bibnamefont {Su}},\
  }\bibfield  {title} {\bibinfo {title} {{Traversing the nucleation-growth
  landscape through heterogeneous random walks}},\ }\href
  {https://doi.org/10.1103/PhysRevE.107.064110} {\bibfield  {journal} {\bibinfo
   {journal} {Phys. Rev. E}\ }\textbf {\bibinfo {volume} {107}},\ \bibinfo
  {pages} {064110} (\bibinfo {year} {2023})}\BibitemShut {NoStop}%
\bibitem [{\citenamefont {Badzey}\ and\ \citenamefont
  {Mohanty}(2005)}]{Badzey2005N}%
  \BibitemOpen
  \bibfield  {author} {\bibinfo {author} {\bibfnamefont {R.~L.}\ \bibnamefont
  {Badzey}}\ and\ \bibinfo {author} {\bibfnamefont {P.}~\bibnamefont
  {Mohanty}},\ }\bibfield  {title} {\bibinfo {title} {{Coherent signal
  amplification in bistable nanomechanical oscillators by stochastic
  resonance}},\ }\href {https://doi.org/10.1038/nature04124} {\bibfield
  {journal} {\bibinfo  {journal} {Nature}\ }\textbf {\bibinfo {volume} {437}},\
  \bibinfo {pages} {995} (\bibinfo {year} {2005})}\BibitemShut {NoStop}%
\bibitem [{\citenamefont {McCann}\ \emph {et~al.}(1999)\citenamefont {McCann},
  \citenamefont {Dykman},\ and\ \citenamefont {Golding}}]{McCann1999N}%
  \BibitemOpen
  \bibfield  {author} {\bibinfo {author} {\bibfnamefont {L.~I.}\ \bibnamefont
  {McCann}}, \bibinfo {author} {\bibfnamefont {M.}~\bibnamefont {Dykman}},\
  and\ \bibinfo {author} {\bibfnamefont {B.}~\bibnamefont {Golding}},\
  }\bibfield  {title} {\bibinfo {title} {{Thermally activated transitions in a
  bistable three-dimensional optical trap}},\ }\href
  {https://doi.org/10.1038/45492} {\bibfield  {journal} {\bibinfo  {journal}
  {Nature}\ }\textbf {\bibinfo {volume} {402}},\ \bibinfo {pages} {785}
  (\bibinfo {year} {1999})}\BibitemShut {NoStop}%
\bibitem [{\citenamefont {Rondin}\ \emph {et~al.}(2017)\citenamefont {Rondin},
  \citenamefont {Gieseler}, \citenamefont {Ricci}, \citenamefont {Quidant},
  \citenamefont {Dellago},\ and\ \citenamefont {Novotny}}]{Rondin2017NN}%
  \BibitemOpen
  \bibfield  {author} {\bibinfo {author} {\bibfnamefont {L.}~\bibnamefont
  {Rondin}}, \bibinfo {author} {\bibfnamefont {J.}~\bibnamefont {Gieseler}},
  \bibinfo {author} {\bibfnamefont {F.}~\bibnamefont {Ricci}}, \bibinfo
  {author} {\bibfnamefont {R.}~\bibnamefont {Quidant}}, \bibinfo {author}
  {\bibfnamefont {C.}~\bibnamefont {Dellago}},\ and\ \bibinfo {author}
  {\bibfnamefont {L.}~\bibnamefont {Novotny}},\ }\bibfield  {title} {\bibinfo
  {title} {{Direct measurement of Kramers turnover with a levitated
  nanoparticle}},\ }\href {https://doi.org/10.1038/nnano.2017.198} {\bibfield
  {journal} {\bibinfo  {journal} {Nat. Nanotechnol.}\ }\textbf {\bibinfo
  {volume} {12}},\ \bibinfo {pages} {1130} (\bibinfo {year}
  {2017})}\BibitemShut {NoStop}%
\bibitem [{\citenamefont {Wu}\ \emph {et~al.}(2009)\citenamefont {Wu},
  \citenamefont {Ghosh}, \citenamefont {Inamdar}, \citenamefont {Lee},
  \citenamefont {Fraser}, \citenamefont {Dill},\ and\ \citenamefont
  {Phillips}}]{Wu2009PRL}%
  \BibitemOpen
  \bibfield  {author} {\bibinfo {author} {\bibfnamefont {D.}~\bibnamefont
  {Wu}}, \bibinfo {author} {\bibfnamefont {K.}~\bibnamefont {Ghosh}}, \bibinfo
  {author} {\bibfnamefont {M.}~\bibnamefont {Inamdar}}, \bibinfo {author}
  {\bibfnamefont {H.~J.}\ \bibnamefont {Lee}}, \bibinfo {author} {\bibfnamefont
  {S.}~\bibnamefont {Fraser}}, \bibinfo {author} {\bibfnamefont
  {K.}~\bibnamefont {Dill}},\ and\ \bibinfo {author} {\bibfnamefont
  {R.}~\bibnamefont {Phillips}},\ }\bibfield  {title} {\bibinfo {title}
  {{Trajectory Approach to Two-State Kinetics of Single Particles on Sculpted
  Energy Landscapes}},\ }\href {https://doi.org/10.1103/PhysRevLett.103.050603}
  {\bibfield  {journal} {\bibinfo  {journal} {Phys. Rev. Lett.}\ }\textbf
  {\bibinfo {volume} {103}},\ \bibinfo {pages} {050603} (\bibinfo {year}
  {2009})}\BibitemShut {NoStop}%
\bibitem [{\citenamefont {Seol}\ \emph {et~al.}(2009)\citenamefont {Seol},
  \citenamefont {Stein},\ and\ \citenamefont {Visscher}}]{Seol2009PRL}%
  \BibitemOpen
  \bibfield  {author} {\bibinfo {author} {\bibfnamefont {Y.}~\bibnamefont
  {Seol}}, \bibinfo {author} {\bibfnamefont {D.~L.}\ \bibnamefont {Stein}},\
  and\ \bibinfo {author} {\bibfnamefont {K.}~\bibnamefont {Visscher}},\
  }\bibfield  {title} {\bibinfo {title} {{Phase Measurements of Barrier
  Crossings in a Periodically Modulated Double-Well Potential}},\ }\href
  {https://doi.org/10.1103/PhysRevLett.103.050601} {\bibfield  {journal}
  {\bibinfo  {journal} {Phys. Rev. Lett.}\ }\textbf {\bibinfo {volume} {103}},\
  \bibinfo {pages} {050601} (\bibinfo {year} {2009})}\BibitemShut {NoStop}%
\bibitem [{\citenamefont {Babi{\ifmmode\check{c}\else\v{c}\fi}}\ \emph
  {et~al.}(2004)\citenamefont {Babi{\ifmmode\check{c}\else\v{c}\fi}},
  \citenamefont {Schmitt}, \citenamefont {Poberaj},\ and\ \citenamefont
  {Bechinger}}]{Babic2004EurLett}%
  \BibitemOpen
  \bibfield  {author} {\bibinfo {author} {\bibfnamefont {D.}~\bibnamefont
  {Babi{\ifmmode\check{c}\else\v{c}\fi}}}, \bibinfo {author} {\bibfnamefont
  {C.}~\bibnamefont {Schmitt}}, \bibinfo {author} {\bibfnamefont
  {I.}~\bibnamefont {Poberaj}},\ and\ \bibinfo {author} {\bibfnamefont
  {C.}~\bibnamefont {Bechinger}},\ }\bibfield  {title} {\bibinfo {title}
  {{Stochastic resonance in colloidal systems}},\ }\href
  {https://doi.org/10.1209/epl/i2004-10055-3} {\bibfield  {journal} {\bibinfo
  {journal} {Europhys. Lett.}\ }\textbf {\bibinfo {volume} {67}},\ \bibinfo
  {pages} {158} (\bibinfo {year} {2004})}\BibitemShut {NoStop}%
\bibitem [{\citenamefont {Simon}\ and\ \citenamefont
  {Libchaber}(1992)}]{Simon1992PRL}%
  \BibitemOpen
  \bibfield  {author} {\bibinfo {author} {\bibfnamefont {A.}~\bibnamefont
  {Simon}}\ and\ \bibinfo {author} {\bibfnamefont {A.}~\bibnamefont
  {Libchaber}},\ }\bibfield  {title} {\bibinfo {title} {{Escape and
  synchronization of a Brownian particle}},\ }\href
  {https://doi.org/10.1103/PhysRevLett.68.3375} {\bibfield  {journal} {\bibinfo
   {journal} {Phys. Rev. Lett.}\ }\textbf {\bibinfo {volume} {68}},\ \bibinfo
  {pages} {3375} (\bibinfo {year} {1992})}\BibitemShut {NoStop}%
\bibitem [{\citenamefont {Yoon}\ \emph {et~al.}(2018)\citenamefont {Yoon},
  \citenamefont {Lee}, \citenamefont {Han}, \citenamefont {Kim}, \citenamefont
  {Ahn}, \citenamefont {Kim},\ and\ \citenamefont {Lee}}]{Yoon2018NC}%
  \BibitemOpen
  \bibfield  {author} {\bibinfo {author} {\bibfnamefont {S.~J.}\ \bibnamefont
  {Yoon}}, \bibinfo {author} {\bibfnamefont {J.}~\bibnamefont {Lee}}, \bibinfo
  {author} {\bibfnamefont {S.}~\bibnamefont {Han}}, \bibinfo {author}
  {\bibfnamefont {C.-K.}\ \bibnamefont {Kim}}, \bibinfo {author} {\bibfnamefont
  {C.~W.}\ \bibnamefont {Ahn}}, \bibinfo {author} {\bibfnamefont {M.-K.}\
  \bibnamefont {Kim}},\ and\ \bibinfo {author} {\bibfnamefont {Y.-H.}\
  \bibnamefont {Lee}},\ }\bibfield  {title} {\bibinfo {title} {{Non-fluorescent
  nanoscopic monitoring of a single trapped nanoparticle via nonlinear point
  sources}},\ }\href {https://doi.org/10.1038/s41467-018-04689-5} {\bibfield
  {journal} {\bibinfo  {journal} {Nat. Commun.}\ }\textbf {\bibinfo {volume}
  {9}},\ \bibinfo {pages} {1} (\bibinfo {year} {2018})}\BibitemShut {NoStop}%
\bibitem [{\citenamefont {Yoon}\ \emph {et~al.}(2020)\citenamefont {Yoon},
  \citenamefont {Song}, \citenamefont {Lee}, \citenamefont {Kim}, \citenamefont
  {Lee},\ and\ \citenamefont {Kim}}]{Yoon2020NanoPhot}%
  \BibitemOpen
  \bibfield  {author} {\bibinfo {author} {\bibfnamefont {S.~J.}\ \bibnamefont
  {Yoon}}, \bibinfo {author} {\bibfnamefont {D.~I.}\ \bibnamefont {Song}},
  \bibinfo {author} {\bibfnamefont {J.}~\bibnamefont {Lee}}, \bibinfo {author}
  {\bibfnamefont {M.-K.}\ \bibnamefont {Kim}}, \bibinfo {author} {\bibfnamefont
  {Y.-H.}\ \bibnamefont {Lee}},\ and\ \bibinfo {author} {\bibfnamefont {C.-K.}\
  \bibnamefont {Kim}},\ }\bibfield  {title} {\bibinfo {title} {{Hopping of
  single nanoparticles trapped in a plasmonic double-well potential}},\ }\href
  {https://doi.org/10.1515/nanoph-2020-0411} {\bibfield  {journal} {\bibinfo
  {journal} {Nanophotonics}\ }\textbf {\bibinfo {volume} {9}},\ \bibinfo
  {pages} {4729} (\bibinfo {year} {2020})}\BibitemShut {NoStop}%
\bibitem [{\citenamefont {Chupeau}\ \emph {et~al.}(2020)\citenamefont
  {Chupeau}, \citenamefont {Gladrow}, \citenamefont {Chepelianskii},
  \citenamefont {Keyser},\ and\ \citenamefont {Trizac}}]{Chupeau2020PNAS}%
  \BibitemOpen
  \bibfield  {author} {\bibinfo {author} {\bibfnamefont {M.}~\bibnamefont
  {Chupeau}}, \bibinfo {author} {\bibfnamefont {J.}~\bibnamefont {Gladrow}},
  \bibinfo {author} {\bibfnamefont {A.}~\bibnamefont {Chepelianskii}}, \bibinfo
  {author} {\bibfnamefont {U.~F.}\ \bibnamefont {Keyser}},\ and\ \bibinfo
  {author} {\bibfnamefont {E.}~\bibnamefont {Trizac}},\ }\bibfield  {title}
  {\bibinfo {title} {{Optimizing Brownian escape rates by potential shaping}},\
  }\href {https://doi.org/10.1073/pnas.1910677116} {\bibfield  {journal}
  {\bibinfo  {journal} {Proc. Natl. Acad. Sci. U.S.A.}\ }\textbf {\bibinfo
  {volume} {117}},\ \bibinfo {pages} {1383} (\bibinfo {year}
  {2020})}\BibitemShut {NoStop}%
\bibitem [{\citenamefont {Jiang}\ \emph {et~al.}(2010)\citenamefont {Jiang},
  \citenamefont {Narushima},\ and\ \citenamefont {Okamoto}}]{Jiang2010NatPhys}%
  \BibitemOpen
  \bibfield  {author} {\bibinfo {author} {\bibfnamefont {Y.}~\bibnamefont
  {Jiang}}, \bibinfo {author} {\bibfnamefont {T.}~\bibnamefont {Narushima}},\
  and\ \bibinfo {author} {\bibfnamefont {H.}~\bibnamefont {Okamoto}},\
  }\bibfield  {title} {\bibinfo {title} {{Nonlinear optical effects in trapping
  nanoparticles with femtosecond pulses}},\ }\href
  {https://doi.org/10.1038/nphys1776} {\bibfield  {journal} {\bibinfo
  {journal} {Nat. Phys.}\ }\textbf {\bibinfo {volume} {6}},\ \bibinfo {pages}
  {1005} (\bibinfo {year} {2010})}\BibitemShut {NoStop}%
\bibitem [{\citenamefont {Zhang}\ \emph {et~al.}(2018)\citenamefont {Zhang},
  \citenamefont {Shen}, \citenamefont {Min}, \citenamefont {Jin}, \citenamefont
  {Jiang}, \citenamefont {Liu}, \citenamefont {Zhu}, \citenamefont {Sheng},
  \citenamefont {Zayats},\ and\ \citenamefont {Yuan}}]{Zhang2018NL}%
  \BibitemOpen
  \bibfield  {author} {\bibinfo {author} {\bibfnamefont {Y.}~\bibnamefont
  {Zhang}}, \bibinfo {author} {\bibfnamefont {J.}~\bibnamefont {Shen}},
  \bibinfo {author} {\bibfnamefont {C.}~\bibnamefont {Min}}, \bibinfo {author}
  {\bibfnamefont {Y.}~\bibnamefont {Jin}}, \bibinfo {author} {\bibfnamefont
  {Y.}~\bibnamefont {Jiang}}, \bibinfo {author} {\bibfnamefont
  {J.}~\bibnamefont {Liu}}, \bibinfo {author} {\bibfnamefont {S.}~\bibnamefont
  {Zhu}}, \bibinfo {author} {\bibfnamefont {Y.}~\bibnamefont {Sheng}}, \bibinfo
  {author} {\bibfnamefont {A.~V.}\ \bibnamefont {Zayats}},\ and\ \bibinfo
  {author} {\bibfnamefont {X.}~\bibnamefont {Yuan}},\ }\bibfield  {title}
  {\bibinfo {title} {{Nonlinearity-Induced Multiplexed Optical Trapping and
  Manipulation with Femtosecond Vector Beams}},\ }\href
  {https://doi.org/10.1021/acs.nanolett.8b01929} {\bibfield  {journal}
  {\bibinfo  {journal} {Nano Lett.}\ }\textbf {\bibinfo {volume} {18}},\
  \bibinfo {pages} {5538} (\bibinfo {year} {2018})}\BibitemShut {NoStop}%
\bibitem [{\citenamefont {Mirzaei-Ghormish}\ \emph {et~al.}(2025)\citenamefont
  {Mirzaei-Ghormish}, \citenamefont {Smalley},\ and\ \citenamefont
  {Camacho}}]{Mirzaei-Ghormish2025PRA}%
  \BibitemOpen
  \bibfield  {author} {\bibinfo {author} {\bibfnamefont {S.}~\bibnamefont
  {Mirzaei-Ghormish}}, \bibinfo {author} {\bibfnamefont {D.}~\bibnamefont
  {Smalley}},\ and\ \bibinfo {author} {\bibfnamefont {R.}~\bibnamefont
  {Camacho}},\ }\bibfield  {title} {\bibinfo {title} {{Nonlinear multistable
  potential traps}},\ }\href {https://doi.org/10.1103/PhysRevA.111.013514}
  {\bibfield  {journal} {\bibinfo  {journal} {Phys. Rev. A}\ }\textbf {\bibinfo
  {volume} {111}},\ \bibinfo {pages} {013514} (\bibinfo {year}
  {2025})}\BibitemShut {NoStop}%
\bibitem [{\citenamefont {Zhou}\ \emph {et~al.}(2020)\citenamefont {Zhou},
  \citenamefont {Qin}, \citenamefont {Qin}, \citenamefont {Yang}, \citenamefont
  {Jiang},\ and\ \citenamefont {Jiang}}]{Zhou2020OL}%
  \BibitemOpen
  \bibfield  {author} {\bibinfo {author} {\bibfnamefont {L.-M.}\ \bibnamefont
  {Zhou}}, \bibinfo {author} {\bibfnamefont {Y.}~\bibnamefont {Qin}}, \bibinfo
  {author} {\bibfnamefont {Y.}~\bibnamefont {Qin}}, \bibinfo {author}
  {\bibfnamefont {Y.}~\bibnamefont {Yang}}, \bibinfo {author} {\bibfnamefont
  {Y.}~\bibnamefont {Jiang}},\ and\ \bibinfo {author} {\bibfnamefont
  {Y.}~\bibnamefont {Jiang}},\ }\bibfield  {title} {\bibinfo {title} {{Fine
  features of optical potential well induced by nonlinearity}},\ }\href
  {https://doi.org/10.1364/OL.412349} {\bibfield  {journal} {\bibinfo
  {journal} {Opt. Lett.}\ }\textbf {\bibinfo {volume} {45}},\ \bibinfo {pages}
  {6266} (\bibinfo {year} {2020})}\BibitemShut {NoStop}%
\bibitem [{\citenamefont {Lee}\ \emph {et~al.}(2012)\citenamefont {Lee},
  \citenamefont {Curran}, \citenamefont {Gibson}, \citenamefont {Tassieri},
  \citenamefont {Heckenberg},\ and\ \citenamefont {Padgett}}]{Lee2012OE}%
  \BibitemOpen
  \bibfield  {author} {\bibinfo {author} {\bibfnamefont {M.~P.}\ \bibnamefont
  {Lee}}, \bibinfo {author} {\bibfnamefont {A.}~\bibnamefont {Curran}},
  \bibinfo {author} {\bibfnamefont {G.~M.}\ \bibnamefont {Gibson}}, \bibinfo
  {author} {\bibfnamefont {M.}~\bibnamefont {Tassieri}}, \bibinfo {author}
  {\bibfnamefont {N.~R.}\ \bibnamefont {Heckenberg}},\ and\ \bibinfo {author}
  {\bibfnamefont {M.~J.}\ \bibnamefont {Padgett}},\ }\bibfield  {title}
  {\bibinfo {title} {{Optical shield: measuring viscosity of turbid fluids
  using optical tweezers}},\ }\href {https://doi.org/10.1364/OE.20.012127}
  {\bibfield  {journal} {\bibinfo  {journal} {Opt. Express}\ }\textbf {\bibinfo
  {volume} {20}},\ \bibinfo {pages} {12127} (\bibinfo {year}
  {2012})}\BibitemShut {NoStop}%
\bibitem [{\citenamefont {Romero-Gonz{\ifmmode\acute{a}\else\'{a}\fi}lez}\
  \emph {et~al.}(2023)\citenamefont
  {Romero-Gonz{\ifmmode\acute{a}\else\'{a}\fi}lez}, \citenamefont
  {Morales-Cruzado}, \citenamefont {de~Lange}, \citenamefont
  {Romero-M{\ifmmode\acute{e}\else\'{e}\fi}ndez},\ and\ \citenamefont
  {P{\ifmmode\acute{e}\else\'{e}\fi}rez-Guti{\ifmmode\acute{e}\else\'{e}\fi}rrez}}]{Romero-Gonzalez2023OptLaserTech}%
  \BibitemOpen
  \bibfield  {author} {\bibinfo {author} {\bibfnamefont {C.}~\bibnamefont
  {Romero-Gonz{\ifmmode\acute{a}\else\'{a}\fi}lez}}, \bibinfo {author}
  {\bibfnamefont {B.}~\bibnamefont {Morales-Cruzado}}, \bibinfo {author}
  {\bibfnamefont {D.~F.}\ \bibnamefont {de~Lange}}, \bibinfo {author}
  {\bibfnamefont {R.}~\bibnamefont
  {Romero-M{\ifmmode\acute{e}\else\'{e}\fi}ndez}},\ and\ \bibinfo {author}
  {\bibfnamefont {F.~G.}\ \bibnamefont
  {P{\ifmmode\acute{e}\else\'{e}\fi}rez-Guti{\ifmmode\acute{e}\else\'{e}\fi}rrez}},\
  }\bibfield  {title} {\bibinfo {title} {{In situ temperature measurement in
  microfluidics using optical tweezers}},\ }\href
  {https://doi.org/10.1016/j.optlastec.2023.109691} {\bibfield  {journal}
  {\bibinfo  {journal} {Opt. Laser Technol.}\ }\textbf {\bibinfo {volume}
  {166}},\ \bibinfo {pages} {109691} (\bibinfo {year} {2023})}\BibitemShut
  {NoStop}%
\bibitem [{\citenamefont {Mao}\ \emph {et~al.}(2025)\citenamefont {Mao},
  \citenamefont {Toftul}, \citenamefont {Balendhran}, \citenamefont {Taha},
  \citenamefont {Kivshar},\ and\ \citenamefont {Kruk}}]{Mao2025LRP}%
  \BibitemOpen
  \bibfield  {author} {\bibinfo {author} {\bibfnamefont {L.}~\bibnamefont
  {Mao}}, \bibinfo {author} {\bibfnamefont {I.}~\bibnamefont {Toftul}},
  \bibinfo {author} {\bibfnamefont {S.}~\bibnamefont {Balendhran}}, \bibinfo
  {author} {\bibfnamefont {M.}~\bibnamefont {Taha}}, \bibinfo {author}
  {\bibfnamefont {Y.}~\bibnamefont {Kivshar}},\ and\ \bibinfo {author}
  {\bibfnamefont {S.}~\bibnamefont {Kruk}},\ }\bibfield  {title} {\bibinfo
  {title} {{Switchable Optical Trapping of Mie-Resonant Phase-Change
  Nanoparticles}},\ }\href {https://doi.org/10.1002/lpor.202400767} {\bibfield
  {journal} {\bibinfo  {journal} {Laser Photonics Rev.}\ }\textbf {\bibinfo
  {volume} {19}},\ \bibinfo {pages} {2400767} (\bibinfo {year}
  {2025})}\BibitemShut {NoStop}%
\bibitem [{\citenamefont {Zhang}(2018)}]{Zhang2018PRAppl}%
  \BibitemOpen
  \bibfield  {author} {\bibinfo {author} {\bibfnamefont {L.}~\bibnamefont
  {Zhang}},\ }\bibfield  {title} {\bibinfo {title} {{Reversals of Orbital
  Angular Momentum Transfer and Radiation Torque}},\ }\href
  {https://doi.org/10.1103/PhysRevApplied.10.034039} {\bibfield  {journal}
  {\bibinfo  {journal} {Phys. Rev. Appl.}\ }\textbf {\bibinfo {volume} {10}},\
  \bibinfo {pages} {034039} (\bibinfo {year} {2018})}\BibitemShut {NoStop}%
\bibitem [{\citenamefont {Gahagan}\ and\ \citenamefont
  {Swartzlander}(1996)}]{Gahagan1996OL}%
  \BibitemOpen
  \bibfield  {author} {\bibinfo {author} {\bibfnamefont {K.~T.}\ \bibnamefont
  {Gahagan}}\ and\ \bibinfo {author} {\bibfnamefont {G.~A.}\ \bibnamefont
  {Swartzlander}},\ }\bibfield  {title} {\bibinfo {title} {{Optical vortex
  trapping of particles}},\ }\href {https://doi.org/10.1364/OL.21.000827}
  {\bibfield  {journal} {\bibinfo  {journal} {Opt. Lett.}\ }\textbf {\bibinfo
  {volume} {21}},\ \bibinfo {pages} {827} (\bibinfo {year} {1996})}\BibitemShut
  {NoStop}%
\bibitem [{\citenamefont {Gahagan}\ and\ \citenamefont
  {Swartzlander}(1999)}]{Gahagan1999JOSAB}%
  \BibitemOpen
  \bibfield  {author} {\bibinfo {author} {\bibfnamefont {K.~T.}\ \bibnamefont
  {Gahagan}}\ and\ \bibinfo {author} {\bibfnamefont {G.~A.}\ \bibnamefont
  {Swartzlander}},\ }\bibfield  {title} {\bibinfo {title} {{Simultaneous
  trapping of low-index and high-index microparticles observed with an
  optical-vortex trap}},\ }\href {https://doi.org/10.1364/JOSAB.16.000533}
  {\bibfield  {journal} {\bibinfo  {journal} {J. Opt. Soc. Am. B}\ }\textbf
  {\bibinfo {volume} {16}},\ \bibinfo {pages} {533} (\bibinfo {year}
  {1999})}\BibitemShut {NoStop}%
\bibitem [{\citenamefont {O'Neil}\ \emph {et~al.}(2002)\citenamefont {O'Neil},
  \citenamefont {MacVicar}, \citenamefont {Allen},\ and\ \citenamefont
  {Padgett}}]{ONeil2002PRL}%
  \BibitemOpen
  \bibfield  {author} {\bibinfo {author} {\bibfnamefont {A.~T.}\ \bibnamefont
  {O'Neil}}, \bibinfo {author} {\bibfnamefont {I.}~\bibnamefont {MacVicar}},
  \bibinfo {author} {\bibfnamefont {L.}~\bibnamefont {Allen}},\ and\ \bibinfo
  {author} {\bibfnamefont {M.~J.}\ \bibnamefont {Padgett}},\ }\bibfield
  {title} {\bibinfo {title} {Intrinsic and {{Extrinsic Nature}} of the
  {{Orbital Angular Momentum}} of a {{Light Beam}}},\ }\href
  {https://doi.org/10.1103/PhysRevLett.88.053601} {\bibfield  {journal}
  {\bibinfo  {journal} {Phys. Rev. Lett.}\ }\textbf {\bibinfo {volume} {88}},\
  \bibinfo {pages} {053601} (\bibinfo {year} {2002})}\BibitemShut {NoStop}%
\bibitem [{\citenamefont
  {Garc{\ifmmode\acute{e}\else\'{e}\fi}s-Ch{\ifmmode\acute{a}\else\'{a}\fi}vez}\
  \emph {et~al.}(2002)\citenamefont
  {Garc{\ifmmode\acute{e}\else\'{e}\fi}s-Ch{\ifmmode\acute{a}\else\'{a}\fi}vez},
  \citenamefont {Volke-Sepulveda}, \citenamefont
  {Ch{\ifmmode\acute{a}\else\'{a}\fi}vez-Cerda}, \citenamefont {Sibbett},\ and\
  \citenamefont {Dholakia}}]{Garces-Chavez2002PRA}%
  \BibitemOpen
  \bibfield  {author} {\bibinfo {author} {\bibfnamefont {V.}~\bibnamefont
  {Garc{\ifmmode\acute{e}\else\'{e}\fi}s-Ch{\ifmmode\acute{a}\else\'{a}\fi}vez}},
  \bibinfo {author} {\bibfnamefont {K.}~\bibnamefont {Volke-Sepulveda}},
  \bibinfo {author} {\bibfnamefont {S.}~\bibnamefont
  {Ch{\ifmmode\acute{a}\else\'{a}\fi}vez-Cerda}}, \bibinfo {author}
  {\bibfnamefont {W.}~\bibnamefont {Sibbett}},\ and\ \bibinfo {author}
  {\bibfnamefont {K.}~\bibnamefont {Dholakia}},\ }\bibfield  {title} {\bibinfo
  {title} {{Transfer of orbital angular momentum to an optically trapped
  low-index particle}},\ }\href {https://doi.org/10.1103/PhysRevA.66.063402}
  {\bibfield  {journal} {\bibinfo  {journal} {Phys. Rev. A}\ }\textbf {\bibinfo
  {volume} {66}},\ \bibinfo {pages} {063402} (\bibinfo {year}
  {2002})}\BibitemShut {NoStop}%
\bibitem [{\citenamefont {Cao}\ \emph {et~al.}(2016{\natexlab{a}})\citenamefont
  {Cao}, \citenamefont {Zhu}, \citenamefont {Lv},\ and\ \citenamefont
  {Ding}}]{Cao2016OE}%
  \BibitemOpen
  \bibfield  {author} {\bibinfo {author} {\bibfnamefont {Y.}~\bibnamefont
  {Cao}}, \bibinfo {author} {\bibfnamefont {T.}~\bibnamefont {Zhu}}, \bibinfo
  {author} {\bibfnamefont {H.}~\bibnamefont {Lv}},\ and\ \bibinfo {author}
  {\bibfnamefont {W.}~\bibnamefont {Ding}},\ }\bibfield  {title} {\bibinfo
  {title} {{Spin-controlled orbital motion in tightly focused high-order
  Laguerre-Gaussian beams}},\ }\href {https://doi.org/10.1364/OE.24.003377}
  {\bibfield  {journal} {\bibinfo  {journal} {Opt. Express}\ }\textbf {\bibinfo
  {volume} {24}},\ \bibinfo {pages} {3377} (\bibinfo {year}
  {2016}{\natexlab{a}})}\BibitemShut {NoStop}%
\bibitem [{\citenamefont {Zhang}\ \emph {et~al.}(2025)\citenamefont {Zhang},
  \citenamefont {Lin}, \citenamefont {Zhuang}, \citenamefont {Lin},
  \citenamefont {Hong}, \citenamefont {Che}, \citenamefont {Zhuo},
  \citenamefont {Li}, \citenamefont {Zhang},\ and\ \citenamefont
  {Zhao}}]{Zhang2025NP}%
  \BibitemOpen
  \bibfield  {author} {\bibinfo {author} {\bibfnamefont {Y.}~\bibnamefont
  {Zhang}}, \bibinfo {author} {\bibfnamefont {Q.}~\bibnamefont {Lin}}, \bibinfo
  {author} {\bibfnamefont {Z.}~\bibnamefont {Zhuang}}, \bibinfo {author}
  {\bibfnamefont {F.}~\bibnamefont {Lin}}, \bibinfo {author} {\bibfnamefont
  {L.}~\bibnamefont {Hong}}, \bibinfo {author} {\bibfnamefont {Z.}~\bibnamefont
  {Che}}, \bibinfo {author} {\bibfnamefont {L.}~\bibnamefont {Zhuo}}, \bibinfo
  {author} {\bibfnamefont {Y.}~\bibnamefont {Li}}, \bibinfo {author}
  {\bibfnamefont {L.}~\bibnamefont {Zhang}},\ and\ \bibinfo {author}
  {\bibfnamefont {D.}~\bibnamefont {Zhao}},\ }\bibfield  {title} {\bibinfo
  {title} {{Dynamics of dual-orbit rotations of nanoparticles induced by
  spin{\textendash}orbit coupling}},\ }\bibfield  {journal} {\bibinfo
  {journal} {Nanophotonics}\ }\href {https://doi.org/10.1515/nanoph-2024-0586}
  {10.1515/nanoph-2024-0586} (\bibinfo {year} {2025})\BibitemShut {NoStop}%
\bibitem [{\citenamefont {Gao}\ \emph {et~al.}(2017)\citenamefont {Gao},
  \citenamefont {Ding}, \citenamefont {Nieto-Vesperinas}, \citenamefont {Ding},
  \citenamefont {Rahman}, \citenamefont {Zhang}, \citenamefont {Lim},\ and\
  \citenamefont {Qiu}}]{Gao2017LSA}%
  \BibitemOpen
  \bibfield  {author} {\bibinfo {author} {\bibfnamefont {D.}~\bibnamefont
  {Gao}}, \bibinfo {author} {\bibfnamefont {W.}~\bibnamefont {Ding}}, \bibinfo
  {author} {\bibfnamefont {M.}~\bibnamefont {Nieto-Vesperinas}}, \bibinfo
  {author} {\bibfnamefont {X.}~\bibnamefont {Ding}}, \bibinfo {author}
  {\bibfnamefont {M.}~\bibnamefont {Rahman}}, \bibinfo {author} {\bibfnamefont
  {T.}~\bibnamefont {Zhang}}, \bibinfo {author} {\bibfnamefont
  {C.}~\bibnamefont {Lim}},\ and\ \bibinfo {author} {\bibfnamefont {C.-W.}\
  \bibnamefont {Qiu}},\ }\bibfield  {title} {\bibinfo {title} {{Optical
  manipulation from the microscale to the nanoscale: fundamentals, advances and
  prospects}},\ }\bibfield  {journal} {\bibinfo  {journal} {Light Sci. Appl.}\
  }\textbf {\bibinfo {volume} {6}},\ \href
  {https://doi.org/10.1038/lsa.2017.39} {10.1038/lsa.2017.39} (\bibinfo {year}
  {2017})\BibitemShut {NoStop}%
\bibitem [{\citenamefont {Ni}\ \emph {et~al.}(2022)\citenamefont {Ni},
  \citenamefont {Liu}, \citenamefont {Chen}, \citenamefont {Hu}, \citenamefont
  {Hu}, \citenamefont {Chen}, \citenamefont {Li}, \citenamefont {Chu},
  \citenamefont {Qiu},\ and\ \citenamefont {Wu}}]{Ni2022NL}%
  \BibitemOpen
  \bibfield  {author} {\bibinfo {author} {\bibfnamefont {J.}~\bibnamefont
  {Ni}}, \bibinfo {author} {\bibfnamefont {S.}~\bibnamefont {Liu}}, \bibinfo
  {author} {\bibfnamefont {Y.}~\bibnamefont {Chen}}, \bibinfo {author}
  {\bibfnamefont {G.}~\bibnamefont {Hu}}, \bibinfo {author} {\bibfnamefont
  {Y.}~\bibnamefont {Hu}}, \bibinfo {author} {\bibfnamefont {W.}~\bibnamefont
  {Chen}}, \bibinfo {author} {\bibfnamefont {J.}~\bibnamefont {Li}}, \bibinfo
  {author} {\bibfnamefont {J.}~\bibnamefont {Chu}}, \bibinfo {author}
  {\bibfnamefont {C.-W.}\ \bibnamefont {Qiu}},\ and\ \bibinfo {author}
  {\bibfnamefont {D.}~\bibnamefont {Wu}},\ }\bibfield  {title} {\bibinfo
  {title} {{Direct Observation of Spin{\textendash}Orbit Interaction of Light
  via Chiroptical Responses}},\ }\href
  {https://doi.org/10.1021/acs.nanolett.2c03266} {\bibfield  {journal}
  {\bibinfo  {journal} {Nano Lett.}\ }\textbf {\bibinfo {volume} {22}},\
  \bibinfo {pages} {9013} (\bibinfo {year} {2022})}\BibitemShut {NoStop}%
\bibitem [{\citenamefont {Triolo}\ \emph {et~al.}(2017)\citenamefont {Triolo},
  \citenamefont {Cacciola}, \citenamefont
  {Patan{\ifmmode\grave{e}\else\`{e}\fi}}, \citenamefont {Saija}, \citenamefont
  {Savasta},\ and\ \citenamefont {Nori}}]{Triolo2017ACSPhot}%
  \BibitemOpen
  \bibfield  {author} {\bibinfo {author} {\bibfnamefont {C.}~\bibnamefont
  {Triolo}}, \bibinfo {author} {\bibfnamefont {A.}~\bibnamefont {Cacciola}},
  \bibinfo {author} {\bibfnamefont {S.}~\bibnamefont
  {Patan{\ifmmode\grave{e}\else\`{e}\fi}}}, \bibinfo {author} {\bibfnamefont
  {R.}~\bibnamefont {Saija}}, \bibinfo {author} {\bibfnamefont
  {S.}~\bibnamefont {Savasta}},\ and\ \bibinfo {author} {\bibfnamefont
  {F.}~\bibnamefont {Nori}},\ }\bibfield  {title} {\bibinfo {title}
  {{Spin-Momentum Locking in the Near Field of Metal Nanoparticles}},\ }\href
  {https://doi.org/10.1021/acsphotonics.7b00436} {\bibfield  {journal}
  {\bibinfo  {journal} {ACS Photonics}\ }\textbf {\bibinfo {volume} {4}},\
  \bibinfo {pages} {2242} (\bibinfo {year} {2017})}\BibitemShut {NoStop}%
\bibitem [{\citenamefont {Tkachenko}\ \emph {et~al.}(2020)\citenamefont
  {Tkachenko}, \citenamefont {Tkachenko}, \citenamefont {Toftul}, \citenamefont
  {Esporlas}, \citenamefont {Maimaiti}, \citenamefont {Maimaiti}, \citenamefont
  {Maimaiti}, \citenamefont {Le~Kien}, \citenamefont {Truong}, \citenamefont
  {Truong}, \citenamefont {Chormaic}, \citenamefont {Chormaic},\ and\
  \citenamefont {Chormaic}}]{Tkachenko2020Optica}%
  \BibitemOpen
  \bibfield  {author} {\bibinfo {author} {\bibfnamefont {G.}~\bibnamefont
  {Tkachenko}}, \bibinfo {author} {\bibfnamefont {G.}~\bibnamefont
  {Tkachenko}}, \bibinfo {author} {\bibfnamefont {I.}~\bibnamefont {Toftul}},
  \bibinfo {author} {\bibfnamefont {C.}~\bibnamefont {Esporlas}}, \bibinfo
  {author} {\bibfnamefont {A.}~\bibnamefont {Maimaiti}}, \bibinfo {author}
  {\bibfnamefont {A.}~\bibnamefont {Maimaiti}}, \bibinfo {author}
  {\bibfnamefont {A.}~\bibnamefont {Maimaiti}}, \bibinfo {author}
  {\bibfnamefont {F.}~\bibnamefont {Le~Kien}}, \bibinfo {author} {\bibfnamefont
  {V.~G.}\ \bibnamefont {Truong}}, \bibinfo {author} {\bibfnamefont {V.~G.}\
  \bibnamefont {Truong}}, \bibinfo {author} {\bibfnamefont {S.~N.}\
  \bibnamefont {Chormaic}}, \bibinfo {author} {\bibfnamefont {S.~N.}\
  \bibnamefont {Chormaic}},\ and\ \bibinfo {author} {\bibfnamefont {S.~N.}\
  \bibnamefont {Chormaic}},\ }\bibfield  {title} {\bibinfo {title}
  {{Light-induced rotation of dielectric microparticles around an optical
  nanofiber}},\ }\href {https://doi.org/10.1364/OPTICA.374441} {\bibfield
  {journal} {\bibinfo  {journal} {Optica}\ }\textbf {\bibinfo {volume} {7}},\
  \bibinfo {pages} {59} (\bibinfo {year} {2020})}\BibitemShut {NoStop}%
\bibitem [{\citenamefont {Valero}\ \emph {et~al.}(2020)\citenamefont {Valero},
  \citenamefont {Kislov}, \citenamefont {Gurvitz}, \citenamefont {Shamkhi},
  \citenamefont {Pavlov}, \citenamefont {Redka}, \citenamefont {Yankin},
  \citenamefont {Zem{\ifmmode\acute{a}\else\'{a}\fi}nek},\ and\ \citenamefont
  {Shalin}}]{Valero2020AdvSci}%
  \BibitemOpen
  \bibfield  {author} {\bibinfo {author} {\bibfnamefont {A.~C.}\ \bibnamefont
  {Valero}}, \bibinfo {author} {\bibfnamefont {D.}~\bibnamefont {Kislov}},
  \bibinfo {author} {\bibfnamefont {E.~A.}\ \bibnamefont {Gurvitz}}, \bibinfo
  {author} {\bibfnamefont {H.~K.}\ \bibnamefont {Shamkhi}}, \bibinfo {author}
  {\bibfnamefont {A.~A.}\ \bibnamefont {Pavlov}}, \bibinfo {author}
  {\bibfnamefont {D.}~\bibnamefont {Redka}}, \bibinfo {author} {\bibfnamefont
  {S.}~\bibnamefont {Yankin}}, \bibinfo {author} {\bibfnamefont
  {P.}~\bibnamefont {Zem{\ifmmode\acute{a}\else\'{a}\fi}nek}},\ and\ \bibinfo
  {author} {\bibfnamefont {A.~S.}\ \bibnamefont {Shalin}},\ }\bibfield  {title}
  {\bibinfo {title} {{Nanovortex-Driven All-Dielectric Optical Diffusion
  Boosting and Sorting Concept for Lab-on-a-Chip Platforms}},\ }\href
  {https://doi.org/10.1002/advs.201903049} {\bibfield  {journal} {\bibinfo
  {journal} {Adv. Sci.}\ }\textbf {\bibinfo {volume} {7}},\ \bibinfo {pages}
  {1903049} (\bibinfo {year} {2020})}\BibitemShut {NoStop}%
\bibitem [{\citenamefont {Wang}\ \emph {et~al.}(2011)\citenamefont {Wang},
  \citenamefont {Schonbrun}, \citenamefont {Steinvurzel},\ and\ \citenamefont
  {Crozier}}]{Wang2011NC}%
  \BibitemOpen
  \bibfield  {author} {\bibinfo {author} {\bibfnamefont {K.}~\bibnamefont
  {Wang}}, \bibinfo {author} {\bibfnamefont {E.}~\bibnamefont {Schonbrun}},
  \bibinfo {author} {\bibfnamefont {P.}~\bibnamefont {Steinvurzel}},\ and\
  \bibinfo {author} {\bibfnamefont {K.~B.}\ \bibnamefont {Crozier}},\
  }\bibfield  {title} {\bibinfo {title} {{Trapping and rotating nanoparticles
  using a plasmonic nano-tweezer with an integrated heat sink - Nature
  Communications}},\ }\href {https://doi.org/10.1038/ncomms1480} {\bibfield
  {journal} {\bibinfo  {journal} {Nat. Commun.}\ }\textbf {\bibinfo {volume}
  {2}},\ \bibinfo {pages} {1} (\bibinfo {year} {2011})}\BibitemShut {NoStop}%
\bibitem [{\citenamefont {Bliokh}\ \emph {et~al.}(2015)\citenamefont {Bliokh},
  \citenamefont
  {Rodr{\ifmmode\acute{\imath}\else\'{\i}\fi}guez-Fortu{\ifmmode\tilde{n}\else\~{n}\fi}o},
  \citenamefont {Nori},\ and\ \citenamefont {Zayats}}]{Bliokh2015NP}%
  \BibitemOpen
  \bibfield  {author} {\bibinfo {author} {\bibfnamefont {K.~Y.}\ \bibnamefont
  {Bliokh}}, \bibinfo {author} {\bibfnamefont {F.~J.}\ \bibnamefont
  {Rodr{\ifmmode\acute{\imath}\else\'{\i}\fi}guez-Fortu{\ifmmode\tilde{n}\else\~{n}\fi}o}},
  \bibinfo {author} {\bibfnamefont {F.}~\bibnamefont {Nori}},\ and\ \bibinfo
  {author} {\bibfnamefont {A.~V.}\ \bibnamefont {Zayats}},\ }\bibfield  {title}
  {\bibinfo {title} {{Spin{\textendash}orbit interactions of light}},\ }\href
  {https://doi.org/10.1038/nphoton.2015.201} {\bibfield  {journal} {\bibinfo
  {journal} {Nat. Photonics}\ }\textbf {\bibinfo {volume} {9}},\ \bibinfo
  {pages} {796} (\bibinfo {year} {2015})}\BibitemShut {NoStop}%
\bibitem [{\citenamefont {Svak}\ \emph {et~al.}(2018)\citenamefont {Svak},
  \citenamefont {Brzobohat{\ifmmode\acute{y}\else\'{y}\fi}}, \citenamefont
  {{\ifmmode\check{S}\else\v{S}\fi}iler}, \citenamefont
  {J{\ifmmode\acute{a}\else\'{a}\fi}kl}, \citenamefont
  {Ka{\ifmmode\check{n}\else\v{n}\fi}ka}, \citenamefont
  {Zem{\ifmmode\acute{a}\else\'{a}\fi}nek},\ and\ \citenamefont
  {Simpson}}]{Svak2018NC}%
  \BibitemOpen
  \bibfield  {author} {\bibinfo {author} {\bibfnamefont {V.}~\bibnamefont
  {Svak}}, \bibinfo {author} {\bibfnamefont {O.}~\bibnamefont
  {Brzobohat{\ifmmode\acute{y}\else\'{y}\fi}}}, \bibinfo {author}
  {\bibfnamefont {M.}~\bibnamefont {{\ifmmode\check{S}\else\v{S}\fi}iler}},
  \bibinfo {author} {\bibfnamefont {P.}~\bibnamefont
  {J{\ifmmode\acute{a}\else\'{a}\fi}kl}}, \bibinfo {author} {\bibfnamefont
  {J.}~\bibnamefont {Ka{\ifmmode\check{n}\else\v{n}\fi}ka}}, \bibinfo {author}
  {\bibfnamefont {P.}~\bibnamefont {Zem{\ifmmode\acute{a}\else\'{a}\fi}nek}},\
  and\ \bibinfo {author} {\bibfnamefont {S.~H.}\ \bibnamefont {Simpson}},\
  }\bibfield  {title} {\bibinfo {title} {{Transverse spin forces and
  non-equilibrium particle dynamics in a circularly polarized vacuum optical
  trap}},\ }\href {https://doi.org/10.1038/s41467-018-07866-8} {\bibfield
  {journal} {\bibinfo  {journal} {Nat. Commun.}\ }\textbf {\bibinfo {volume}
  {9}},\ \bibinfo {pages} {5453} (\bibinfo {year} {2018})}\BibitemShut
  {NoStop}%
\bibitem [{\citenamefont {Brzobohat{\ifmmode\acute{y}\else\'{y}\fi}}\ \emph
  {et~al.}(2023)\citenamefont {Brzobohat{\ifmmode\acute{y}\else\'{y}\fi}},
  \citenamefont {Ducha{\ifmmode\check{n}\else\v{n}\fi}}, \citenamefont
  {J{\ifmmode\acute{a}\else\'{a}\fi}kl}, \citenamefont
  {Je{\ifmmode\check{z}\else\v{z}\fi}ek}, \citenamefont
  {{\ifmmode\check{S}\else\v{S}\fi}iler}, \citenamefont
  {Zem{\ifmmode\acute{a}\else\'{a}\fi}nek},\ and\ \citenamefont
  {Simpson}}]{Brzobohaty2023NC}%
  \BibitemOpen
  \bibfield  {author} {\bibinfo {author} {\bibfnamefont {O.}~\bibnamefont
  {Brzobohat{\ifmmode\acute{y}\else\'{y}\fi}}}, \bibinfo {author}
  {\bibfnamefont {M.}~\bibnamefont {Ducha{\ifmmode\check{n}\else\v{n}\fi}}},
  \bibinfo {author} {\bibfnamefont {P.}~\bibnamefont
  {J{\ifmmode\acute{a}\else\'{a}\fi}kl}}, \bibinfo {author} {\bibfnamefont
  {J.}~\bibnamefont {Je{\ifmmode\check{z}\else\v{z}\fi}ek}}, \bibinfo {author}
  {\bibfnamefont {M.}~\bibnamefont {{\ifmmode\check{S}\else\v{S}\fi}iler}},
  \bibinfo {author} {\bibfnamefont {P.}~\bibnamefont
  {Zem{\ifmmode\acute{a}\else\'{a}\fi}nek}},\ and\ \bibinfo {author}
  {\bibfnamefont {S.~H.}\ \bibnamefont {Simpson}},\ }\bibfield  {title}
  {\bibinfo {title} {{Synchronization of spin-driven limit cycle oscillators
  optically levitated in vacuum}},\ }\href
  {https://doi.org/10.1038/s41467-023-41129-5} {\bibfield  {journal} {\bibinfo
  {journal} {Nat. Commun.}\ }\textbf {\bibinfo {volume} {14}},\ \bibinfo
  {pages} {5441} (\bibinfo {year} {2023})}\BibitemShut {NoStop}%
\bibitem [{\citenamefont {Arita}\ \emph {et~al.}(2023)\citenamefont {Arita},
  \citenamefont {Simpson}, \citenamefont {Bruce}, \citenamefont {Wright},
  \citenamefont {Zem{\ifmmode\acute{a}\else\'{a}\fi}nek},\ and\ \citenamefont
  {Dholakia}}]{Arita2023CommunPhys}%
  \BibitemOpen
  \bibfield  {author} {\bibinfo {author} {\bibfnamefont {Y.}~\bibnamefont
  {Arita}}, \bibinfo {author} {\bibfnamefont {S.~H.}\ \bibnamefont {Simpson}},
  \bibinfo {author} {\bibfnamefont {G.~D.}\ \bibnamefont {Bruce}}, \bibinfo
  {author} {\bibfnamefont {E.~M.}\ \bibnamefont {Wright}}, \bibinfo {author}
  {\bibfnamefont {P.}~\bibnamefont {Zem{\ifmmode\acute{a}\else\'{a}\fi}nek}},\
  and\ \bibinfo {author} {\bibfnamefont {K.}~\bibnamefont {Dholakia}},\
  }\bibfield  {title} {\bibinfo {title} {{Cooling the optical-spin driven limit
  cycle oscillations of a levitated gyroscope}},\ }\href
  {https://doi.org/10.1038/s42005-023-01336-4} {\bibfield  {journal} {\bibinfo
  {journal} {Commun. Phys.}\ }\textbf {\bibinfo {volume} {6}},\ \bibinfo
  {pages} {238} (\bibinfo {year} {2023})}\BibitemShut {NoStop}%
\bibitem [{\citenamefont {Arzola}\ \emph {et~al.}(2019)\citenamefont {Arzola},
  \citenamefont {Chv{\ifmmode\acute{a}\else\'{a}\fi}tal}, \citenamefont
  {J{\ifmmode\acute{a}\else\'{a}\fi}kl},\ and\ \citenamefont
  {Zem{\ifmmode\acute{a}\else\'{a}\fi}nek}}]{Arzola2019SciRep}%
  \BibitemOpen
  \bibfield  {author} {\bibinfo {author} {\bibfnamefont {A.~V.}\ \bibnamefont
  {Arzola}}, \bibinfo {author} {\bibfnamefont {L.}~\bibnamefont
  {Chv{\ifmmode\acute{a}\else\'{a}\fi}tal}}, \bibinfo {author} {\bibfnamefont
  {P.}~\bibnamefont {J{\ifmmode\acute{a}\else\'{a}\fi}kl}},\ and\ \bibinfo
  {author} {\bibfnamefont {P.}~\bibnamefont
  {Zem{\ifmmode\acute{a}\else\'{a}\fi}nek}},\ }\bibfield  {title} {\bibinfo
  {title} {{Spin to orbital light momentum conversion visualized by particle
  trajectory}},\ }\href {https://doi.org/10.1038/s41598-019-40475-z} {\bibfield
   {journal} {\bibinfo  {journal} {Sci. Rep.}\ }\textbf {\bibinfo {volume}
  {9}},\ \bibinfo {pages} {4127} (\bibinfo {year} {2019})}\BibitemShut
  {NoStop}%
\bibitem [{\citenamefont {Tzarouchis}\ and\ \citenamefont
  {Sihvola}(2018)}]{Tzarouchis2018AS}%
  \BibitemOpen
  \bibfield  {author} {\bibinfo {author} {\bibfnamefont {D.}~\bibnamefont
  {Tzarouchis}}\ and\ \bibinfo {author} {\bibfnamefont {A.}~\bibnamefont
  {Sihvola}},\ }\bibfield  {title} {\bibinfo {title} {{Light Scattering by a
  Dielectric Sphere: Perspectives on the Mie Resonances}},\ }\href
  {https://doi.org/10.3390/app8020184} {\bibfield  {journal} {\bibinfo
  {journal} {Appl. Sci.}\ }\textbf {\bibinfo {volume} {8}},\ \bibinfo {pages}
  {184} (\bibinfo {year} {2018})}\BibitemShut {NoStop}%
\bibitem [{\citenamefont {Ashkin}\ and\ \citenamefont
  {Dziedzic}(1977)}]{Ashkin1977PRL}%
  \BibitemOpen
  \bibfield  {author} {\bibinfo {author} {\bibfnamefont {A.}~\bibnamefont
  {Ashkin}}\ and\ \bibinfo {author} {\bibfnamefont {J.~M.}\ \bibnamefont
  {Dziedzic}},\ }\bibfield  {title} {\bibinfo {title} {{Observation of
  Resonances in the Radiation Pressure on Dielectric Spheres}},\ }\href
  {https://doi.org/10.1103/PhysRevLett.38.1351} {\bibfield  {journal} {\bibinfo
   {journal} {Phys. Rev. Lett.}\ }\textbf {\bibinfo {volume} {38}},\ \bibinfo
  {pages} {1351} (\bibinfo {year} {1977})}\BibitemShut {NoStop}%
\bibitem [{\citenamefont {Cao}\ \emph {et~al.}(2016{\natexlab{b}})\citenamefont
  {Cao}, \citenamefont {Mao}, \citenamefont {Gao}, \citenamefont {Ding},\ and\
  \citenamefont {Qiu}}]{Cao2016Nanoscale}%
  \BibitemOpen
  \bibfield  {author} {\bibinfo {author} {\bibfnamefont {T.}~\bibnamefont
  {Cao}}, \bibinfo {author} {\bibfnamefont {L.}~\bibnamefont {Mao}}, \bibinfo
  {author} {\bibfnamefont {D.}~\bibnamefont {Gao}}, \bibinfo {author}
  {\bibfnamefont {W.}~\bibnamefont {Ding}},\ and\ \bibinfo {author}
  {\bibfnamefont {C.-W.}\ \bibnamefont {Qiu}},\ }\bibfield  {title} {\bibinfo
  {title} {{Fano resonant Ge2Sb2Te5 nanoparticles realize switchable lateral
  optical force}},\ }\href {https://doi.org/10.1039/C5NR08804F} {\bibfield
  {journal} {\bibinfo  {journal} {Nanoscale}\ }\textbf {\bibinfo {volume}
  {8}},\ \bibinfo {pages} {5657} (\bibinfo {year}
  {2016}{\natexlab{b}})}\BibitemShut {NoStop}%
\bibitem [{\citenamefont {Kislov}\ \emph {et~al.}(2021)\citenamefont {Kislov},
  \citenamefont {Gurvitz}, \citenamefont {Bobrovs}, \citenamefont {Pavlov},
  \citenamefont {Redka}, \citenamefont
  {Marqu{\ifmmode\acute{e}\else\'{e}\fi}s}, \citenamefont {Ginzburg},\ and\
  \citenamefont {Shalin}}]{Kislov2021APR}%
  \BibitemOpen
  \bibfield  {author} {\bibinfo {author} {\bibfnamefont {D.~A.}\ \bibnamefont
  {Kislov}}, \bibinfo {author} {\bibfnamefont {E.~A.}\ \bibnamefont {Gurvitz}},
  \bibinfo {author} {\bibfnamefont {V.}~\bibnamefont {Bobrovs}}, \bibinfo
  {author} {\bibfnamefont {A.~A.}\ \bibnamefont {Pavlov}}, \bibinfo {author}
  {\bibfnamefont {D.~N.}\ \bibnamefont {Redka}}, \bibinfo {author}
  {\bibfnamefont {M.~I.}\ \bibnamefont
  {Marqu{\ifmmode\acute{e}\else\'{e}\fi}s}}, \bibinfo {author} {\bibfnamefont
  {P.}~\bibnamefont {Ginzburg}},\ and\ \bibinfo {author} {\bibfnamefont
  {A.~S.}\ \bibnamefont {Shalin}},\ }\bibfield  {title} {\bibinfo {title}
  {{Multipole Engineering of Attractive{-}Repulsive and Bending Optical
  Forces}},\ }\href {https://doi.org/10.1002/adpr.202100082} {\bibfield
  {journal} {\bibinfo  {journal} {Adv. Photonics Res.}\ }\textbf {\bibinfo
  {volume} {2}},\ \bibinfo {pages} {2100082} (\bibinfo {year}
  {2021})}\BibitemShut {NoStop}%
\bibitem [{\citenamefont {Lepeshov}\ \emph {et~al.}(2023)\citenamefont
  {Lepeshov}, \citenamefont {Meyer}, \citenamefont {Maurer}, \citenamefont
  {Romero-Isart},\ and\ \citenamefont {Quidant}}]{Lepeshov2023PRL}%
  \BibitemOpen
  \bibfield  {author} {\bibinfo {author} {\bibfnamefont {S.}~\bibnamefont
  {Lepeshov}}, \bibinfo {author} {\bibfnamefont {N.}~\bibnamefont {Meyer}},
  \bibinfo {author} {\bibfnamefont {P.}~\bibnamefont {Maurer}}, \bibinfo
  {author} {\bibfnamefont {O.}~\bibnamefont {Romero-Isart}},\ and\ \bibinfo
  {author} {\bibfnamefont {R.}~\bibnamefont {Quidant}},\ }\bibfield  {title}
  {\bibinfo {title} {{Levitated Optomechanics with Meta-Atoms}},\ }\href
  {https://doi.org/10.1103/PhysRevLett.130.233601} {\bibfield  {journal}
  {\bibinfo  {journal} {Phys. Rev. Lett.}\ }\textbf {\bibinfo {volume} {130}},\
  \bibinfo {pages} {233601} (\bibinfo {year} {2023})}\BibitemShut {NoStop}%
\bibitem [{\citenamefont {Xu}\ \emph {et~al.}(2020)\citenamefont {Xu},
  \citenamefont {Nieto-Vesperinas}, \citenamefont {Qiu}, \citenamefont {Liu},
  \citenamefont {Gao}, \citenamefont {Zhang},\ and\ \citenamefont
  {Li}}]{Xu2020LPR}%
  \BibitemOpen
  \bibfield  {author} {\bibinfo {author} {\bibfnamefont {X.}~\bibnamefont
  {Xu}}, \bibinfo {author} {\bibfnamefont {M.}~\bibnamefont
  {Nieto-Vesperinas}}, \bibinfo {author} {\bibfnamefont {C.-W.}\ \bibnamefont
  {Qiu}}, \bibinfo {author} {\bibfnamefont {X.}~\bibnamefont {Liu}}, \bibinfo
  {author} {\bibfnamefont {D.}~\bibnamefont {Gao}}, \bibinfo {author}
  {\bibfnamefont {Y.}~\bibnamefont {Zhang}},\ and\ \bibinfo {author}
  {\bibfnamefont {B.}~\bibnamefont {Li}},\ }\bibfield  {title} {\bibinfo
  {title} {{Kerker-Type Intensity-Gradient Force of Light}},\ }\href
  {https://doi.org/10.1002/lpor.201900265} {\bibfield  {journal} {\bibinfo
  {journal} {Laser Photonics Rev.}\ }\textbf {\bibinfo {volume} {14}},\
  \bibinfo {pages} {1900265} (\bibinfo {year} {2020})}\BibitemShut {NoStop}%
\bibitem [{\citenamefont {Chen}\ \emph {et~al.}(2015)\citenamefont {Chen},
  \citenamefont {Liu}, \citenamefont {Zi},\ and\ \citenamefont
  {Lin}}]{Chen2015ACSnano}%
  \BibitemOpen
  \bibfield  {author} {\bibinfo {author} {\bibfnamefont {H.}~\bibnamefont
  {Chen}}, \bibinfo {author} {\bibfnamefont {S.}~\bibnamefont {Liu}}, \bibinfo
  {author} {\bibfnamefont {J.}~\bibnamefont {Zi}},\ and\ \bibinfo {author}
  {\bibfnamefont {Z.}~\bibnamefont {Lin}},\ }\bibfield  {title} {\bibinfo
  {title} {{Fano Resonance-Induced Negative Optical Scattering Force on
  Plasmonic Nanoparticles}},\ }\href {https://doi.org/10.1021/nn506835j}
  {\bibfield  {journal} {\bibinfo  {journal} {ACS Nano}\ }\textbf {\bibinfo
  {volume} {9}},\ \bibinfo {pages} {1926} (\bibinfo {year} {2015})}\BibitemShut
  {NoStop}%
\bibitem [{\citenamefont {Zhou}\ \emph {et~al.}(2023)\citenamefont {Zhou},
  \citenamefont {Zhang}, \citenamefont {Xu}, \citenamefont {Nieto-Vesperinas},
  \citenamefont {Yan}, \citenamefont {Li}, \citenamefont {Gao}, \citenamefont
  {Zhang},\ and\ \citenamefont {Yao}}]{Zhou2023LPR}%
  \BibitemOpen
  \bibfield  {author} {\bibinfo {author} {\bibfnamefont {Y.}~\bibnamefont
  {Zhou}}, \bibinfo {author} {\bibfnamefont {Y.}~\bibnamefont {Zhang}},
  \bibinfo {author} {\bibfnamefont {X.}~\bibnamefont {Xu}}, \bibinfo {author}
  {\bibfnamefont {M.}~\bibnamefont {Nieto-Vesperinas}}, \bibinfo {author}
  {\bibfnamefont {S.}~\bibnamefont {Yan}}, \bibinfo {author} {\bibfnamefont
  {M.}~\bibnamefont {Li}}, \bibinfo {author} {\bibfnamefont {W.}~\bibnamefont
  {Gao}}, \bibinfo {author} {\bibfnamefont {Y.}~\bibnamefont {Zhang}},\ and\
  \bibinfo {author} {\bibfnamefont {B.}~\bibnamefont {Yao}},\ }\bibfield
  {title} {\bibinfo {title} {{Optical Forces on Multipoles Induced by the
  Belinfante Spin Momentum}},\ }\href {https://doi.org/10.1002/lpor.202300245}
  {\bibfield  {journal} {\bibinfo  {journal} {Laser Photonics Rev.}\ }\textbf
  {\bibinfo {volume} {17}},\ \bibinfo {pages} {2300245} (\bibinfo {year}
  {2023})}\BibitemShut {NoStop}%
\bibitem [{\citenamefont {Nan}\ \emph {et~al.}(2023)\citenamefont {Nan},
  \citenamefont
  {Rodr{\ifmmode\acute{\imath}\else\'{\i}\fi}guez-Fortu{\ifmmode\tilde{n}\else\~{n}\fi}o},
  \citenamefont {Yan}, \citenamefont {Kingsley-Smith}, \citenamefont {Ng},
  \citenamefont {Yao}, \citenamefont {Yan},\ and\ \citenamefont
  {Xu}}]{Nan2023NC}%
  \BibitemOpen
  \bibfield  {author} {\bibinfo {author} {\bibfnamefont {F.}~\bibnamefont
  {Nan}}, \bibinfo {author} {\bibfnamefont {F.~J.}\ \bibnamefont
  {Rodr{\ifmmode\acute{\imath}\else\'{\i}\fi}guez-Fortu{\ifmmode\tilde{n}\else\~{n}\fi}o}},
  \bibinfo {author} {\bibfnamefont {S.}~\bibnamefont {Yan}}, \bibinfo {author}
  {\bibfnamefont {J.~J.}\ \bibnamefont {Kingsley-Smith}}, \bibinfo {author}
  {\bibfnamefont {J.}~\bibnamefont {Ng}}, \bibinfo {author} {\bibfnamefont
  {B.}~\bibnamefont {Yao}}, \bibinfo {author} {\bibfnamefont {Z.}~\bibnamefont
  {Yan}},\ and\ \bibinfo {author} {\bibfnamefont {X.}~\bibnamefont {Xu}},\
  }\bibfield  {title} {\bibinfo {title} {{Creating tunable lateral optical
  forces through multipolar interplay in single nanowires}},\ }\href
  {https://doi.org/10.1038/s41467-023-42076-x} {\bibfield  {journal} {\bibinfo
  {journal} {Nat. Commun.}\ }\textbf {\bibinfo {volume} {14}},\ \bibinfo
  {pages} {1} (\bibinfo {year} {2023})}\BibitemShut {NoStop}%
\bibitem [{\citenamefont {Kiselev}\ \emph {et~al.}(2020)\citenamefont
  {Kiselev}, \citenamefont {Kiselev}, \citenamefont {Achouri}, \citenamefont
  {Martin},\ and\ \citenamefont {Martin}}]{Kiselev2020OE}%
  \BibitemOpen
  \bibfield  {author} {\bibinfo {author} {\bibfnamefont {A.}~\bibnamefont
  {Kiselev}}, \bibinfo {author} {\bibfnamefont {A.}~\bibnamefont {Kiselev}},
  \bibinfo {author} {\bibfnamefont {K.}~\bibnamefont {Achouri}}, \bibinfo
  {author} {\bibfnamefont {O.~J.~F.}\ \bibnamefont {Martin}},\ and\ \bibinfo
  {author} {\bibfnamefont {O.~J.~F.}\ \bibnamefont {Martin}},\ }\bibfield
  {title} {\bibinfo {title} {{Multipole interplay controls optical forces and
  ultra-directional scattering}},\ }\href {https://doi.org/10.1364/OE.400387}
  {\bibfield  {journal} {\bibinfo  {journal} {Opt. Express}\ }\textbf {\bibinfo
  {volume} {28}},\ \bibinfo {pages} {27547} (\bibinfo {year}
  {2020})}\BibitemShut {NoStop}%
\bibitem [{\citenamefont {Rockstuhl}\ and\ \citenamefont
  {Herzig}(2004)}]{Rockstuhl2004OL}%
  \BibitemOpen
  \bibfield  {author} {\bibinfo {author} {\bibfnamefont {C.}~\bibnamefont
  {Rockstuhl}}\ and\ \bibinfo {author} {\bibfnamefont {H.~P.}\ \bibnamefont
  {Herzig}},\ }\bibfield  {title} {\bibinfo {title} {{Wavelength-dependent
  optical force on elliptical silver cylinders at plasmon resonance}},\ }\href
  {https://doi.org/10.1364/OL.29.002181} {\bibfield  {journal} {\bibinfo
  {journal} {Opt. Lett.}\ }\textbf {\bibinfo {volume} {29}},\ \bibinfo {pages}
  {2181} (\bibinfo {year} {2004})}\BibitemShut {NoStop}%
\bibitem [{\citenamefont {Stilgoe}\ \emph {et~al.}(2008)\citenamefont
  {Stilgoe}, \citenamefont {Nieminen}, \citenamefont
  {Kn{\ifmmode\ddot{o}\else\"{o}\fi}ner}, \citenamefont {Heckenberg},\ and\
  \citenamefont {Rubinsztein-Dunlop}}]{Stilgoe2008OE}%
  \BibitemOpen
  \bibfield  {author} {\bibinfo {author} {\bibfnamefont {A.~B.}\ \bibnamefont
  {Stilgoe}}, \bibinfo {author} {\bibfnamefont {T.~A.}\ \bibnamefont
  {Nieminen}}, \bibinfo {author} {\bibfnamefont {G.}~\bibnamefont
  {Kn{\ifmmode\ddot{o}\else\"{o}\fi}ner}}, \bibinfo {author} {\bibfnamefont
  {N.~R.}\ \bibnamefont {Heckenberg}},\ and\ \bibinfo {author} {\bibfnamefont
  {H.}~\bibnamefont {Rubinsztein-Dunlop}},\ }\bibfield  {title} {\bibinfo
  {title} {{The effect of Mie resonances on trapping in optical tweezers}},\
  }\href {https://doi.org/10.1364/OE.16.015039} {\bibfield  {journal} {\bibinfo
   {journal} {Opt. Express}\ }\textbf {\bibinfo {volume} {16}},\ \bibinfo
  {pages} {15039} (\bibinfo {year} {2008})}\BibitemShut {NoStop}%
\bibitem [{\citenamefont {Duan}\ \emph {et~al.}(2022)\citenamefont {Duan},
  \citenamefont {Bruce}, \citenamefont {Li},\ and\ \citenamefont
  {Dholakia}}]{Duan2022PRA}%
  \BibitemOpen
  \bibfield  {author} {\bibinfo {author} {\bibfnamefont {X.-Y.}\ \bibnamefont
  {Duan}}, \bibinfo {author} {\bibfnamefont {G.~D.}\ \bibnamefont {Bruce}},
  \bibinfo {author} {\bibfnamefont {F.}~\bibnamefont {Li}},\ and\ \bibinfo
  {author} {\bibfnamefont {K.}~\bibnamefont {Dholakia}},\ }\bibfield  {title}
  {\bibinfo {title} {{Asymmetric longitudinal optical binding force between two
  identical dielectric particles with electric and magnetic dipolar
  responses}},\ }\href {https://doi.org/10.1103/PhysRevA.106.013108} {\bibfield
   {journal} {\bibinfo  {journal} {Phys. Rev. A}\ }\textbf {\bibinfo {volume}
  {106}},\ \bibinfo {pages} {013108} (\bibinfo {year} {2022})}\BibitemShut
  {NoStop}%
\bibitem [{\citenamefont {Wagner}\ \emph {et~al.}(2014)\citenamefont {Wagner},
  \citenamefont {Lipinski},\ and\ \citenamefont
  {Wiemann}}]{Wagner2014JNanopartRes}%
  \BibitemOpen
  \bibfield  {author} {\bibinfo {author} {\bibfnamefont {T.}~\bibnamefont
  {Wagner}}, \bibinfo {author} {\bibfnamefont {H.-G.}\ \bibnamefont
  {Lipinski}},\ and\ \bibinfo {author} {\bibfnamefont {M.}~\bibnamefont
  {Wiemann}},\ }\bibfield  {title} {\bibinfo {title} {{Dark field nanoparticle
  tracking analysis for size characterization of plasmonic and non-plasmonic
  particles}},\ }\href {https://doi.org/10.1007/s11051-014-2419-x} {\bibfield
  {journal} {\bibinfo  {journal} {J. Nanopart. Res.}\ }\textbf {\bibinfo
  {volume} {16}},\ \bibinfo {pages} {2419} (\bibinfo {year}
  {2014})}\BibitemShut {NoStop}%
\bibitem [{\citenamefont {Bohren}\ and\ \citenamefont
  {Huffman}(1984)}]{Bohren1984}%
  \BibitemOpen
  \bibfield  {author} {\bibinfo {author} {\bibfnamefont {C.~F.}\ \bibnamefont
  {Bohren}}\ and\ \bibinfo {author} {\bibfnamefont {D.~R.}\ \bibnamefont
  {Huffman}},\ }\href@noop {} {\emph {\bibinfo {title} {Absorption and
  Scattering of Light by {{Small Particles}}}}},\ Vol.~\bibinfo {volume} {1}\
  (\bibinfo  {publisher} {Wiley},\ \bibinfo {year} {1984})\BibitemShut
  {NoStop}%
\bibitem [{\citenamefont {Kramers}(1940)}]{Kramers1940}%
  \BibitemOpen
  \bibfield  {author} {\bibinfo {author} {\bibfnamefont {H.~A.}\ \bibnamefont
  {Kramers}},\ }\bibfield  {title} {\bibinfo {title} {{Brownian motion in a
  field of force and the diffusion model of chemical reactions}},\ }\href
  {https://doi.org/10.1016/S0031-8914(40)90098-2} {\bibfield  {journal}
  {\bibinfo  {journal} {Physica}\ }\textbf {\bibinfo {volume} {7}},\ \bibinfo
  {pages} {284} (\bibinfo {year} {1940})}\BibitemShut {NoStop}%
\bibitem [{\citenamefont {Landauer}\ and\ \citenamefont
  {Swanson}(1961)}]{Landauer1961PhysRev}%
  \BibitemOpen
  \bibfield  {author} {\bibinfo {author} {\bibfnamefont {R.}~\bibnamefont
  {Landauer}}\ and\ \bibinfo {author} {\bibfnamefont {J.~A.}\ \bibnamefont
  {Swanson}},\ }\bibfield  {title} {\bibinfo {title} {{Frequency Factors in the
  Thermally Activated Process}},\ }\href
  {https://doi.org/10.1103/PhysRev.121.1668} {\bibfield  {journal} {\bibinfo
  {journal} {Phys. Rev.}\ }\textbf {\bibinfo {volume} {121}},\ \bibinfo {pages}
  {1668} (\bibinfo {year} {1961})}\BibitemShut {NoStop}%
\bibitem [{\citenamefont {Mel'nikov}(1991)}]{Melnikov1991PhysRep}%
  \BibitemOpen
  \bibfield  {author} {\bibinfo {author} {\bibfnamefont {V.~I.}\ \bibnamefont
  {Mel'nikov}},\ }\bibfield  {title} {\bibinfo {title} {{The Kramers problem:
  Fifty years of development}},\ }\href
  {https://doi.org/10.1016/0370-1573(91)90108-X} {\bibfield  {journal}
  {\bibinfo  {journal} {Phys. Rep.}\ }\textbf {\bibinfo {volume} {209}},\
  \bibinfo {pages} {1} (\bibinfo {year} {1991})}\BibitemShut {NoStop}%
\bibitem [{\citenamefont {Novotny}\ and\ \citenamefont
  {Hecht}(2012)}]{Novotny2012}%
  \BibitemOpen
  \bibfield  {author} {\bibinfo {author} {\bibfnamefont {L.}~\bibnamefont
  {Novotny}}\ and\ \bibinfo {author} {\bibfnamefont {B.}~\bibnamefont
  {Hecht}},\ }\href@noop {} {\emph {\bibinfo {title} {Principles of
  Nano-Optics}}},\ \bibinfo {edition} {2nd}\ ed.\ (\bibinfo  {publisher}
  {Cambridge University Press},\ \bibinfo {address} {Cambridge},\ \bibinfo
  {year} {2012})\BibitemShut {NoStop}%
\bibitem [{\citenamefont {Collett}(2005)}]{Collett2005}%
  \BibitemOpen
  \bibfield  {author} {\bibinfo {author} {\bibfnamefont {E.}~\bibnamefont
  {Collett}},\ }\href@noop {} {\emph {\bibinfo {title} {Field {{Guide}} to
  {{Polarization}}}}},\ Vol.~\bibinfo {volume} {1}\ (\bibinfo  {publisher}
  {SPIE},\ \bibinfo {year} {2005})\BibitemShut {NoStop}%
\bibitem [{\citenamefont {Yu}\ \emph {et~al.}(2019)\citenamefont {Yu},
  \citenamefont {Jiang}, \citenamefont {Chen}, \citenamefont {Liu},\ and\
  \citenamefont {Lin}}]{Yu2019PRA}%
  \BibitemOpen
  \bibfield  {author} {\bibinfo {author} {\bibfnamefont {X.}~\bibnamefont
  {Yu}}, \bibinfo {author} {\bibfnamefont {Y.}~\bibnamefont {Jiang}}, \bibinfo
  {author} {\bibfnamefont {H.}~\bibnamefont {Chen}}, \bibinfo {author}
  {\bibfnamefont {S.}~\bibnamefont {Liu}},\ and\ \bibinfo {author}
  {\bibfnamefont {Z.}~\bibnamefont {Lin}},\ }\bibfield  {title} {\bibinfo
  {title} {{Approach to fully decomposing an optical force into conservative
  and nonconservative components}},\ }\href
  {https://doi.org/10.1103/PhysRevA.100.033821} {\bibfield  {journal} {\bibinfo
   {journal} {Phys. Rev. A}\ }\textbf {\bibinfo {volume} {100}},\ \bibinfo
  {pages} {033821} (\bibinfo {year} {2019})}\BibitemShut {NoStop}%
\bibitem [{\citenamefont {Jiang}\ \emph {et~al.}(2015)\citenamefont {Jiang},
  \citenamefont {Ng},\ and\ \citenamefont {Lin}}]{Jiang2015arXiv}%
  \BibitemOpen
  \bibfield  {author} {\bibinfo {author} {\bibfnamefont {Y.}~\bibnamefont
  {Jiang}}, \bibinfo {author} {\bibfnamefont {J.}~\bibnamefont {Ng}},\ and\
  \bibinfo {author} {\bibfnamefont {Z.}~\bibnamefont {Lin}},\ }\bibfield
  {title} {\bibinfo {title} {{Ab initio derivation of multipolar expansion of
  optical force}},\ }\bibfield  {journal} {\bibinfo  {journal} {arXiv}\ }\href
  {https://doi.org/10.48550/arXiv.1512.04201} {10.48550/arXiv.1512.04201}
  (\bibinfo {year} {2015}),\ \Eprint {https://arxiv.org/abs/1512.04201}
  {1512.04201} \BibitemShut {NoStop}%
\bibitem [{\citenamefont {Chen}\ \emph {et~al.}(2011)\citenamefont {Chen},
  \citenamefont {Ng}, \citenamefont {Lin},\ and\ \citenamefont
  {Chan}}]{Chen2011NP}%
  \BibitemOpen
  \bibfield  {author} {\bibinfo {author} {\bibfnamefont {J.}~\bibnamefont
  {Chen}}, \bibinfo {author} {\bibfnamefont {J.}~\bibnamefont {Ng}}, \bibinfo
  {author} {\bibfnamefont {Z.}~\bibnamefont {Lin}},\ and\ \bibinfo {author}
  {\bibfnamefont {C.~T.}\ \bibnamefont {Chan}},\ }\bibfield  {title} {\bibinfo
  {title} {{Optical pulling force}},\ }\href
  {https://doi.org/10.1038/nphoton.2011.153} {\bibfield  {journal} {\bibinfo
  {journal} {Nat. Photonics}\ }\textbf {\bibinfo {volume} {5}},\ \bibinfo
  {pages} {531} (\bibinfo {year} {2011})}\BibitemShut {NoStop}%
\bibitem [{\citenamefont {Toftul}\ \emph {et~al.}(2023)\citenamefont {Toftul},
  \citenamefont {Fedorovich}, \citenamefont {Kislov}, \citenamefont {Frizyuk},
  \citenamefont {Koshelev}, \citenamefont {Kivshar},\ and\ \citenamefont
  {Petrov}}]{Toftul2023PRL}%
  \BibitemOpen
  \bibfield  {author} {\bibinfo {author} {\bibfnamefont {I.}~\bibnamefont
  {Toftul}}, \bibinfo {author} {\bibfnamefont {G.}~\bibnamefont {Fedorovich}},
  \bibinfo {author} {\bibfnamefont {D.}~\bibnamefont {Kislov}}, \bibinfo
  {author} {\bibfnamefont {K.}~\bibnamefont {Frizyuk}}, \bibinfo {author}
  {\bibfnamefont {K.}~\bibnamefont {Koshelev}}, \bibinfo {author}
  {\bibfnamefont {Y.}~\bibnamefont {Kivshar}},\ and\ \bibinfo {author}
  {\bibfnamefont {M.}~\bibnamefont {Petrov}},\ }\bibfield  {title} {\bibinfo
  {title} {{Nonlinearity-Induced Optical Torque}},\ }\href
  {https://doi.org/10.1103/PhysRevLett.130.243802} {\bibfield  {journal}
  {\bibinfo  {journal} {Phys. Rev. Lett.}\ }\textbf {\bibinfo {volume} {130}},\
  \bibinfo {pages} {243802} (\bibinfo {year} {2023})}\BibitemShut {NoStop}%
\bibitem [{\citenamefont {Toftul}\ \emph {et~al.}(2025)\citenamefont {Toftul},
  \citenamefont {Petrov}, \citenamefont {Quidant},\ and\ \citenamefont
  {Kivshar}}]{Toftul2025ACSPhot}%
  \BibitemOpen
  \bibfield  {author} {\bibinfo {author} {\bibfnamefont {I.}~\bibnamefont
  {Toftul}}, \bibinfo {author} {\bibfnamefont {M.}~\bibnamefont {Petrov}},
  \bibinfo {author} {\bibfnamefont {R.}~\bibnamefont {Quidant}},\ and\ \bibinfo
  {author} {\bibfnamefont {Y.}~\bibnamefont {Kivshar}},\ }\bibfield  {title}
  {\bibinfo {title} {{Optical Supertorque Induced by Mie-Resonant Modes}},\
  }\bibfield  {journal} {\bibinfo  {journal} {ACS Photonics}\ }\textbf
  {\bibinfo {volume} {2025}},\ \href
  {https://doi.org/10.1021/acsphotonics.5c00134} {10.1021/acsphotonics.5c00134}
  (\bibinfo {year} {2025})\BibitemShut {NoStop}%
\bibitem [{\citenamefont {Toftul}\ \emph {et~al.}(2026)\citenamefont {Toftul},
  \citenamefont {Golat}, \citenamefont
  {Rodr{\ifmmode\acute{\imath}\else\'{\i}\fi}guez-Fortu{\ifmmode\tilde{n}\else\~{n}\fi}o},
  \citenamefont {Nori}, \citenamefont {Kivshar},\ and\ \citenamefont
  {Bliokh}}]{Toftul2026RMP}%
  \BibitemOpen
  \bibfield  {author} {\bibinfo {author} {\bibfnamefont {I.}~\bibnamefont
  {Toftul}}, \bibinfo {author} {\bibfnamefont {S.}~\bibnamefont {Golat}},
  \bibinfo {author} {\bibfnamefont {F.~J.}\ \bibnamefont
  {Rodr{\ifmmode\acute{\imath}\else\'{\i}\fi}guez-Fortu{\ifmmode\tilde{n}\else\~{n}\fi}o}},
  \bibinfo {author} {\bibfnamefont {F.}~\bibnamefont {Nori}}, \bibinfo {author}
  {\bibfnamefont {Y.}~\bibnamefont {Kivshar}},\ and\ \bibinfo {author}
  {\bibfnamefont {K.~Y.}\ \bibnamefont {Bliokh}},\ }\bibfield  {title}
  {\bibinfo {title} {{Radiation forces and torques in optics and acoustics}},\
  }\href {https://doi.org/10.1103/7zdw-p26g} {\bibfield  {journal} {\bibinfo
  {journal} {Rev. Mod. Phys.}\ }\textbf {\bibinfo {volume} {98}},\ \bibinfo
  {pages} {025002} (\bibinfo {year} {2026})}\BibitemShut {NoStop}%
\bibitem [{\citenamefont {Jiang}\ \emph {et~al.}(2016)\citenamefont {Jiang},
  \citenamefont {Chen}, \citenamefont {Ng},\ and\ \citenamefont
  {Lin}}]{Jiang2016arXiv}%
  \BibitemOpen
  \bibfield  {author} {\bibinfo {author} {\bibfnamefont {Y.}~\bibnamefont
  {Jiang}}, \bibinfo {author} {\bibfnamefont {J.}~\bibnamefont {Chen}},
  \bibinfo {author} {\bibfnamefont {J.}~\bibnamefont {Ng}},\ and\ \bibinfo
  {author} {\bibfnamefont {Z.}~\bibnamefont {Lin}},\ }\bibfield  {title}
  {\bibinfo {title} {{Decomposition of optical force into conservative and
  nonconservative components}},\ }\bibfield  {journal} {\bibinfo  {journal}
  {arXiv}\ }\href {https://doi.org/10.48550/arXiv.1604.05138}
  {10.48550/arXiv.1604.05138} (\bibinfo {year} {2016}),\ \Eprint
  {https://arxiv.org/abs/1604.05138} {1604.05138} \BibitemShut {NoStop}%
\bibitem [{\citenamefont {Toftul}(2026)}]{zenodo_archive}%
  \BibitemOpen
  \bibfield  {author} {\bibinfo {author} {\bibfnamefont {I.}~\bibnamefont
  {Toftul}},\ }\href {https://doi.org/10.5281/ZENODO.16907811} {\bibinfo
  {title} {github.com/toftul/quadrupole-optical-force-and-particle-hopping:
  Codes and experimental data}} (\bibinfo {year} {2026})\BibitemShut {NoStop}%
\bibitem [{\citenamefont {Taha}\ \emph {et~al.}(2023)\citenamefont {Taha},
  \citenamefont {Balendhran}, \citenamefont {Sherrell}, \citenamefont
  {Kirkwood}, \citenamefont {Wen}, \citenamefont {Wang}, \citenamefont {Meng},
  \citenamefont {Bullock}, \citenamefont {Crozier},\ and\ \citenamefont
  {Sciacca}}]{Taha2023Apr}%
  \BibitemOpen
  \bibfield  {author} {\bibinfo {author} {\bibfnamefont {M.}~\bibnamefont
  {Taha}}, \bibinfo {author} {\bibfnamefont {S.}~\bibnamefont {Balendhran}},
  \bibinfo {author} {\bibfnamefont {P.~C.}\ \bibnamefont {Sherrell}}, \bibinfo
  {author} {\bibfnamefont {N.}~\bibnamefont {Kirkwood}}, \bibinfo {author}
  {\bibfnamefont {D.}~\bibnamefont {Wen}}, \bibinfo {author} {\bibfnamefont
  {S.}~\bibnamefont {Wang}}, \bibinfo {author} {\bibfnamefont {J.}~\bibnamefont
  {Meng}}, \bibinfo {author} {\bibfnamefont {J.}~\bibnamefont {Bullock}},
  \bibinfo {author} {\bibfnamefont {K.~B.}\ \bibnamefont {Crozier}},\ and\
  \bibinfo {author} {\bibfnamefont {L.}~\bibnamefont {Sciacca}},\ }\bibfield
  {title} {\bibinfo {title} {{Infrared modulation via near-room-temperature
  phase transitions of vanadium oxides {\&} core{\textendash}shell
  composites}},\ }\href {https://doi.org/10.1039/D2TA09753B} {\bibfield
  {journal} {\bibinfo  {journal} {J. Mater. Chem. A}\ }\textbf {\bibinfo
  {volume} {11}},\ \bibinfo {pages} {7629} (\bibinfo {year}
  {2023})}\BibitemShut {NoStop}%
\bibitem [{\citenamefont {Wolf}(1959)}]{Wolf1959}%
  \BibitemOpen
  \bibfield  {author} {\bibinfo {author} {\bibfnamefont {E.}~\bibnamefont
  {Wolf}},\ }\bibfield  {title} {\bibinfo {title} {{Electromagnetic diffraction
  in optical systems. I. An integral representation of the image field}},\
  }\href {https://doi.org/10.1098/rspa.1959.0199} {\bibfield  {journal}
  {\bibinfo  {journal} {Proc. Roy. Soc. A}\ }\textbf {\bibinfo {volume}
  {253}},\ \bibinfo {pages} {349} (\bibinfo {year} {1959})}\BibitemShut
  {NoStop}%
\bibitem [{\citenamefont {Richards}\ and\ \citenamefont
  {Wolf}(1959)}]{RichardsWolf1959}%
  \BibitemOpen
  \bibfield  {author} {\bibinfo {author} {\bibfnamefont {B.}~\bibnamefont
  {Richards}}\ and\ \bibinfo {author} {\bibfnamefont {E.}~\bibnamefont
  {Wolf}},\ }\bibfield  {title} {\bibinfo {title} {{Electromagnetic diffraction
  in optical systems. II. Structure of the image field in an aplanatic
  system}},\ }\href {https://doi.org/10.1098/rspa.1959.0200} {\bibfield
  {journal} {\bibinfo  {journal} {Proc. Roy. Soc. A}\ }\textbf {\bibinfo
  {volume} {253}},\ \bibinfo {pages} {358} (\bibinfo {year}
  {1959})}\BibitemShut {NoStop}%
\bibitem [{\citenamefont {Lenton}\ \emph {et~al.}(2020)\citenamefont {Lenton},
  \citenamefont {Nieminen}, \citenamefont {Loke}, \citenamefont {Stilgoe},
  \citenamefont {Hu}, \citenamefont {Kn{\ifmmode\ddot{o}\else\"{o}\fi}ner},
  \citenamefont {Bra{\ifmmode\acute{n}\else\'{n}\fi}czyk}, \citenamefont
  {Heckenberg},\ and\ \citenamefont {Rubinsztein-Dunlop}}]{Lenton2020OTT}%
  \BibitemOpen
  \bibfield  {author} {\bibinfo {author} {\bibfnamefont {I.~C.~D.}\
  \bibnamefont {Lenton}}, \bibinfo {author} {\bibfnamefont {T.~A.}\
  \bibnamefont {Nieminen}}, \bibinfo {author} {\bibfnamefont {V.~L.~Y.}\
  \bibnamefont {Loke}}, \bibinfo {author} {\bibfnamefont {A.~B.}\ \bibnamefont
  {Stilgoe}}, \bibinfo {author} {\bibfnamefont {Y.}~\bibnamefont {Hu}},
  \bibinfo {author} {\bibfnamefont {G.}~\bibnamefont
  {Kn{\ifmmode\ddot{o}\else\"{o}\fi}ner}}, \bibinfo {author} {\bibfnamefont
  {A.~M.}\ \bibnamefont {Bra{\ifmmode\acute{n}\else\'{n}\fi}czyk}}, \bibinfo
  {author} {\bibfnamefont {N.~R.}\ \bibnamefont {Heckenberg}},\ and\ \bibinfo
  {author} {\bibfnamefont {H.}~\bibnamefont {Rubinsztein-Dunlop}},\ }\href@noop
  {} {\bibinfo {title} {Optical tweezers toolbox}},\ \bibinfo {howpublished}
  {\url{https://github.com/ilent2/ott}} (\bibinfo {year} {2020})\BibitemShut
  {NoStop}%
\bibitem [{\citenamefont {Mizuyama}(2018)}]{Mizuyama2018COMSOLBlog}%
  \BibitemOpen
  \bibfield  {author} {\bibinfo {author} {\bibfnamefont {Y.}~\bibnamefont
  {Mizuyama}},\ }\href
  {https://www.comsol.com/blogs/the-nonparaxial-gaussian-beam-formula-for-simulating-wave-optics/}
  {\bibinfo {title} {{The Nonparaxial Gaussian Beam Formula for Simulating Wave
  Optics}}},\ \bibinfo {howpublished} {COMSOL Blog} (\bibinfo {year}
  {2018})\BibitemShut {NoStop}%
\bibitem [{COM(2024)}]{COMSOL63WaveOptics}%
  \BibitemOpen
  \href
  {https://doc.comsol.com/6.3/doc/com.comsol.help.woptics/woptics_ug_optics.6.87.html}
  {\emph {\bibinfo {title} {{Wave Optics Module User's Guide, COMSOL
  Multiphysics 6.3}}}},\ \bibinfo {organization} {COMSOL AB},\ \bibinfo
  {address} {Stockholm, Sweden} (\bibinfo {year} {2024}),\ \bibinfo {note}
  {section ``Gaussian Beams as Background Fields and Input
  Fields''}\BibitemShut {NoStop}%
\bibitem [{tof(2026)}]{toftul2026github_nonparax}%
  \BibitemOpen
  \href {https://github.com/toftul/nonparax/tree/main} {\bibinfo {title}
  {{toftul/nonparax: Exact non-paraxial Gaussian beams: white paper and Python
  reference code}}} (\bibinfo {year} {2026}),\ \bibinfo {note} {[Online;
  accessed 24. Sep. 2026]}\BibitemShut {NoStop}%
\bibitem [{Note1()}]{Note1}%
  \BibitemOpen
  \bibinfo {note} {Octupole terms were derived with the help of Claude Code
  Opus 5 model developed by Anthropic}\BibitemShut {NoStop}%
\bibitem [{\citenamefont {Xu}\ \emph {et~al.}(2024)\citenamefont {Xu},
  \citenamefont {{Nieto-Vesperinas}}, \citenamefont {Zhou}, \citenamefont
  {Zhang}, \citenamefont {Li}, \citenamefont {{Rodr{\'i}guez-Fortu{\~n}o}},
  \citenamefont {Yan},\ and\ \citenamefont {Yao}}]{Xu2024NC}%
  \BibitemOpen
  \bibfield  {author} {\bibinfo {author} {\bibfnamefont {X.}~\bibnamefont
  {Xu}}, \bibinfo {author} {\bibfnamefont {M.}~\bibnamefont
  {{Nieto-Vesperinas}}}, \bibinfo {author} {\bibfnamefont {Y.}~\bibnamefont
  {Zhou}}, \bibinfo {author} {\bibfnamefont {Y.}~\bibnamefont {Zhang}},
  \bibinfo {author} {\bibfnamefont {M.}~\bibnamefont {Li}}, \bibinfo {author}
  {\bibfnamefont {F.~J.}\ \bibnamefont {{Rodr{\'i}guez-Fortu{\~n}o}}}, \bibinfo
  {author} {\bibfnamefont {S.}~\bibnamefont {Yan}},\ and\ \bibinfo {author}
  {\bibfnamefont {B.}~\bibnamefont {Yao}},\ }\bibfield  {title} {\bibinfo
  {title} {Gradient and curl optical torques},\ }\href
  {https://doi.org/10.1038/s41467-024-50440-8} {\bibfield  {journal} {\bibinfo
  {journal} {Nat Commun}\ }\textbf {\bibinfo {volume} {15}},\ \bibinfo {pages}
  {6230} (\bibinfo {year} {2024})}\BibitemShut {NoStop}%
\bibitem [{\citenamefont {Wei}\ and\ \citenamefont
  {Rodr{\ifmmode\acute{\imath}\else\'{\i}\fi}guez-Fortu{\ifmmode\tilde{n}\else\~{n}\fi}o}(2022)}]{Wei2022PRB}%
  \BibitemOpen
  \bibfield  {author} {\bibinfo {author} {\bibfnamefont {L.}~\bibnamefont
  {Wei}}\ and\ \bibinfo {author} {\bibfnamefont {F.~J.}\ \bibnamefont
  {Rodr{\ifmmode\acute{\imath}\else\'{\i}\fi}guez-Fortu{\ifmmode\tilde{n}\else\~{n}\fi}o}},\
  }\bibfield  {title} {\bibinfo {title} {{Optical multipolar torque in
  structured electromagnetic fields: On helicity gradient torque, quadrupolar
  torque, and spin of the field gradient}},\ }\href
  {https://doi.org/10.1103/PhysRevB.105.125424} {\bibfield  {journal} {\bibinfo
   {journal} {Phys. Rev. B}\ }\textbf {\bibinfo {volume} {105}},\ \bibinfo
  {pages} {125424} (\bibinfo {year} {2022})}\BibitemShut {NoStop}%
\bibitem [{\citenamefont {Pesce}\ \emph {et~al.}(2020)\citenamefont {Pesce},
  \citenamefont {Jones}, \citenamefont {Marag{\`o}},\ and\ \citenamefont
  {Volpe}}]{Pesce2020EPJP}%
  \BibitemOpen
  \bibfield  {author} {\bibinfo {author} {\bibfnamefont {G.}~\bibnamefont
  {Pesce}}, \bibinfo {author} {\bibfnamefont {P.~H.}\ \bibnamefont {Jones}},
  \bibinfo {author} {\bibfnamefont {O.~M.}\ \bibnamefont {Marag{\`o}}},\ and\
  \bibinfo {author} {\bibfnamefont {G.}~\bibnamefont {Volpe}},\ }\bibfield
  {title} {\bibinfo {title} {Optical tweezers: Theory and practice},\ }\href
  {https://doi.org/10.1140/epjp/s13360-020-00843-5} {\bibfield  {journal}
  {\bibinfo  {journal} {Eur. Phys. J. Plus}\ }\textbf {\bibinfo {volume}
  {135}},\ \bibinfo {pages} {949} (\bibinfo {year} {2020})}\BibitemShut
  {NoStop}%
\bibitem [{\citenamefont {Jacobs}(2013)}]{Jacobs2013}%
  \BibitemOpen
  \bibfield  {author} {\bibinfo {author} {\bibfnamefont {K.}~\bibnamefont
  {Jacobs}},\ }\href@noop {} {\emph {\bibinfo {title} {Stochastic {{Processes}}
  for {{Physicists}}: {{Understanding Noisy Systems}}}}},\ Vol.~\bibinfo
  {volume} {1}\ (\bibinfo  {publisher} {Cambridge University Press},\ \bibinfo
  {year} {2013})\BibitemShut {NoStop}%
\end{thebibliography}%

\end{document}